\documentclass[sigconf]{acmart}
\usepackage{xcolor,colortbl}

\usepackage{booktabs} 
\usepackage[section]{placeins}
\usepackage{amsmath}
\usepackage{amsfonts}
\usepackage{amssymb}
\usepackage{graphicx}
\usepackage{url}
\usepackage{subfig}
\usepackage[subtle]{savetrees}

\newcommand{\goes}[1]{\xrightarrow{#1}}

\newcommand{\sig}{\mathbf{s}}
\newcommand{\disc}{d}

\newtheorem{problem}{Problem}

\setcopyright{none}

\acmConference[]{}{}{} 
\acmYear{}
\copyrightyear{}

\RequirePackage[normalem]{ulem} 
\RequirePackage{color}\definecolor{RED}{rgb}{1,0,0}\definecolor{BLUE}{rgb}{0,0,1} 

\newcommand{\newstuff}[1]{#1}

\begin{document}
\title[Synthesizing Stealthy Reprogramming Attacks on Cardiac Devices]{Synthesizing Stealthy Reprogramming Attacks \\ on Cardiac Devices}

\author{Nicola Paoletti}
\affiliation{\institution{Department of Computer Science, \\Royal~Holloway, University of London, UK}}

\author{Zhihao Jiang}
\affiliation{\institution{School of Information Science and Technology, ShanghaiTech University, China}}

\author{Md Ariful Islam}
\affiliation{\institution{Department of Computer Science,\\Texas Tech University, USA}}

\author{Houssam~Abbas, Rahul Mangharam}
\affiliation{\institution{Department of \\Electrical and Systems Engineering, University of Pennsylvania, USA}}

\author{Shan Lin}
\affiliation{\institution{Department of Electrical and Computer Engineering, Stony Brook University, USA}}

\author{Zachary~Gruber, Scott A.~Smolka}
\affiliation{\institution{Department of Computer Science,\\Stony Brook University, USA}}
%
%
\renewcommand{\shortauthors}{N. Paoletti et al.}

\begin{abstract}
An Implantable Cardioverter Defibrillator (ICD) is a medical device used for the detection of potentially fatal cardiac arrhythmia and their treatment through the delivery of electrical shocks intended to restore normal heart rhythm.  An ICD \emph{reprogramming attack} seeks to alter the device's parameters to induce unnecessary shocks and, even more egregious, prevent required therapy.  In this paper, we present a formal approach for the synthesis of ICD reprogramming attacks that are both \emph{effective}, i.e., lead to fundamental changes in the required therapy, and \emph{stealthy}, i.e., involve minimal changes to the nominal ICD parameters.  We focus on the \emph{discrimination algorithm} underlying Boston Scientific devices (one of the principal ICD manufacturers) and formulate the synthesis problem as one of multi-objective optimization.  Our solution technique is based on an Optimization Modulo Theories encoding of the problem and allows us to derive device parameters that are optimal with respect to the effectiveness-stealthiness tradeoff (i.e., lie along the corresponding Pareto front).  To the best of our knowledge, our work is the first to derive systematic ICD reprogramming attacks designed to maximize therapy disruption while minimizing detection.  To evaluate our technique, we employ an extensive dataset of synthetic EGMs (cardiac signals), each generated with a prescribed arrhythmia, allowing us to synthesize attacks tailored to the victim's cardiac condition.  Our approach readily generalizes to unseen signals, representing the unknown EGM of the victim patient.
\end{abstract}

\maketitle              

\section{Introduction}

An \emph{Implantable Cardioverter Defibrillator} (ICD) is a medical device for the detection and treatment of potentially fatal heart conditions such as ventricular tachycardia (VT) and ventricular fibrillation (VF).
ICDs run embedded software that processes intracardiac signals, called \emph{electrograms} (EGMs), to detect arrhythmias and deliver appropriate therapy in the form of electrical shocks.  ICD software implements so-called \emph{discrimination algorithms} which comprise multiple discrimination criteria (discriminators) for the detection and classification of arrhythmia episodes based on the analysis of EGM features such as ventricular intervals and signal morphology.  

ICD discriminators feature a number of programmable parameters that, if adequately configured, ensure minimal rates of mis-classification and inappropriate/missed therapy~\cite{moss2012reduction}. In contrast, wrongly configured parameters can result in unnecessary shocks, which are painful and damage the cardiac tissue, and even worse can prevent required therapy, leading to sudden cardiac death. 

An ICD \emph{reprogramming attack} is one that alters the device's parameters to induce mis-classification and inappropriate therapy. 
Reprogramming attacks can significantly compromise patient safety, with high-profile patients being obvious targets (e.g.\ former US Vice President Cheney had his pacemaker's wireless access disabled to prevent assassination attempts~\cite{peterson2013yes}). Seminal work by Halperin et al.~\cite{halperin2008pacemakers} demonstrated that ICDs can be accessed and reprogrammed by unauthorized users using off-the-shelf software radios. More recently, over half a million cardiac devices have been recalled by the FDA for security risks related to wireless communication~\cite{fda_recall}, \newstuff{and researchers managed to gain control of a pacemaker/ICD by exploiting vulnerabilities in the device's remote monitoring infrastructure~\cite{rios_black_hat}}. These incidents confirm that vulnerabilities in implantable cardiac devices exist, and a thorough investigation of cyber-attacks on ICDs is needed to improve device safety and security.

In this paper, we present a formal approach for the automated synthesis of ICD reprogramming attacks that are both \textit{effective}, i.e., lead to fundamental changes in the required therapy, and \textit{stealthy}, i.e., involve minimal changes to the nominal ICD parameters.  Stealthy attacks are therefore difficult to detect and even if detected, would most likely be attributed to a clinician's error in configuring the device. 

We follow a model-based approach, as the attacks are not evaluated on the actual hardware but on a model of the ICD algorithm.  We focus on the \textit{Rhythm ID} algorithm implemented in Boston Scientific ICDs (one of the principal ICD manufacturers), which was compiled from device manuals and the medical literature~\cite{bs_reference_guide1,zanker2016tachycardia}. 
The discriminators used and computations performed by Rhythm ID are also found in the algorithms of the three other major ICD manufacturers. Thus, focusing on Rhythm ID does not limit the applicability of our approach.

\begin{figure}
\centering
\includegraphics[width=.9\columnwidth]{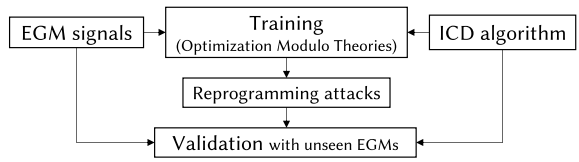}
\caption{Overview of our method for synthesis of stealthy reprogramming attacks on ICDs.}
\label{fig:overview}
\vspace*{-.3cm}
\end{figure}

Our method, illustrated in Figure~\ref{fig:overview}, synthesizes device parameters that are optimal with respect to the effectiveness-stealthiness tradeoff (i.e., lie along the corresponding Pareto front). We formulate this synthesis problem as one of multi-objective optimization, and solve it using \textit{optimization modulo theories} (OMT) techniques~\cite{bjorner2015nuz}, an extension of SMT for finding models that optimize given objectives. 
OMT is uniquely suited to solve this problem, because the problem is combinatorial in nature (parameters can be configured from a finite set of values), and is also constrained by the behavior of the ICD algorithm, which can be adequately encoded as SMT constraints. 

The synthesized reprogramming attacks yield optimal effectiveness and stealthiness with respect to a set of \textit{training EGM signals}.  We employ the method of~\cite{jiang2016silico} to generate synthetic EGMs with prescribed arrhythmia.  This allows the attacker to synthesize malicious parameters tailored to the victim's cardiac condition. 


In summary, our main contributions are the following.
\begin{itemize}
\item We introduce, to the best of our knowledge, the first method for the derivation of systematic reprogramming attacks on cardiac devices designed to maximize therapy disruption while minimizing the likelihood of detection.
\item We formulate the problem of synthesizing malicious parameters as a multi-objective optimization problem.
\item We present a method, based on OMT techniques and an efficient SMT encoding of the ICD algorithm, for precisely solving this optimization problem. 
\item We evaluate the method by synthesizing attacks tailored to 19 different arrhytmias (i.e., \emph{condition-specific} attacks), as well as more generic attacks (\emph{condition-agnostic}) that are suitable when the attacker has little knowledge of the victim's condition. Our results demonstrate that arrhythmogenic conditions are particularly vulnerable as only minor changes to the detection thresholds are sufficient to prevent the required therapy.
\item We show that our approach is suitable for real-world attacks as it readily generalizes to unseen signals (i.e., \emph{test EGMs}), representing the unknown EGMs of the victim patient. 
\end{itemize}

\section{Background}\label{sec:background}
\begin{figure}
\centering
\includegraphics[width=\columnwidth]{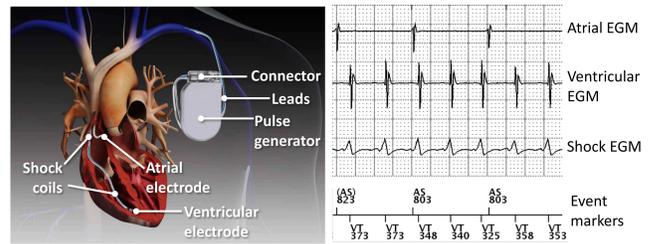}
\caption{\textbf{Left:} illustration of a dual-chamber ICD.
Original picture by Zenh\"{a}usern \& Partner, CC BY-SA 3.0 CH. 
\textbf{Right:} sensed atrial, ventricular and shock electrograms. Event markers label sensed impulses (AS: atrial, VT: ventricular tachycardia) and corresponding intervals in milliseconds.}\label{fig:icd_pic}
\vspace*{-.5cm}
\end{figure}

ICDs are battery-powered devices implanted under the pectoral muscles in the chest and connected to the cardiac muscle through one (in single-chamber ICDs) or two (dual-chamber) leads that sense the electrical activity of the heart and deliver life-saving electrical defibrillation shocks when dangerous arrhythmia is detected (see Figure~\ref{fig:icd_pic}). Shocks are delivered through shocking coils located along the ventricular lead. To improve the battery lifetime and the discomfort to the patient, modern ICDs first attempt a so-called anti-tachycardia pacing (ATP), consisting of a burst of low-voltage impulses to the ventricle, resorting to a high-energy shock only if ATP fails. ICDs also incorporate the functionality of pacemakers, i.e., they detect slow heart rhythm and correct it by delivering low-voltage electrical impulses, but in this work, we focus only on the component responsible for detecting and terminating tachycardia.

Sensed electrical signals are called \textit{intracardiac electrograms (EGMs)}, which in a dual-chamber ICD are of three types (see Figure~\ref{fig:icd_pic}): \textit{atrial and ventricular EGMs}, describing the local, near-field electrical activity in the right atrium and ventricle, respectively; and the \textit{shock EGM}, a far-field signal that gives a global view of the electrical activity, measured from the shock coil to the ICD can. 

ICD discrimination algorithms are responsible for detecting tachycardia episodes and initiating adequate therapy based on the sensed EGMs. These algorithms are embedded in the device and employ signal-processing methods such as peak detection to identify cardiac events; viz.\ electrical activation of the atria and ventricles (heart beats).  Therapy delivery depends on a number of discrimination criteria, or \emph{discriminators}, used to distinguish between potentially fatal Ventricular Tachy-arrhythmias (VT) and non-fatal Supra-Ventricular Tachy-arrhythmias (SVTs). 

Since an ICD only has three signals, there are a limited number of features that can be used as discriminators.  Atrial rate, ventricular rate, and far-field ventricular morphology are the core features that all major ICD manufacturers employ; see~\cite{interv_EP} for further details on the physiological meaning of these features.  To generalize to a large variety of physiological conditions and to avoid "over-fitting" the algorithm to known conditions, device manufacturers have adopted simple decision tree-like structures and simple discriminators to distinguish between SVT and VT.

\subsection{ICD Discrimination Algorithm}
\begin{figure}
\centering
\includegraphics[width = .95\columnwidth]{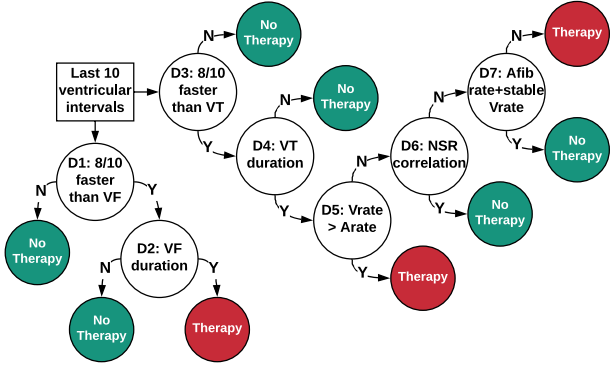}
\caption{Discrimination tree of the Boston Scientific \textit{Rhythm ID} algorithm. White nodes denote discrimination criteria. Any sequence of decisions eventually leads to either delivering (red) or not delivering (green) the therapy.}
\label{fig:rhythmID} 
\vspace*{-.7cm}
\end{figure}

Figure~\ref{fig:rhythmID} illustrates the \textit{Rhythm ID} algorithm implemented in Boston Scientific (BSc) ICDs.  The algorithm consists of a number of discriminators arranged in a decision tree-like structure, where each discriminator depends on one or more programmable parameters.  Leaves of the tree determine whether or not therapy is delivered during the current heart cycle. 

The parameters of the algorithm are given in Table~\ref{tbl:BS_params}. 
We consider the description of the Rhythm ID algorithm by Jiang et al.~\cite{jiang2016silico}, where the authors provided a MATLAB implementation of the algorithm based on the manufacturer's manuals and the medical literature~\cite{bs_reference_guide1,zanker2016tachycardia}. This implementation faithfully captures the behavior of the Rhythm ID algorithm, as it was validated by demonstrating  conformance to a BSc commercial ICD device on 11 test cases.
%
%
The  algorithm and its discriminators, described next, are executed at each ventricular event, which marks the end of the corresponding heart cycle. 

\newcommand{\mypara}[1]{\vspace*{3pt} \noindent \textbf{#1}}
\mypara{D1, 8/10 faster that VF}: this discriminator is true iff at least eight out of the last ten ventricular intervals (i.e., the time between two consecutive ventricular beats) are shorter than the programmable threshold $\mathrm{VF_{th}}$. This discriminator detects the onset of arrhythmia (VF in this case), as a high ventricular rate is a strong indication of VF.  If \textbf{D1} is true, therapy is delivered only if the VF episode is sustained.  To check if VF persists, the algorithm starts the so-called \emph{VF duration timer}, as described in discriminator~\textbf{D2}.

\mypara{D2, VFduration}: when in VF duration mode, the algorithm checks that at least six out of the last ten ventricular intervals are below $\mathrm{VF_{th}}$, and that the last interval is below $\mathrm{VF_{th}}$.  If this criterion is not met, the algorithm exits the VF duration mode as the episode did not persist, and thus requires no therapy.  If this criterion stays true for the entire VF duration (parameter $\mathrm{VFdur}$), then therapy is given.

\mypara{D3, 8/10 faster that VT}: this criterion is analogous to \textbf{D1}, but uses the VT threshold $\mathrm{VT_{th}}$.

\mypara{D4, VTduration}: this criterion is analogous to \textbf{D2}, but uses the VT threshold $\mathrm{VT_{th}}$ and the duration parameter $\mathrm{VTdur}$.  The difference with \textbf{D2} is that in this case, therapy is not given immediately at the end of the duration timer; rather, the algorithm ensures that the episode is not mistaken for SVT, as illustrated below.

\mypara{D5, V rate $>$ A rate}: it is true iff over the last ten heart cycles, the average ventricular rate is at least 10 BPM faster the average atrial rate.  If true, \textbf{D5} indicates that tachycardia originated in the ventricles and thus must be treated.  Otherwise, the algorithm inspects \textbf{D6} and \textbf{D7}.

\mypara{D6, NSR correlation}: this criterion, also called \textit{Rhythm Match}, compares the morphology of the far-field shock EGM with that of a pre-computed normal sinus rhythm (NSR) template.  The two signals being similar suggests that the arrhythmia originated in the atria, indicating SVT (no therapy).  In particular, for at least three out of the last ten heart cycles, the two signals should have a so-called feature correlation coefficient (FCC) greater than parameter $\mathrm{NSRcor_{th}}$.  The FCC is computed by looking at the voltages of the two signals at prescribed time-points. See~\cite{bs_reference_guide1} for more details on the computation of the FCC.


\mypara{D7, AFib rate and stable Vrate}: if \textbf{D6} does not hold, \textbf{D7} makes the final decision on the therapy.  In particular, the device diagnoses SVT if at least six out of the last ten atrial intervals are shorter than threshold $\mathrm{AFib_{th}}$ (suggesting that the tachycardia originated in the atria) and the ventricular rhythm is stable, i.e., the last ten ventricular intervals have variance below parameter $\mathrm{stb}$.  Otherwise, VT is diagnosed and therapy is initiated.


\begin{table}
\centering
\setlength{\tabcolsep}{0.3em}
\begin{footnotesize}
\begin{tabular}{llp{.47\columnwidth}}
{Name} & {Description} & \textbf{Nominal} (Programmable)\\ \hline
$\mathrm{VF_{th}}$ (BPM) & VF detection threshold & $\bf 200$ ($110:5:210,$ $220:10:250$)\\
$\mathrm{VT_{th}}$ (BPM) & VT detection threshold & $\bf 160$ ($90:5:210,220$)\\
$\mathrm{AFib_{th}}$ (BPM) & AFib detection threshold & $\bf 170$ ($100:10:300$) \\
$\mathrm{VFdur}$ (s) & Sustained VF duration & $\bf 1.0$ ($1:0.5:5, 6:1:15$)\\
$\mathrm{VTdur}$ (s) & Sustained VT duration & $\bf 2.5$ ($1:0.5:5, 6:1:15,$ $20:5:30$)\\
$\mathrm{NSRcor_{th}}$ &  Rhythm Match score & $\bf 0.94$ ($0.7:0.01:0.96$)\\
$\mathrm{stb}$ (ms$^2$) & Stability score & $\bf 20$ ($6:2:32,35:5:60,70:10:120$) \\ \hline
\end{tabular}
\end{footnotesize}
\caption{Parameters of the Rhythm ID algorithm, including nominal and programmable values~\cite{bs_reference_guide1}. AFib: atrial fibrillation. $n:k:m$ denotes the sequence $n,n+k, n+2k, \ldots, m$. Thresholds are programmed in BPM (beats per minute) but the algorithm employs the corresponding time duration.}\label{tbl:BS_params}
\vspace*{-.8cm}
\end{table}

The algorithm presented in Figure~\ref{fig:rhythmID} considers two tachycardia zones (VF and VT branches).  BSc ICDs can actually be configured to work with one, two, or three zones.  With three zones, the algorithm would have an additional branch (called VT-1) \emph{with discriminators identical to those found in the VT branch} but with different parameters (lower detection rate and longer VT duration).  We focus on two zones because it is the most common configuration, and the two-zone attack can easily be extended to handle one more or one less zone. 

BSc ICDs support setting a separate post-therapy configuration of the parameters to check if therapy was successful.  This is not part of the discrimination algorithm we consider because our reprogramming attacks are not concerned with post-therapy analysis.  We could have easily incorporated the post-therapy phase, as it uses the same discriminators described above but with possibly different parameter values.

We reiterate that discriminators \textbf{D1}--\textbf{D7}, or slight variations thereof, are found in other ICD manufacturers' algorithms.  Thus, the attack-synthesis method presented below apply to other devices as well.

\subsection{Generation of Synthetic EGMs}\label{sect:synth_EGMs}
Discrimination algorithms utilize two elements of EGMs for feature extraction: timing of atrial and ventricular events, and morphology of far-field ventricular events.  Jiang et al.~\cite{jiang2016silico} have developed a heart model that can generate realistic synthetic EGMs that can be used to evaluate the safety and efficacy of discrimination algorithms.

The timing of heart events is generated by a timed-automata model of the electrical conduction system of the heart~\cite{jiang2012cyber}, which allows simulating cardiac dynamics under different parameter settings. 
The morphology of far-field ventricular events is sampled from a large database of real patient EGM records~\cite{AAEL}.  EGM signals are then synthesized by overlaying the sampled EGM morphology templates on the sequence of cardiac events generated by the timed model. 

Finally, different heart conditions are reproduced by running the timed-automaton model on different parameters. For example, a generic SVT condition has ventricular intervals in the range of $[280,530]$~ms; then, EGMs for a specific SVT condition are synthesized by uniformly sampling parameters from a sub-interval of this range.

Jiang et al.\ generated synthetic EGMs for the 19 heart conditions specified in the RIGHT clinical trial~\cite{RIGHT}, a trial designed to evaluate the BSc discrimination algorithm.  The validity and faithfulness of these EGMs were validated by electrophysiologists. In this paper, we therefore use the same synthetic EGM dataset. The signals are open-loop (i.e., fixed): our attack model does not require closed-loop modeling, as explained in Section~\ref{sec:attack_model}.

\vspace*{-.1cm}\section{Attack model}\label{sec:attack_model}
We present a model-based approach to synthesizing reprogramming attacks on ICDs, as the attacks are not evaluated on the actual physical device but on a model of the device.  The BSc algorithm model that we consider was compiled from device manuals and the medical literature, and faithfully reproduces the behavior of the real device in terms of arrhythmia discrimination and therapy, as discussed in Section~\ref{sec:background}.
We focus on reprogramming attacks where the attacker manipulates the parameter values of the victim's ICD with the aim of causing harm while going undetected.  These two objectives are respectively called the \textit{effectiveness} and \textit{stealthiness} of the attack, and are formalized in Section~\ref{sec:prob}.  

An attack is effective when it compromises the decision of the discrimination algorithm in such a way that the a required therapy is prevented (e.g., during VF), or an inappropriate therapy is introduced (e.g., during SVT). Our attack model is concerned with inducing at least one compromised decision, which suffices to cause adverse or even fatal effects: depriving a patient of treatment for VF can lead to sudden cardiac death, while inappropriate shocks can result in damaging heart-tissue remodeling and cause significant psychological distress~\cite{jiang2016silico}.
Note that the unaltered parameters can themselves have a low rate of inappropriate or missed therapy~\cite{RIGHT}, which is, however, negligible compared to that for malicious parameters.

In our attack model, stealthiness depends on the clinician's ability to detect the attack.  We are therefore interested in finding malicious parameters that exhibit small deviations from the clinical settings of the victim's ICD, changes that are difficult for the clinician to notice or that can be mistaken for human error. 
Indeed, the victim has no means to monitor their ICD parameters outside of clinic, and upon experiencing unusual activity by the ICD, s/he will likely seek medical aid rather than suspect a cyber-attack.  Hence, the in-clinic setting is of primary interest. 
Moreover, the victim will likely be unable detect the attacker on the spot, because a successful malicious reprogramming does not typically induce adverse outcomes immediately but with some delay, depending on the frequency that the victim experiences arrhythmia and the probability that the reprogrammed parameters mis-classify that arrhythmia.

Reprogramming attacks are synthesized in an offline \textit{training phase}, which allows the attacker to obtain malicious parameters with optimal effectiveness and stealthiness with respect to a set of training EGM signals. 
Such parameters are derived by solving a multi-objective optimization problem over a set of logical constraints describing the behavior of the discrimination algorithm over the training signals. We solve the problem through SMT-based techniques that are guaranteed to find parameters attaining the exact optimal effectiveness-stealthiness front (see Sections~\ref{sec:prob} and~\ref{sec:encoding}). 


The malicious parameters synthesized in the training phase are \emph{validated using a disjoint test dataset}. In this way, we can evaluate how the attack generalizes with previously unseen signals, which mimic the unknown EGM of the victim. 

We assume that the attacker has no knowledge of the victim's ICD parameters, and thus that their best strategy is to train the attack by assuming that the default (unaltered) parameters correspond to the nominal values (see Table~\ref{tbl:BS_params}). Therefore, the stealthiness computed under nominal parameters might deviate from that under the actual victim's parameters.  However, this discrepancy is limited by the fact that condition- or patient-specific parameters tend to be close to the nominal ones, which are considered safe for any kind of arrhythmia requiring an ICD~\cite{moss2012reduction}. I.e., nominal parameters provide a good estimate of the victim's parameters. 

Due to limited availability of real patient signals, 
we choose to work with \textit{synthetic EGMs}.  We remark, however, that our approach supports both.  The automated EGM generation method of Section~\ref{sect:synth_EGMs} gives the attacker a crucial advantage: if the attacker knows that the victim is affected by a specific arrhythmia, then it can tailor the attack to the victim in question by generating a training dataset of synthetic signals with that arrhythmia. 
In our evaluation, we consider training datasets tailored to specific conditions (\textit{condition-specific attacks}) as well as more generic datasets that include signals for different arrhythmias (\textit{condition-agnostic attacks}).  The latter are suitable when the attacker has little knowledge of the victim's condition.

Open-loop EGM signals are adequate for our purposes because 
successful attacks do not affect the signals in a significant way: when the attack prevents a required shock for an EGM with arrhythmia, the arrhythmia persists and the EGM is unaffected; when the attack introduces inappropriate shocks during an already normal heart rhythm, the EGM is also unaffected, as shocks restore the electrical activity of the heart to normal sinus rhythm.

\vspace*{-.15cm}\paragraph{Real-world attacks.} We discuss additional assumptions that would make our model-based method suitable to real-world attacks using radio signals via software-defined radios. 

Firstly, the attacker must know the ICD model of the victim, so that it can select the appropriate discrimination algorithm to use in the training phase. The ICD model can be revealed by sending discovery signals to the device (as shown in~\cite{halperin2008pacemakers}), or from the victim's medical records. To change the parameter settings, the attacker also must know the communication protocol of the ICD, which can be reverse-engineered as also shown in~\cite{halperin2008pacemakers}. 
In our work, we focus on a single discrimination algorithm. Due, however, to the universality of discriminators, our approach can be easily adapted to other algorithms.

Secondly, the radio antenna transmitting the attack signals must be physically close to the victim. To do so, the attacker could approach the victim (e.g., in a crowded space) or hide/disguise the transmitter and leave it running in proximity of the victim.


\vspace*{-.15cm}\paragraph{Countermeasures.} Previous studies have proposed methods for preventing attacks on implantable medical devices, but to date, none of these have been put in place by device manufacturers. A solution is securing device accesses through an authentication token (smart card, NFC device, etc.) that shares a secret key with the device. The patient would provide these credentials to grant the clinician access to the device. To further secure the authentication, the key could be derived from some of the patient's biometrics, such as the electrocardiogram~\cite{xu2011imdguard}. In emergency situations where the token might not be available, one could restrict access from devices only at very close proximity, as done in~\cite{rasmussen2009proximity}. Finally, a simple detection method would be notifying the patient with a beep whenever a communication happens with the device~\cite{halperin2008pacemakers}.

\section{Problem Formulation}\label{sec:prob}

We formalize the synthesis of reprogramming attacks, (which corresponds to the training phase), as a multi-objective optimization problem that seeks to derive ICD parameters achieving two main (and contrasting) objectives: \textit{effectiveness}, i.e., the attack must maximize therapy disruption; and \textit{stealthiness}, that is, the attack must be difficult to detect. 


For a set $X$, $X^*$ denotes the Kleene closure of $X$.  For a sequence $\mathbf{x} \in X^*$, $|\mathbf{x}|$ denotes its length and, for $k = 0, \ldots, |\sigma| -1$, $\mathbf{x}[k] \in X$ denotes its $k+1$-st element. 
Let $\mathsf{Sig} \subseteq {\mathbb{R}^m}^*$ be the set of $m$-dimensional, finite-length, discrete-time \textit{cardiac signals}.  For signal $\sig \in \mathsf{Sig}$, $\sig[k]$ correspond to the values of atrial, ventricular and shock EGMs ($m=3$) at the $k+1$-st sample of the signal. 

Parameters are tuples $\mathbf{p}=(p_1,\ldots,p_n)$, where $p_i \in P_i$ is the value of the $i$-th ICD parameter, and $P_i$ is its finite domain (for each parameter there is a finite set of programmable values -- see Table~\ref{tbl:BS_params}). We denote with $\mathbb{P} = \bigtimes_{i=1}^n P_i$ the set of possible parameterizations.

From an abstract viewpoint, we can characterize a {\it discrimination algorithm} as a function $\disc: \mathbb{P} \goes{} (\mathsf{Sig} \goes{} \mathbb{B}^*)$, where $\mathbb{B}^*$ is the set of Boolean sequences. For parameters $\mathbf{p}\in \mathbb{P}$ and signal $\sig \in \mathsf{Sig}$, $\disc(\mathbf{p})(\sig)$ is a Boolean-valued sequence called a \textit{therapy signal}, with as many elements as the number of heart cycles in $\sig$.  For $k < |\disc(\mathbf{p})(\sig)|$, $\disc(\mathbf{p})(\sig)[k]$ is true if the ICD requires delivering therapy at the $k$-th heart cycle, and is false otherwise.  Recall from Section~\ref{sec:background} that the discrimination algorithm is only invoked at each ventricular event (corresponding to the end of the heart cycle), and thus, intermediate time points between two ventricular events are not relevant to studying therapy decisions. 
Note that we do not consider ICD parameters that affect the detection of ventricular events, meaning that the length of the therapy signal $\disc(\mathbf{p})(\sig)$ stays the same for any $\mathbf{p}\in \mathbb{P}$.


\paragraph{Effectiveness.} 
Let $\mathbf{p}^* = (p^*_1,\ldots,p^*_n) \in \mathbb{P}$ be the default parameters of ICD algorithm $\disc$, and $\mathbf{p} =(p_1,\ldots,p_n) \in \mathbb{P}$ be an attack parameter. The effectiveness of $\mathbf{p}$ is evaluated over an input dataset of signals $S \subseteq \mathsf{Sig}$ (either training or test dataset), and is denoted by $f_e(\mathbf{p},S)$.

Following our description of the attack model, we define effectiveness as the proportion of signals in $S$ where the attack prevents the ICD from delivering any therapy when, without the attack, it would deliver some, and forces the ICD to deliver some therapy when, without the attack, it would deliver none:

\begin{equation}\label{eq:cnt_rth}
f_e(\mathbf{p},S)=\frac{1}{|S|}\cdot \sum_{\sig \in S} I\left(R_{th}(\disc,\mathbf{p},\sig) \neq R_{th}(\disc,\mathbf{p}^*,\sig) \right),
\end{equation}
where $I$ is the indicator function, and $R_{th}(\disc,\mathbf{p},\sig)$ is the \emph{therapy reachability} value, describing whether or not therapy is administered at any point for signal $\sig$ and parameters $\mathbf{p}$:
\begin{equation}\label{eq:rth}
R_{th}(\disc,\mathbf{p},\sig) = \bigvee_{k=0}^{|\disc(\mathbf{p})(\sig)|-1} \disc(\mathbf{p})(\sig)[k].
\end{equation}

Therapy reachability is motivated by the fact that we employ synthetic EGMs reflecting a number of arrhythmogenic (VF/VT-like) and non-arrhythmogenic (SVT-like) situations, with the former requiring device-delivered therapy and the latter requiring that such therapy not be delivered.  We deem an attack successful on an EGM if the EGM is mis-classified in this manner. 

In practice, attacks that prevent therapy during VF or VT can be fatal (these arrhythmias can lead to sudden cardiac death~\cite{jiang2016silico}) and thus are more dangerous than attacks introducing unnecessary therapy during SVT. In our definition of effectiveness, these two cases are given the same importance to avoid excessive bias towards attacks preventing therapy. Also, VT/VF-like and SVT-like EGMs should never occur in the same set of training or test data, \newstuff{because attacks that can both prevent therapy (for VT/VF) and introduce unnecessary therapy (for SVT) are clearly impossible.}

\begin{figure}
\centering
\includegraphics[width=\columnwidth, trim=0 1.5cm 0cm 1cm, clip=true]{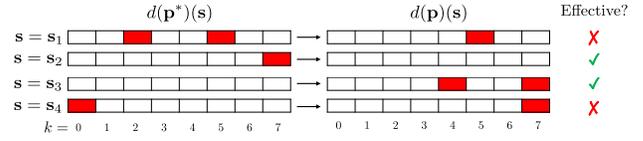}
\caption{Example of attack effectiveness. Left: therapy signal before the attack. Right: after the attack. Red cells mark heart cycles where therapy is given.}
\label{fig:eff_exmpl}
\end{figure}

Consider the example of Figure~\ref{fig:eff_exmpl} showing a set of signals $S = \{ \sig_1, \ldots \sig_4\}$ of length 8 and the corresponding therapy signals for the default ($\disc(\mathbf{p}^*)(\sig)$) and reprogrammed ($\disc(\mathbf{p})(\sig)$) parameters. For signal $\sig_1$, two therapy episodes occur at cycles $k=2$ and $k=5$, respectively. In this case, the attack is not effective as it manages to prevent only one of the two therapies. In contrast, the attack is effective for $\sig_2$ (therapy prevented) and $\sig_3$ (therapy introduced). For $\sig_4$, the attack only delays the therapy so it is not considered successful. The overall effectiveness of the attack is thus $2/4 = 0.5$.  

\paragraph{Stealthiness.} An attack is considered stealthy when the deviation between the reprogrammed parameters $\mathbf{p}$ and the default parameters $\mathbf{p}^*$ is small. To capture this deviation, we introduce a measure of \emph{parameter distance} that we seek to minimize to achieve optimal stealthiness. Since ICD parameters can be only programmed to a finite set of values, we quantify the distance between two parameters as the number of programmable values separating them. 

For $i=1,\ldots,n$, let $P_i = \left\lbrace p^i_{1}, \ldots, p^i_{n_i} \right\rbrace$ be the programmable values for the $i$-th ICD parameters. W.l.o.g. assume that the values $p^i_{1}, \ldots, p^i_{n_i}$ are ordered. Rewrite the default parameters as $\mathbf{p}^*=\left(p^1_{I^*_1},\ldots,p^n_{I^*_n}\right)$ and the attack parameters as $\mathbf{p}=\left(p^1_{I_1},\ldots,p^n_{I_n}\right)$, i.e., $I^*_i$ is the index of the element of $P_i$ corresponding to the value of the $i$-th parameter in $\mathbf{p}^*$. $I_i$ is defined in an analogous way for $\mathbf{p}$. Then, the distance between $\mathbf{p}$ and $\mathbf{p}^*$ is defined as: 

\vspace*{-5pt}
\begin{equation}\label{eq:param_dist}
f_s(\mathbf{p}) = \max_{i=1,\ldots,n} \left| I_i - I^*_i \right|.
\end{equation}

\begin{figure}
\centering
\includegraphics[width=.8\columnwidth]{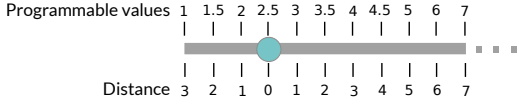}
\caption{Illustration of programmable values and corresponding distance for parameter $\mathrm{VTdur}$. The green circle corresponds to the default setting (zero distance).}
\label{fig:exmpl_stealth}
\vspace*{-.5cm}
\end{figure}

We explain (\ref{eq:param_dist}) with an example. Suppose that the $i$-th parameter is $\mathrm{VTdur}$ from Table~\ref{tbl:BS_params}, which can be programmed to any value in the set $P_i = \{1, 1.5, \ldots, 5, 6, \ldots, 15,20$, $\ldots,30\}$. We set  $\mathbf{p}^*$ using the nominal value of $2.5$ for $\mathrm{VTdur}$, which corresponds to the \mbox{4-th} element of $P_i$. Hence, $I^*_i = 4$. Consider attack parameters $\mathbf{p}$ where $\mathrm{VTdur}$ is set to $4.5$, i.e., the 8-th value of $P_i$ ($I_i = 8$). The distance relative to $\mathrm{VTdur}$ is the number of programmable values separating the default setting ($2.5$) and the attack ($4.5$), which is given by $\left| I_i - I^*_i \right| = \left| 8 - 4 \right| = 4$. Indeed, the two are separated by four programmable values ($3, 3.5, 4, 4.5$). The overall distance is the maximum separation over all ICD parameters. See Figure~\ref{fig:exmpl_stealth}.

This notion of distance assumes that parameters are equipped with a linear order, which is the case for all numeric parameters of the BSc ICD algorithm. For categorical parameters, one could either assign the same distance to all categories different from the nominal one, or repeat the synthesis for each category.

\paragraph{Optimal stealthy attacks.} We formulate the synthesis of stealthy reprogramming attacks as a multi-objective optimization problem, where we seek to optimize effectiveness and stealthiness (maximize $f_e$ and minimize $f_s$) of the parameters w.r.t.\ a set of training EGMs. Multi-objective optimization allows one to derive the optimal trade-off between multiple, possibly contrasting objectives, implying that we do not need to assume any weight or priority ordering for the objectives. The result of this analysis is a so-called \emph{Pareto front}, i.e., a set of non-dominated points in the objective space of possible effectiveness and parameter distance values. 

\begin{problem}[Reprogramming attack synthesis]\label{prob:repr}
For effectiveness objective $f_e$ and distance objective $f_s$, training set of signals $S \subseteq \mathsf{Sig}$, find the set $\mathbf{P}$ of Pareto-optimal parameters, i.e.:
\begin{multline}
\small
\mathbf{P} = \{ \mathbf{p} \in  \mathbb{P} \mid \not \exists \mathbf{p}' \in  \mathbb{P}. \ ( f_e(\mathbf{p}',S) > f_e(\mathbf{p},S) \wedge f_s(\mathbf{p}') \leq f_s(\mathbf{p}) ) \ \vee  \\
 ( f_e(\mathbf{p}',S) \geq f_e(\mathbf{p},S) \wedge f_s(\mathbf{p}') < f_s(\mathbf{p})  )\}.
\end{multline}
\end{problem}

Consider for instance two parameters $\mathbf{p}_1$ and  $\mathbf{p}_2$, such that for some $S$, $f_e(\mathbf{p}_1,S) = 0.5$, $f_e(\mathbf{p}_2,S) = 0.7$, $f_s(\mathbf{p}_1) = 5$, and $f_s(\mathbf{p}_2) = 5$. $\mathbf{p}_2$ has better effectiveness than $\mathbf{p}_1$ and same distance, so $\mathbf{p}_2$ dominates $\mathbf{p}_1$, meaning that $\mathbf{p}_1$ cannot be in the Pareto-optimal front. $\mathbf{p}_2$ is in the Pareto-optimal front if there are no parameters that dominate it. 

To quantify how well the attacks generalize to unseen data, we introduce a validation score defined as the average deviation of the attack effectiveness between training and test data.  

Given a training set $S$, a set of Pareto-optimal parameters $\mathbf{P}$ with respect to $S$, and a test set $S'$, we define the validation score as:
$\sum_{\mathbf{p} \in \mathbf{P}} (f_e(\mathbf{p},S') - f_e(\mathbf{p},S))/|\mathbf{P}|$.
Positive values indicate that the parameters $\mathbf{P}$ have better performance with unseen data than with training data, whereas negative values imply the opposite. Note that the validation score need not consider stealthiness because this is independent of the signals. 

\section{OMT Encoding}\label{sec:encoding}
\renewcommand{\implies}{\Rightarrow}
In this section, we present a solution method for the reprogramming attack synthesis problem (Problem~\ref{prob:repr}). 
We formalize the behavior of the BSc discrimination algorithm in the framework of Satisfiability Modulo Theories (SMT)~\cite{barrett2009satisfiability}, within which the ICD algorithm is described as a set of first-order formulas over some (decidable) background theory. 
Parameters are represented as uninterpreted constants in the SMT encoding, and parameter synthesis corresponds to finding a satisfiable assignment to those constants, i.e., a so-called model. 
In particular, we formulate Problem~\ref{prob:repr} as an Optimization Modulo Theories (OMT) problem, i.e., an extension of SMT for finding models that optimize given objectives~\cite{bjorner2015nuz}. 

The synthesis of optimal reprogramming attacks is difficult as it entails solving a combinatorial multi-objective optimization problem (non-continuous, non-convex) constrained by the behavior of the discrimination algorithm, which cannot be captured by simple (in)equality constraints. Therefore, classical optimization methods such as linear or convex programming are not suitable, while nonlinear optimization techniques such as genetic algorithms would provide only sub-optimal solutions. 
In contrast, OMT is uniquely suited to solve this problem, as the ICD algorithm can be adequately encoded as SMT constraints and the parameters found by OMT are guaranteed to be optimal.

Since we are interested in analyzing the behavior of the algorithm offline over a fixed set of EGM signals, we can pre-compute for each signal the non-linear operations underlying some of the discriminators, such as the Rhythm Match score. This allows us to encode the problem over the decidable theory of quantifier-free linear integer real arithmetic (SMT QF\_LIRA). Importantly, we pre-compute only the operations that are not affected by the ICD parameters, meaning that our encoding accounts for all possible behaviors induced by different parametrizations. 

W.l.o.g.\ assume that the training dataset $S$ is indexed. The behavior of the algorithm for the $j$-th signal is described by a sequence of symbolic states $s_{j,0}, \ldots s_{j,N_j}$, one for each cardiac cycle, where $N_j$ is the number of cycles in the $j$-th signal. The evolution of the discrimination algorithm over the training signals is characterized by the following formula (inspired by bounded model checking~\cite{biere1999symbolic}):

\small \vspace*{-5pt}
\begin{equation}\label{eq:BMC-like_eq}
\mathsf{paramRanges} \wedge \bigwedge_{j = 1}^{|S|} \left( \mathsf{Init}(s_{j,0}) \wedge \bigwedge_{k=0}^{N_j-1} T(k,s_{j,k},s_{j,k+1}) \right)
\end{equation}
\normalsize
where $\mathsf{paramRanges}$ is a predicate describing the programmable values of the ICD parameters (see Table~\ref{tbl:BS_params}); $\mathsf{Init}(s_{j,0})$ is the predicate for constraining the initial state of the algorithm, and $T(k,s_{j,k},s_{j,k+1})$ is the transition relation determining from the current state and heart cycle, the admissible states of the algorithm at the next cycle. In our case, the transition relation is deterministic, i.e., for fixed $s_{j,k}$ and $k$, there exists only one state $s_{j,k+1}$ such that $T(k,s_{j,k},s_{j,k+1})$ holds. In (\ref{eq:BMC-like_eq}), states $s_{j,k}$ are implicitly existentially quantified. 

In the BSc algorithm, the state $s_{j,k}$ for the $j$-th signal and $k$-th heart cycle is represented by 

\small \vspace*{-10pt}
\begin{equation*}
s_{j,k} \overset{\bf def}{=} \mathsf{(VFd_{j,k}, VTd_{j,k}, tVF_{j,k}, tVT_{j,k})} \in \mathbb{B}\times \mathbb{B}\times \mathbb{Z^{\geq}}\times \mathbb{Z^{\geq}},\end{equation*}
\normalsize
where $\sf VFd_{j,k}$ and $\sf VTd_{j,k}$ tell whether or not the algorithm is, respectively, in the VF duration and VT duration mode, with $\sf tVF_{j,k}, tVT_{j,k}$ being the clocks that keep track of time spent in the respective modes. The clocks are digital ($\in \mathbb{Z^{\geq}}$) and measure the time in milliseconds. Note that the value of the therapy signal is not part of the state but, as we shall see, is encoded as a state predicate. 

For any signal $j$, the initial state of the algorithm is given by the following $\mathsf{Init}$ predicate 

\small \vspace*{-10pt}
$$\mathsf{Init}(s_{j,0}) = {\sf \neg VFd_{j,k} \ \wedge \ \neg VTd_{j,k} \ \wedge \ tVF_{j,k}=0 \ \wedge \ tVT_{j,k}=0},$$
\normalsize
indicating that the algorithm is in neither duration mode and that the clocks are set to zeros.

An example path of the BSc algorithm encoding is given below. $s \goes{k} s'$ denotes a transition between states $s$ and $s'$ at the $k$-th heart cycle, i.e., such that $T(k,s,s')$ holds. 

\small \vspace*{-10pt}
\begin{multline*}
\ldots \goes{12} (\bot,\bot,0,0) \goes{13} (\bot,\top,0,0) \goes{14} (\bot,\top,0,309) \goes{15} \ldots \\
\ldots \goes{25} (\bot,\top,0, 2317) \goes{26} (\bot,\bot,0, 0).
\end{multline*}
\normalsize
The transition at $k=13$ marks the start of VT duration ($\sf VTd$ passes from $\bot$ to $\top$). The algorithm stays in VT duration for 13 more heart cycles during which the episode persists, until it reaches the end of the timer: at the start of the 26-th cycle the VT clock evaluates to $\sf tVF = 2317$, but at the end of the cycle, the clock would exceed the VT duration parameter which, in this example, is set to the nominal value $\mathrm{VTdur} = 2500$ milliseconds\footnote{To produce a concrete path, we must fix an interpretation for the ICD parameters.}. At this point, it delivers therapy and resets the VT clock, going back to state $(\bot,\bot,0,0)$. 

\paragraph{Transition relation.} The transition relation encodes the behavior of the discrimination algorithm presented in Section~\ref{sec:background}. 
For the sake of simplicity, we omit the signal index from the equations below. 

\small 
\begin{align}
&T(k,s_k,s_{k+1}) = \nonumber\\
& \ \left( \sf{\left( VFstart_k \wedge (\neg VFd_k \vee VFend_k)  \right) \implies VFd_{k+1}} \right) \label{eq:SMT_BS1}\\
& \ \wedge \left( \sf{\left( VTstart_k \wedge (\neg VTd_k \vee VTend_k)  \right) \implies VTd_{k+1}} \right) \label{eq:SMT_BS2}\\
& \ \wedge \left( \sf{\left( \neg VFd_k \wedge \neg VFstart_k  \right) \implies \neg VFd_{k+1}} \right) \label{eq:SMT_BS3}\\
& \ \wedge \left( \sf{\left( \neg VTd_k \wedge \neg VTstart_k  \right) \implies \neg VTd_{k+1}} \right) \label{eq:SMT_BS4}\\
& \ \wedge \left( \sf{ \left(VFd_k  \wedge \neg VFstart_k \wedge VFend_k \right) \implies \neg VFd_{k+1}} \right) \label{eq:SMT_BS5}\\
& \ \wedge \left( \sf{ \left(VTd_k  \wedge \neg VTstart_k \wedge VTend_k \right) \implies \neg VTd_{k+1}} \right) \label{eq:SMT_BS6}\\
& \ \wedge ( \sf{ (VFd_k  \wedge \neg VFend_k) \implies} \sf{ (VFd_{k+1} \wedge tVF_{k+1} = tVF_k + \mathit{Vint}_k )} ) \label{eq:SMT_BS7}\\
& \ \wedge ( \sf{ (VTd_k  \wedge \neg VTend_k) \implies} \sf{ (VTd_{k+1} \wedge tVT_{k+1} = tVT_k + \mathit{Vint}_k )} ) \label{eq:SMT_BS8}\\
& \ \wedge \left( \sf{ (\neg VFd_k  \vee VFend_k) \implies tVF_{k+1} = 0 } \right) \label{eq:SMT_BS9}\\
& \ \wedge \left( \sf{ (\neg VTd_k  \vee VTend_k) \implies tVT_{k+1} = 0 } \right) \label{eq:SMT_BS10}
\end{align}
\normalsize
\eqref{eq:SMT_BS1} establishes that VF duration starts in the next state ($\sf VFd_{k+1}$ holds) when predicate $\sf VFstart_k$ holds and we are not in VF duration ($\sf \neg VFd_k$) or the current VF duration mode just ended ($\sf VFend_k$). 
\eqref{eq:SMT_BS2} is the analogous for the VT zone. Predicate ${\sf VFstart_k}$ encodes the first discriminator of the BSc algorithm (last 8/10 ventricular intervals faster than $\mathrm{VF_{th}}$), and is defined by:

\small \vspace*{-10pt}
\begin{align}
{\sf VFstart_k} = & \left( \sum_{n=0}^9 \mathsf{ite}(\mathit{Vint}_{k-n}<\mathrm{VF_{th}},1,0) \right) \geq 8
\end{align}
\normalsize
where $\mathsf{ite}$ is the if-then-else function, and $\mathit{Vint}_{k}$ is the duration of the ventricular interval for the $k$-th cycle. Ventricular intervals are pre-computed from the input signals and thus have fixed interpretation in the SMT encoding.  Predicate $\sf VFend_k$ is defined as:

\small \vspace*{-10pt}
\begin{align}
{\sf VFend_k} = & \ {\sf VFclkOver_k}  \vee \neg {\sf VFpersist_k}, \text{ where}\\
{\sf VFclkOver_k} = & \ {\sf tVF_k} + \mathit{Vint}_{k} \geq \mathrm{VFdur}\\
{\sf VFpersist_k} = & \left( \sum_{n=1}^9 \mathsf{ite}(\mathit{Vint}_{k-n}<\mathrm{VF_{th}},1,0) \right) \geq 5  \wedge \mathit{Vint}_{k} <\mathrm{VF_{th}}.
\end{align}
\normalsize
\noindent $\sf VFend_k$ is true when the episode does not persist (${\sf VFpersist_k}$ encodes the second BSc discriminator), or when the duration expires, i.e., ${\sf VFclkOver_k}$ holds (${\sf tVF_k} + \mathit{Vint}_{k}$ is the time spent in VF duration at the end of the $k$-th cycle). 
$\mathrm{VF_{th}}$ and $\mathrm{VFdur}$ are uninterpreted constants representing the (unknown) ICD parameters to synthesize for the VF detection threshold and VF duration, respectively. 
Predicates ${\sf VTstart_k}$, ${\sf VFend_k}$, ${\sf VTpersist_k}$ and ${\sf VTclkOver_k}$ are defined in an analogous way for the VT zone.
\eqref{eq:SMT_BS3} tells that if we are not in VF duration and $\sf VFstart_k$ does not hold, then VF duration cannot start in the next state. \eqref{eq:SMT_BS4} is the analogous of (\ref{eq:SMT_BS3}) for the VT zone. 
\eqref{eq:SMT_BS5} and~\eqref{eq:SMT_BS6} handle the situation when the algorithm exits from the VF and VT duration modes, respectively, and a new duration cannot start because no new episode is detected (i.e., in the case of VF, $\sf \neg VFstart_k$ holds). \eqref{eq:SMT_BS7} and~\eqref{eq:SMT_BS8} consider the opposite situation that the algorithm stays in the VF/VT duration mode, in which case the corresponding clock is updated.
\eqref{eq:SMT_BS9} and~\eqref{eq:SMT_BS10} express that the VF and VT duration clocks are set to zero when the algorithm is outside the corresponding duration mode, or the mode has just ended. 

Finally, we introduce the predicate $\sf Th_k$ indicating whether or not therapy is administered at the $k$-th cycle, in this way providing a symbolic representation of the therapy signal, i.e., for signal $\sig$ and parameters $\mathbf{p}$, $\sf Th_k$ corresponds to $\disc(\mathbf{p})(\sig)[k]$.

\small \vspace*{-10pt}
\begin{multline}
\sf{Th_k = (VFd_k \wedge VFpersist_k \wedge VFclkOver_k) \vee} \\
\sf{\left(VTd_k \wedge VTpersist_k \wedge VTclkOver_k \wedge \left(D5_k \vee \neg (D6_k \vee D7_k)\right) \right)}.
\end{multline}
\normalsize
The formula captures the discrimination tree presented in Section~\ref{sec:background}. ${\it D5_k}, {\sf D6_k, D7_k}$ encode the last three BSc discriminators. $\it D5_k$ is true if the average ventricular rate is at least 10 BPM faster the average atrial rate. $\it D5_k$ does not depend on any ICD parameter and thus, is pre-computed for improving efficiency. $\sf D6_k$ and $\sf D7_k$ are given by: 

\small \vspace*{-10pt}
\begin{align}
{\sf D6_k} = & \left( \sum_{n=0}^9 \mathsf{ite}(\mathit{FCC}_{k-n}\geq\mathrm{NSRcor_{th}},1,0) \right) \geq 3  \\
{\sf D7_k} = & \left( \sum_{n=0}^9 \mathsf{ite}(\mathit{Aint}_{k'-n}< \mathrm{AFib_{th}} ,1,0) \right) \geq 6  \wedge \mathit{Vvar}_k \leq  \mathrm{stb}
\end{align}
\normalsize
where $\mathit{FCC}_{k}$ and $\mathit{Vvar}_{k}$ are pre-computed constants, respectively indicating the Rhythm Match score and the variance of the last 10 ventricular intervals. $\mathit{Aint}_{k'-n}$ is the pre-computed duration of the $(k'-n)$-th atrial interval, where $k'$ is the number of atrial intervals occurred within $k$ heart cycles. $\mathrm{NSRcor_{th}}$, $\mathrm{AFib_{th}}$ and $\mathrm{stb}$ are the symbolic encoding of the corresponding ICD parameters. 

\paragraph{Effectiveness and stealthiness encoding.} We show how to encode effectiveness maximization as a MaxSMT problem. For each signal $j$, we define the following soft constraint:

\small \vspace*{-5pt}
\begin{equation}\label{eq:eff_smt}
\mathsf{effective}_j = \left( \mathit{Rth}^*_j = \neg \bigvee_{k=0}^{N_j-1} {\sf Th}_k \right),
\end{equation}
\normalsize
where $\mathit{Rth}^*_j$ is the therapy reachability value (telling whether or not therapy is administered at any point) for signal $j$ and default parameters. $\mathit{Rth}^*_j$ can be pre-computed for efficiency. $\bigvee_{k=0}^{N_j-1} {\sf Th}_k$ represents the therapy reachability value for the attack parameters, and thus, $\mathsf{effective}_j$ is true if the attack disrupts the default therapy. Note that maximizing the effectiveness $f_e$ defined in (\ref{eq:cnt_rth}) is equivalent to maximizing the number of $\mathsf{effective}_j$ constraints satisfied. Hence the MaxSMT formulation.

Parameter distance is encoded as an uninterpreted integer constant to minimize, $\sf dist$. Recall that we measure distance between two parameters as the number of programmable values separating them, and that in BSc ICDs, any parameter has a finite number of numeric programmable values. It follows that $\sf dist$ has a finite domain, i.e.\ $\sf dist \in \{0, 1, \ldots, dist_{\max}\}$
\footnote{$\mathsf{dist_{\max}} = \max_{i=1,\ldots,n} \max\left\lbrace n_i - I_i^*, I_i^* - 1\right\rbrace$, where $n_i$ is the number of programmable values for the $i$-th parameter and $I_i^*$ is the index of its default value.}. 


We encode $\sf dist$ in an implicit way, that is, we do not add constraints for (\ref{eq:param_dist}) but we restrict the parameter domains conditioned on the distance value as follows:

\small \vspace*{-5pt}
\begin{equation}\label{eq:stealth_encoding}
\bigwedge_{s = 0}^{\sf dist_{\max}} \mathsf{dist} \leq s \implies \left(  \bigwedge_{i=1}^n p^i_{L} \leq  \mathsf{P}_i \leq p^i_{U} \right),
\end{equation}
\normalsize
where $\mathsf{P}_i$ is the SMT encoding of the $i$-th parameter, $L = \max\left\lbrace I_i^* - s, 1\right\rbrace$, and $U = \min\left\lbrace I_i^* + s, n_i\right\rbrace$. In other words, $p^i_{L}$ is the $s$-th closest left neighbor of $\mathsf{P}_i$'s default value, $p^i_{U}$ is its $s$-th closest right neighbor. Therefore, $p^i_{L} \leq  \mathsf{P}_i \leq p^i_{U}$ restricts the domain of $\mathsf{P}_i$ to values with distance at most $s$, from which the correctness of (\ref{eq:stealth_encoding}) follows. 

To clarify this encoding, below is shown part of the concrete instantiation of (\ref{eq:stealth_encoding}) relative to parameter $\mathrm{VTdur}$: 

\small 
\begin{align*}
&\left( \mathsf{dist} \leq 0 \implies  \left( \ldots \wedge 2500 \leq \mathrm{VTdur} \leq 2500 \wedge \ldots \right) \right) \wedge \\
&\left( \mathsf{dist} \leq 1 \implies  \left( \ldots \wedge 2000 \leq \mathrm{VTdur} \leq 3000 \wedge \ldots \right) \right) \wedge \\
&\left( \mathsf{dist} \leq 2 \implies  \left( \ldots \wedge 1500 \leq \mathrm{VTdur} \leq 3500 \wedge \ldots \right) \right) \wedge \ldots \\
\end{align*}
\normalsize

\vspace*{-13pt}
\paragraph{Synthesis of Pareto-optimal attacks.} The OMT solver returns the set of Pareto-optimal objective values, i.e., the set of all $(s,e)$ pairs such that $s = f_s(\mathbf{p})$ and $e = f_e(\mathbf{p},S)$ for some Pareto-optimal parameter $\mathbf{p} \in \mathbf{P}$ w.r.t.\ training set $S$. For each $(s,e)$, the solver computes a witness $\mathbf{p}'$ yielding that Pareto-optimal objective value. The synthesized parameters is the set of all such $\mathbf{p}'$. This implies that we synthesize a subset of $\mathbf{P}$ since the witness might not be unique, but do not exclude any $(s,e)$ in the space of Pareto-optimal objectives.

%
%
%
%

\section{Results and discussion}\label{sec:results}
We apply our method to the synthesis of condition-specific attacks. We employ synthetic EGMs for 19 different conditions, generated through the method of Section~\ref{sect:synth_EGMs}, and synthesize Pareto-optimal parameters using a training set of 100 signals for each condition. We validate the attacks with test sets of 50 signals per condition (disjoint from the training sets). In our experiments, we found that the performance with unseen test signals stays relatively constant for any training set size larger than 40; see Figure~8 in the Appendix.  Thus, 100 training signals provide a sufficiently complete representation of the signal space.  All EGMs have a duration of 30 seconds, but their lengths -- given by the number of heart cycles -- vary depending on the ventricular interval duration.

These 19 arrythmias can be broadly classified in two categories: VT and SVT. The former consists of arrythmias where the majority of signals require ICD therapy (based on the nominal parameters), and thus, it covers both VT and VF. The latter includes conditions where most of the signals do not require therapy. In particular, we have 8 VT and 11 SVT conditions, with all VT signals requiring therapy at some point and all SVT signals not requiring any therapy. 


We also synthesize condition-agnostic attacks, suitable when the attacker has little knowledge of the victim. Specifically, we consider two attacks for generic VT and SVT conditions, using training sets of 200 EGMs derived by randomly choosing among the 8 VT-like conditions and the 11 SVT conditions, respectively. We validate the two attacks with disjoint test sets of 100 signals. 

The method for generating synthetic EGMs was implemented in MATLAB. For parameter synthesis, we used the z3 SMT solver~\cite{bjorner2015nuz}.

\begin{table}
\centering
\begin{footnotesize}
\setlength{\tabcolsep}{0.3em}
\begin{tabular}{rl | llll | ll}
\multicolumn{2}{r|}{\textbf{Condition}} & \textbf{Effectiveness} & \textbf{Distance} & $|\mathbf{P}|$ & \textbf{V. score} & \textbf{Time} & $|\sigma|$\\ \hline
1 & SVT & 0.338 [0.02,0.87] & 15.5 [13,18] & 6 & -0.0217 & 776 & 57.59 [53,62]\\
2 & SVT & 0.397 [0.04,0.92] & 15.5 [13,18] & 6 & -0.0433 & 459 & 58.19 [55,63]\\
3 & VT & 0.497 [0.01,1.00] & 6.583 [1,13] & 12 & -0.0033 & 4776 & 90.48 [81,100]\\
4 & VT & 0.561 [0.01,1.00] & 9.583 [4,16] & 12 & 0.0025& 8208 & 84.64 [74,95]\\
5 & SVT & 0.505 [0.01,1.00] & 9.154 [1,17] & 13 & -0.0523& 1894 & 64.3 [58,70]\\
6 & SVT & 0.298 [0.03,0.55] & 10 [4,18] & 9 & 0.02 & 455 & 61.03 [54,73]\\
7 & VT & 0.504 [0.01,1.00] & 9.357 [2,16] & 14 & -0.0593 & 5270 & 84.36 [75,96]\\
8 & SVT & 0.170 [0.01,0.48] & 9.5 [7,12] & 6 & -0.05 & 460 & 48.64 [42, 57]\\
9 & SVT & 0 [0,0] & 0 [0,0] & 1 & 0 & 279 & 47.72 [44,51]\\
10 & VT & 0.565 [0.01,1.00] & 7.091 [2,13] & 11 & -0.0518 & 4739 & 89.34 [80,102]\\
11 & SVT & 0.033 [0.01,0.06] & 11 [10,12] & 3 & -0.0267 & 343 & 45.87 [43,52]\\
12 & SVT & 0.326 [0.01,0.75] & 11.385 [3,18] & 13 & -0.0077 & 876 & 59.39 [54,66]\\
13 & SVT & 0.084 [0.01,0.20] & 16 [14,18] & 5 & -0.036 & 363 & 50.38 [46,56]\\
14 & SVT & 0.067 [0.01,0.16] & 15.333 [12,18] & 6 & -0.01 & 539 & 52.01 [48,59]\\
15 & SVT & 0.498 [0.01,0.92] & 13.5 [11,16] & 6 & 0.0083 & 374 & 51.23 [36,60]\\
16 & VT & 0.468 [0.02,0.99] & 6 [1,11] & 11 & -0.0064 & 4419 & 89.06 [80,100]\\
17 & VT & 0.490 [0.05,1.00] & 10.6 [6,16] & 10 & -0.004 & 2699 & 84.82 [75,95]\\
18 & VT & 0.517 [0.04,1.00] & 10.7 [6,16] & 10 & -0.009 & 2489 & 84.45 [75,95]\\
19 & VT & 0.506 [0.04,1.00] & 10.6 [6,16] & 10 & -0.02 & 2812 & 84.87 [75,96]\\ \hline
\end{tabular}
\end{footnotesize}
\caption{Statistics for Pareto-optimal condition-specific parameters. Effectiveness and parameter distance are in the form $\mu [m,M]$ (mean $\mu$, minimum $m$, maximum $M$ objective function value for all solutions). $|\mathbf{P}|$ is the number of Pareto-optimal solutions. \textbf{V. score} is the validation score. \textbf{Time} is the runtime in seconds. $|\sigma|$ is the length of the training signals, in the form $\mu [m,M]$ (mean, min, max).}
\label{tbl:overall_results}
\vspace*{-.9cm}
\end{table}

\begin{figure*}
\centering
\subfloat[Cond. 2 (SVT)]{\includegraphics[width = .18\textwidth]{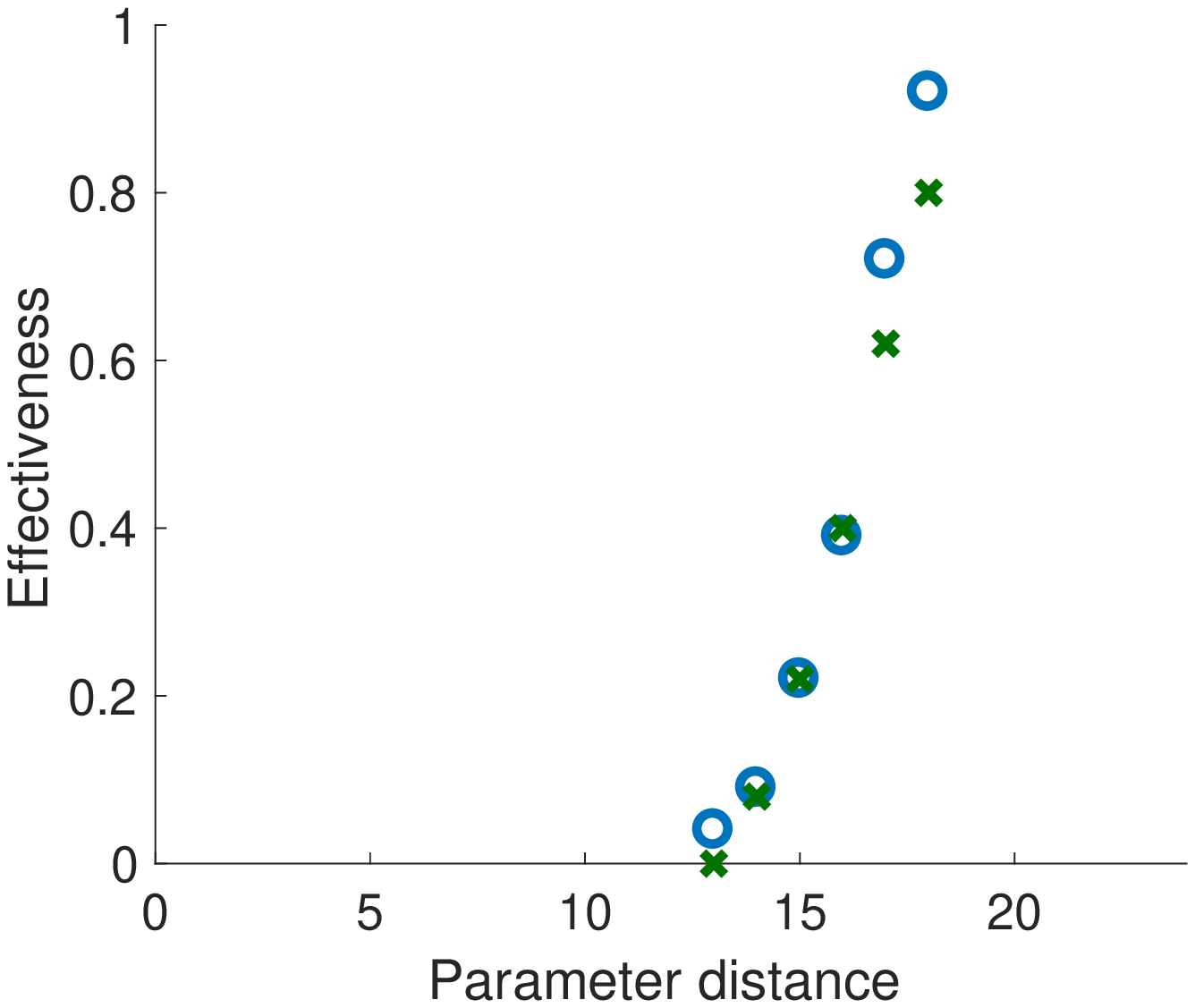}} \hfill 
\subfloat[Cond. 5 (SVT)]{\includegraphics[width = .18\textwidth]{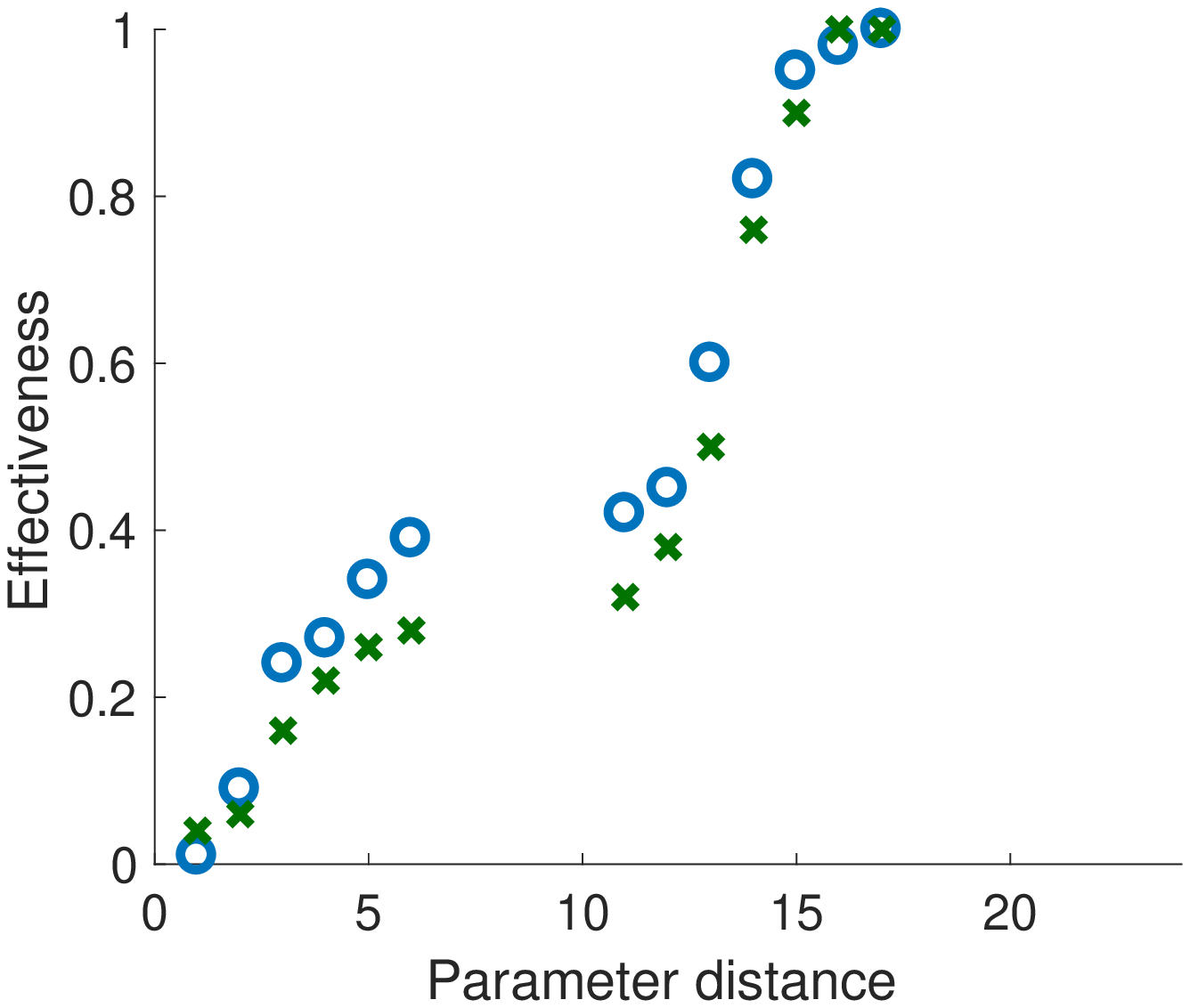}} \hfill 
\subfloat[Cond. 10 (VT)]{\includegraphics[width = .18\textwidth]{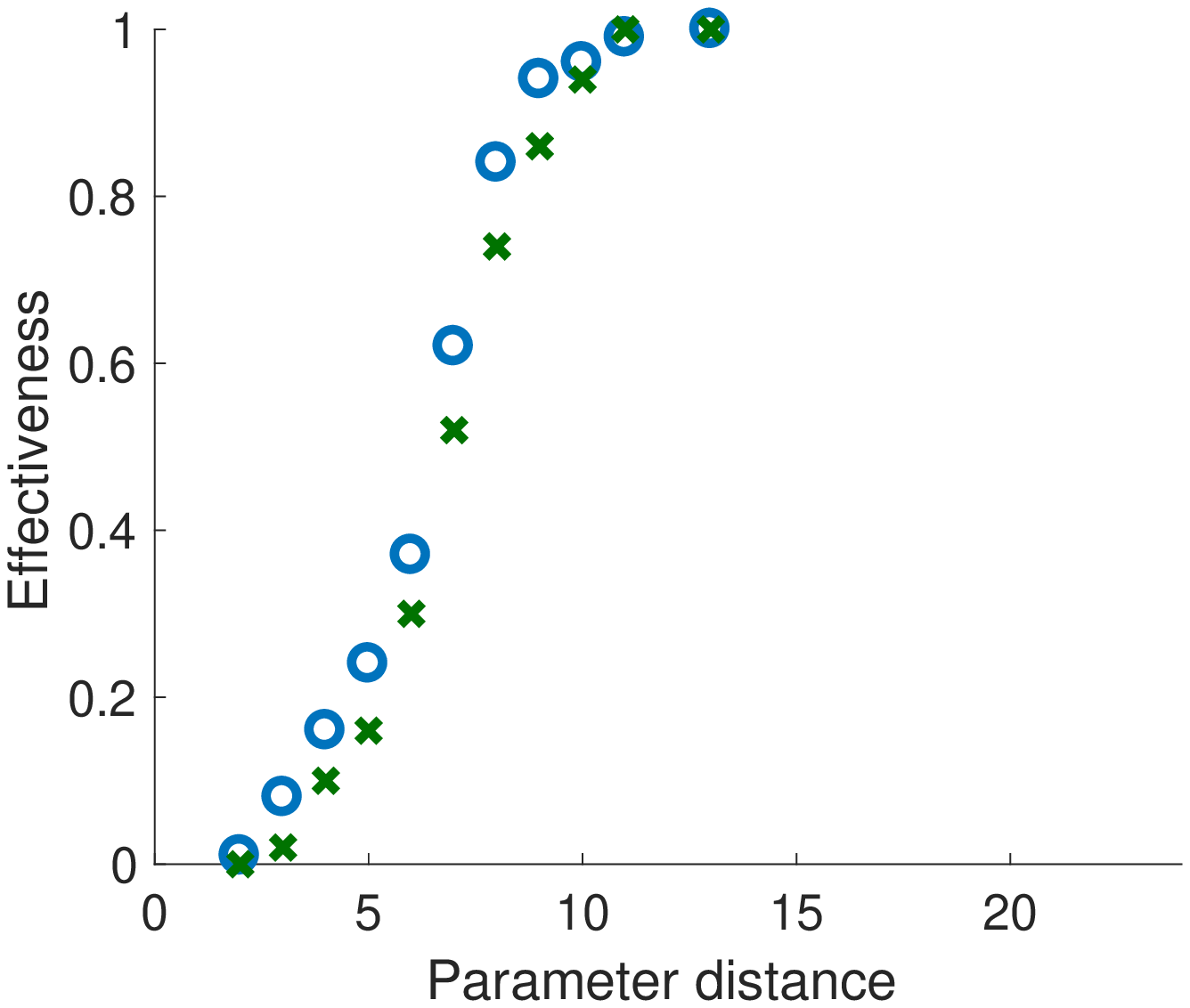}} \hfill 
\subfloat[Cond. 11 (SVT)]{\includegraphics[width = .18\textwidth]{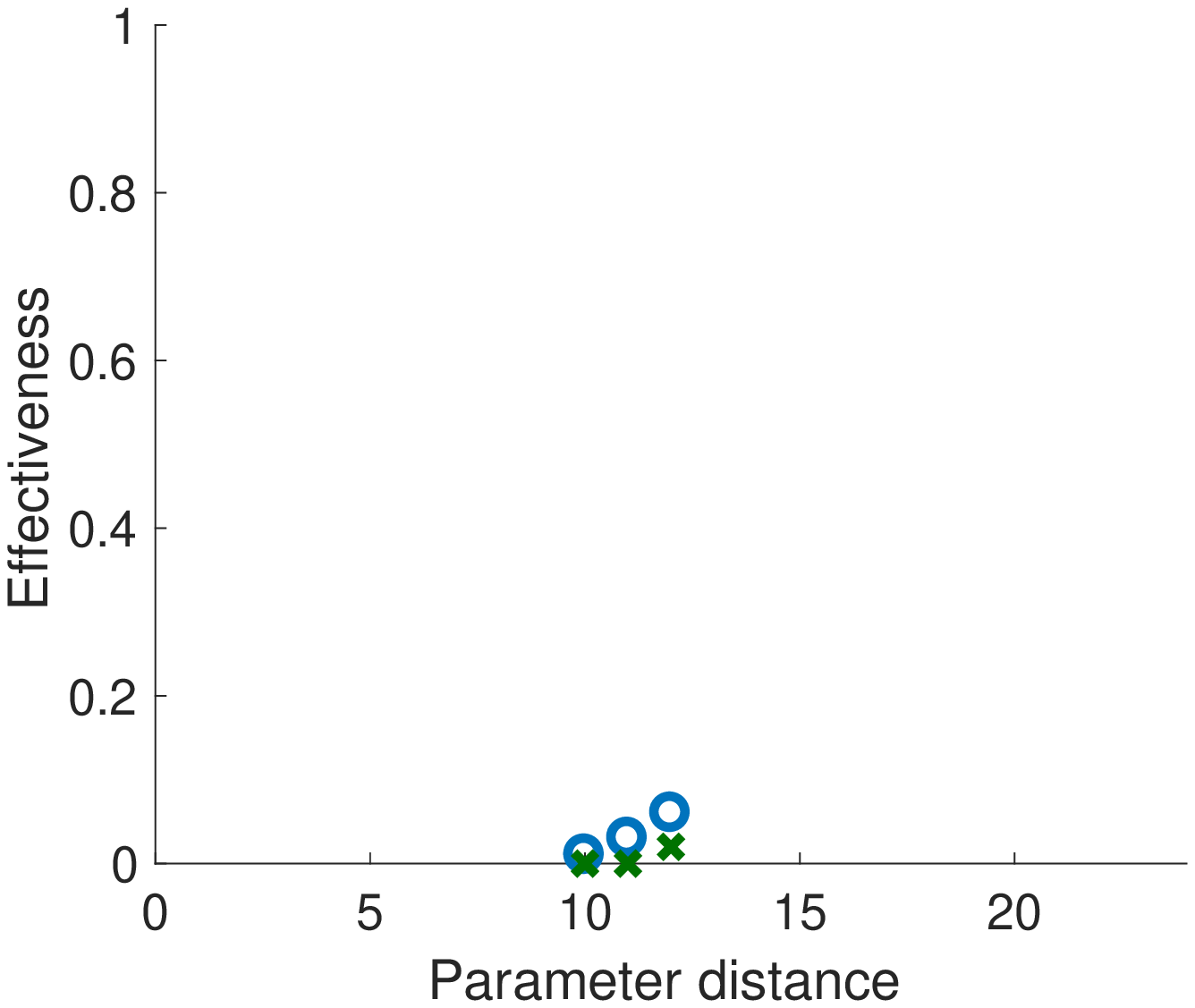}} \hfill 
\subfloat[Cond. 17 (VT)]{\includegraphics[width = .18\textwidth]{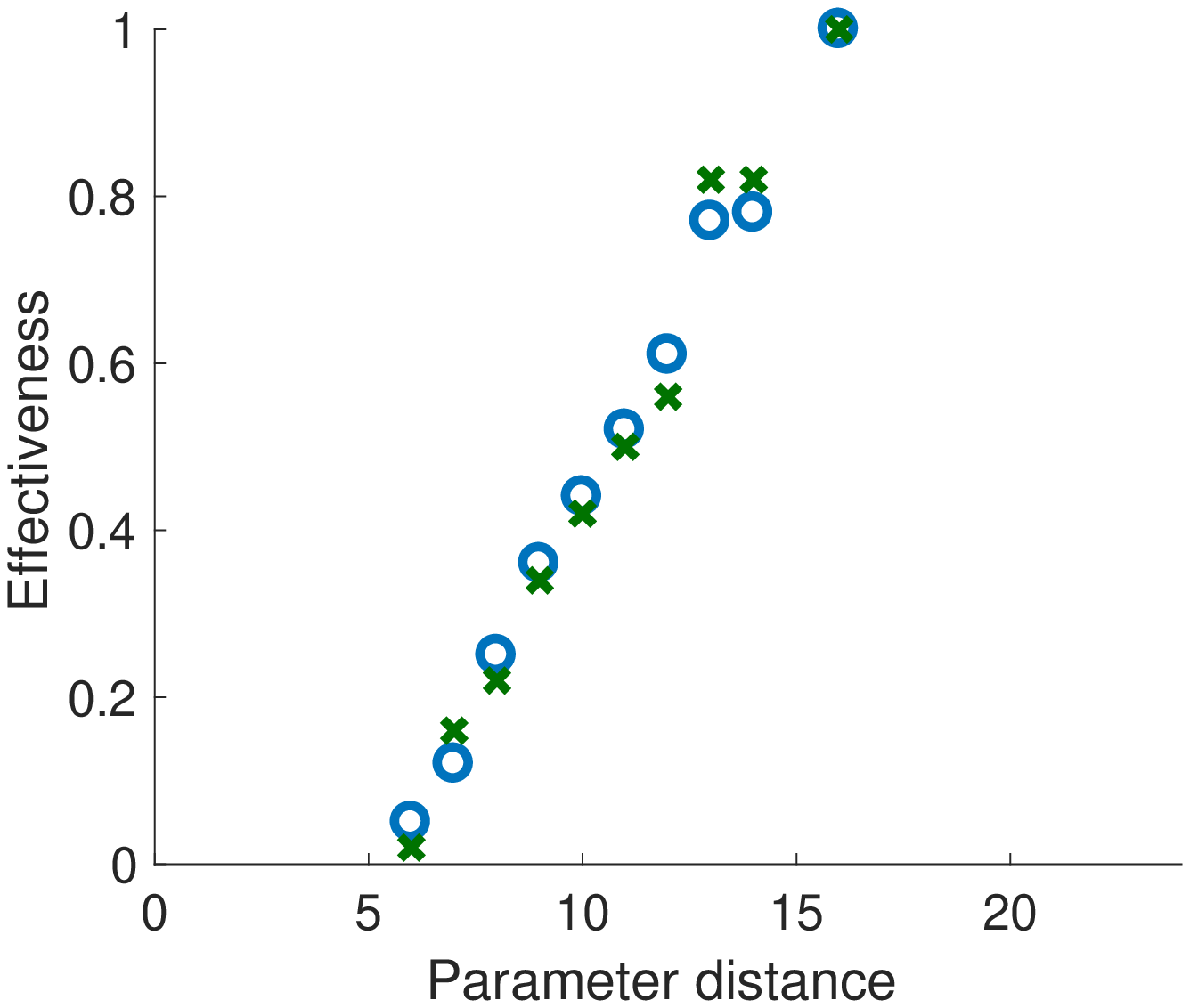}} 
\caption{Pareto fronts for condition-specific reprogramming attacks. Blue dots represent the Pareto front obtained with training signals. Green crosses indicate the effectiveness of the synthesized parameters on the test signals. Only a selection of representative conditions are included, see Figure~7 in the appendix for the full set of plots.}
\label{fig:pareto_fronts_selection}
\vspace*{-.2cm}
\end{figure*}

\vspace*{-4pt}
\paragraph{Condition-specific attacks} Table~\ref{tbl:overall_results} provides statistics on the synthesized Pareto-optimal attacks. Figure~\ref{fig:pareto_fronts_selection} shows the Pareto-optimal fronts for a selection of representative conditions (see Figure~7 in the appendix for the full set). 
The synthesized parameters for all conditions are in Tables~4-21 of the appendix. 

Remarkably, the synthesized attacks attain validation scores that are either positive or very close to zero, indicating that the attacks generalize well with unseen data and, thus, would have comparable effectiveness when applied to the unknown EGM of the victim. 

As visible in Table~\ref{tbl:overall_results}, our method can derive effective attacks for all VT conditions, since the corresponding Pareto fronts always contain a parametrization able to affect the therapy of all training signals (effectiveness 1), with the exception of condition 16 where the maximum effectiveness is 0.99. However, not all attacks on VT conditions are comparably stealthy (see Figure~\ref{fig:pareto_fronts_selection}). For instance, for condition 10 a parameter distance of 7 ensures that the attack is effective with half of the training signals, while for condition 17, the same effectiveness level is obtained only at a distance of 11 from the nominal parameters (worse stealthiness). 

In contrast, attacks on SVT conditions are not all equally successful. For condition 5 we can find parameters with 100\% effectiveness as well as stealthier attacks that e.g.\ are able to affect almost 40\% of the signals with a distance of only 5. For conditions 2 and 15 we obtain parameters with nearly 100\% effectiveness but with poor stealthiness (the minimal distance of a Pareto-optimal attack is 13 and 11, respectively). Some EGMs turned out to be difficult to attack. Specifically, for condition 11 the strongest attack affects only 6\% of the signals and, for condition 9, no Pareto-optimal attacks exist but the trivial one that leaves the nominal parameters unchanged.

The reason why VT conditions are easier to attack is that it takes only a minor increase to the VT and VF detection thresholds (parameters $\mathrm{VF_{th}}$ and $\mathrm{VT_{th}}$) to make the ICD mis-classify a tachyarrhythmia episode. 
On the other hand, $\mathrm{VF_{th}}$ and $\mathrm{VT_{th}}$ must be reprogrammed to very low values for the ICD to classify a slow heart rate as VT or VF to induce unnecessary therapy. This is not always possible because in SVT conditions, the heart rate is often below the lowest programmable values for $\mathrm{VF_{th}}$ (110 BPM) and $\mathrm{VT_{th}}$ (90 BPM), which explains why, for instance, no attack parameters exist that can affect condition~9.  We remark that these results are \emph{provably correct} because OMT is \emph{guaranteed} to find Pareto-optimal attack parameters, when they exist.

Besides increasing $\mathrm{VF_{th}}$ and $\mathrm{VT_{th}}$, the attacks on VT conditions synthesized by our method tend to increase the VF and VT durations ($\mathrm{VFdur}$ and $\mathrm{VTdur}$) thus reducing the probability that the ICD classifies an episode as sustained, which is a necessary condition for delivering therapy. For instance, the most effective attack for condition 10 has $\mathrm{VF_{th}} = 250$ BPM, $\mathrm{VT_{th}} = 205$ BPM, $\mathrm{VFdur}=10$~s, and $\mathrm{VTdur} = 13$ s, against nominal values of 200, 160, 1, and 2.5, respectively. 
For some VT conditions, the attacks also tamper with the VT zone-related parameters: they decrease the Rhythm Match score $\mathrm{NSRcor_{th}}$ and the detection threshold for atrial fibrillation $\mathrm{AFib_{th}}$, while increasing the stability score (parameter $\mathrm{stb}$), making discriminators D6 and D7 more likely to be satisfied thus tricking the ICD into classifying the episode as SVT. Indeed, for condition 10 the most effective reprogramming has $\mathrm{NSRcor_{th}} = 0.86$ (nominal 0.94), $\mathrm{AFib_{th}} = 100$ BPM (nominal 170), and $\mathrm{stb} = 35$ (nominal 20). 

\begin{figure}
\centering
\includegraphics[width=.9\columnwidth]{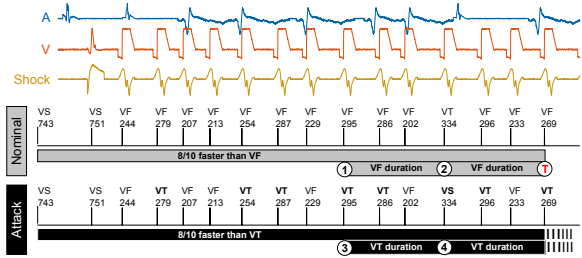}
\caption{Execution of BSc discrimination algorithm with nominal and attack parameters on atrial (A), ventricular (V), and shock EGMs from condition 10, at the start of a VF episode. Markers are: VF - sensed ventricular fibrillation, VT - tachycardia, and VS - normal rate. Intervals are in milliseconds. See text for a detailed explanation.}
\label{fig:attack_EGM_example}
\vspace*{-.5cm}
\end{figure}

Figure~\ref{fig:attack_EGM_example} compares nominal and reprogrammed parameters over an execution of the BSc algorithm at the start of a VF episode, using an EGM from condition 10. With nominal parameters, VF duration starts after the last 8/10 ventricular intervals faster than VF (see marker 1 in Fig.~\ref{fig:attack_EGM_example}) and ends after an interval is found below the VF threshold (see marker 2). A new VF duration can start right away, ending this time with a therapy~(marker T). 

In this example, the reprogramming sets $\mathrm{VF_{th}} = 240$ BPM (250 ms), $\mathrm{VF_{th}} = 185$ BPM (325 ms), and $\mathrm{VTdur} = 7$~s (corresponding to parameter \#6 in Table~12 of the appendix). With the higher VF threshold, the attack leads to marking the VF episode as VT, triggering VT duration (marker 3). VT duration ends with one interval found below the reprogrammed VT threshold (marker 4). A new VT duration can start right away, but therapy is prevented due to the long $\mathrm{VTdur}$. 

Attacks on SVT conditions follow the opposite strategy. They tend to keep $\mathrm{VF_{th}}$, $\mathrm{VT_{th}}$, $\mathrm{VFdur}$ and $\mathrm{VTdur}$ to the minimum programmable values, thus increasing the probability that slow heart rhythms are classified as a sustained tachyarrhythmia episode. An example is condition 5, for which the most effective attack has $\mathrm{VF_{th}} = 115$ BPM, $\mathrm{VT_{th}} = 110$ BPM, $\mathrm{VFdur}=1$ s and $\mathrm{VTdur} = 1$ s. Such an attack is 100\% successful regardless the other VT zone-related parameters, while for other SVT conditions we also need to increase the Rhythm Match threshold. In contrast, the parameters of discriminator D7, $\mathrm{AFib_{th}}$ and $\mathrm{stb}$, appear to have little effect. 

\paragraph{Condition-agnostic attacks} Pareto fronts for the condition-agnos\-tic attacks on VT and SVT, hereafter referred to as VT attack and SVT attack, are shown in Figure~\ref{fig:pareto_fronts_mixed}. The corresponding parameters are available in Tables~22 and~23 of the appendix.
These attacks attain very good validation scores, comparable to the condition-specific scores, suggesting that our method can generalize well also with heterogeneous arrhythmias. 
The Pareto front for the VT attack has a similar profile to those for the condition-specific attacks: the effectiveness is poor for parameter distance below 5, it has a sharp increase between distance 5 and 10, growing slowly after that up to reaching 100\% success at distance 16. The attack strategy is the same discussed for the condition-specific case: as the parameter distance grows, our method finds parameters with gradually higher values for $\mathrm{VF_{th}}$, $\mathrm{VT_{th}}$, $\mathrm{VFdur}$, $\mathrm{VTdur}$ and $\mathrm{stb}$, and lower values for $\mathrm{NSRcor_{th}}$ and $\mathrm{Afib_{th}}$. 

On the other hand, the parameters for the SVT attack reach a maximum effectiveness of 49\% at distance 18, compatibly with the fact that condition-specific attacks are reasonably successful only for a subset of SVT conditions. The attack strategy confirms our previous discussion, with the synthesized parameters having minimal values of $\mathrm{VF_{th}}$, $\mathrm{VT_{th}}$, $\mathrm{VFdur}$ and $\mathrm{VTdur}$.

\begin{figure}
\centering
\subfloat[VT. Validation score: -0.0032]{\includegraphics[width = .2\textwidth]{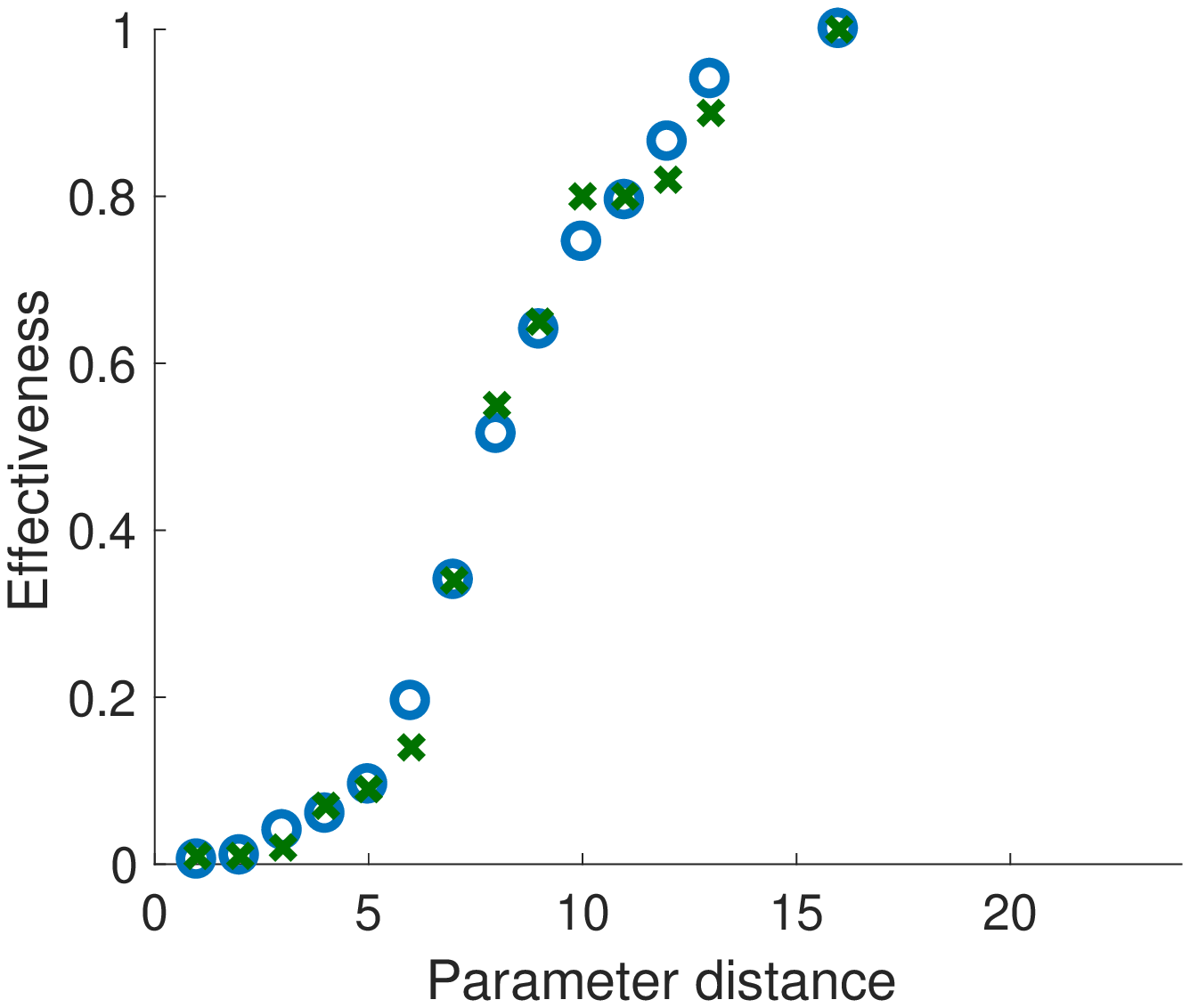}} \hfill
\subfloat[SVT. Validation score: 0.0179]{\includegraphics[width = .2\textwidth]{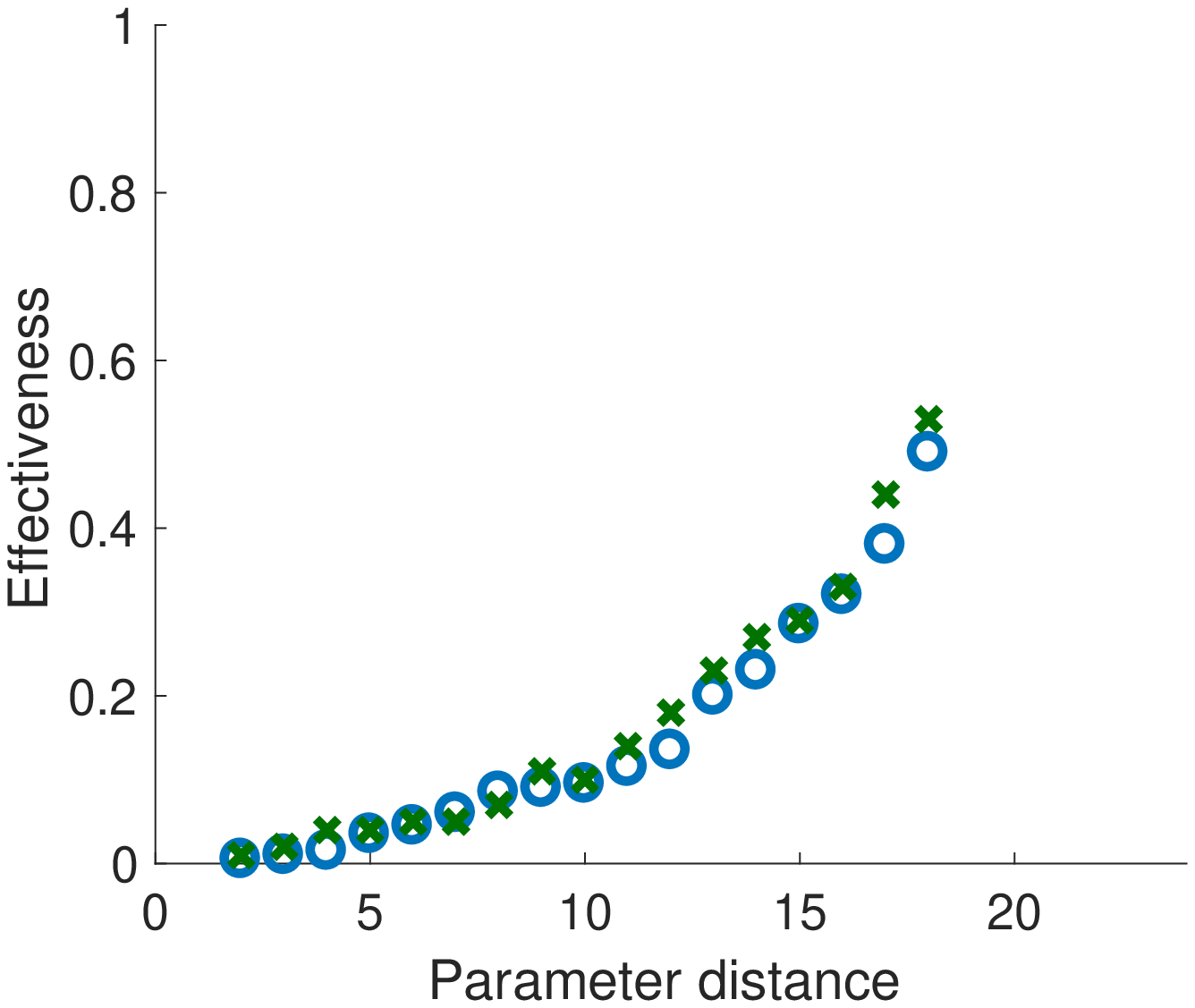}} 
\caption{Pareto fronts for condition-agnostic reprogramming attacks.  Legend is as in Figure~\ref{fig:pareto_fronts_selection}.}
\label{fig:pareto_fronts_mixed}
\vspace*{-.5cm}
\end{figure}

\paragraph{Performance and adequacy} 
Performance results for the synthesis of condition-specific attacks, reported in Table~\ref{tbl:overall_results}, show that VT conditions are more computationally demanding than SVT ones, with runtimes ranging from 2489 to 8208 seconds versus a range of 279 to 1894 seconds for SVT.  The reason is that VT conditions are characterized by shorter ventricular intervals, leading to more heart beats for the same EGM duration and thus, to longer signals. The path length and the number of training signals are indeed the main factors affecting the complexity of OMT-based synthesis.

We demonstrate the adequacy of our approach by showing that the parameters synthesized through OMT comfortably outperform those found by a random search (RS).  For this purpose, we ran RS for each condition and for the same runtime of OMT, and compared the area under the curve (AUC) of the Pareto fronts obtained with OMT and RS, with both training and test EGMs.  Higher AUC values imply better performance.  We remark that the parameters found by OMT are guaranteed to be Pareto-optimal with respect to training EGMs, and so RS (or any other optimization method) cannot have better performance on the training data. Indeed, RS yields AUC values strictly less than OMT for all conditions except 18 and 19, for which RS and OMT produced the same Pareto fronts (see Table~24 in the Appendix for the full set of AUC values).  With test data, OMT outperforms RS on 11 conditions, while the opposite happens only for three conditions.  For the remaining conditions, OMT and RS have equal AUC values.  These results confirm that OMT exhibits superior performance also with regard to unseen signals. 

\section{Related work}\label{sec:related}
The work of Halperin and colleagues~\cite{halperin2008pacemakers} was the first to show that ICDs can be accessed and reprogrammed by unauthorized users using off-the-shelf hardware. Unlike our work, however, they did not provide automated methods to derive stealthy attacks, nor did they consider the problem of tailoring the attacks based on the heart condition.  Rather, they simply showed that they can disable all therapies, an attack which is easily detectable. 
Other examples are spoofing attacks on ICDs~\cite{kune2013ghost}, attacks on insulin pumps and glucose monitors~\cite{li2011hijacking}, and on electrocardiogram-based biometrics~\cite{eberz2017broken}. 

The work of Jiang et al.~\cite{jiang2016silico} offers a model-based approach to analyze the accuracy ICD algorithms by conducting in-silico trials on synthetic cardiac signals. 
Our work relies on~\cite{jiang2016silico} for the generation of synthetic EGMs and the reverse-engineering of the Boston Scientific algorithm, but tackles the fundamentally different problem of designing stealthy attacks on ICDs, introducing a novel formalization of attack synthesis as multi-objective optimization, and an SMT-based encoding for its solution. 
In~\cite{KMPP15}, an SMT-based method is presented for the synthesis of cardiac pacemaker parameters that ensure safe heart rhythm and maximize robustness to parameter deviations. There are substantial differences between~\cite{KMPP15} and our work, both in the problem under study (improving the pacemaker function versus compromising the ICD function), and in the kind of devices considered: pacemakers help correct slow heart rhythms through low-voltage electrical pulses, and thus follow completely different algorithms from ICDs.

Related research include model-based methods for attack detection and identification in cyber-physical systems~\cite{pasqualetti2013attack,tiwari2014safety,hei2015patient} and methods for secure state estimation, i.e., for reconstructing the plant state from attack-prone sensor data~\cite{pajic2014robustness,fawzi2014secure}, some of which employ, as we do, SMT-based techniques to this purpose~\cite{shoukry2017secure}. 

SAT-based software verification techniques are applied in~\cite{inverso2018sat} for the synthesis of spoofing-attacks on control systems. 

\section{Conclusions}\label{sec:conclusion}
The lives of millions of patients rely on the correct functioning of implantable cardiac devices. As demonstrated by recent security-related recalls, vulnerabilities in these devices exist, which can be exploited to put the patient's safety in jeopardy through the malicious reprogramming of the device. 

Motivated by the need to improve the security, safety and robustness of such devices, we presented the first framework for systematically synthesizing reprogramming attacks on ICDs designed to maximize therapy disruption while minimizing detection.  Such attacks can therefore be tailored to the victim's physiology through condition-specific synthetic cardiac signals. 

Our approach builds on automated reasoning methods (OMT) that allowed us to synthesize malicious parameters that precisely attain Pareto-optimal performance w.r.t.\ stealthiness and effectiveness criteria. We demonstrated that such attacks are particularly effective in preventing therapy in the presence of VT and VF, and that they readily generalize to unseen signals.  This makes our approach suitable for real-world attacks.

For future work, we plan to evaluate synthesized attacks on a real ICD device, building on the hardware testbed for cardiac pacemakers of~\cite{jiang2014heart}. We will also investigate spoofing attacks on EGM sensors and techniques for making ICD discrimination algorithms more resilient to such attacks.

\bibliographystyle{ACM-Reference-Format}
\bibliography{ICD_attack_compact}


\begin{thebibliography}{29}


\ifx \showCODEN    \undefined \def \showCODEN     #1{\unskip}     \fi
\ifx \showDOI      \undefined \def \showDOI       #1{#1}\fi
\ifx \showISBNx    \undefined \def \showISBNx     #1{\unskip}     \fi
\ifx \showISBNxiii \undefined \def \showISBNxiii  #1{\unskip}     \fi
\ifx \showISSN     \undefined \def \showISSN      #1{\unskip}     \fi
\ifx \showLCCN     \undefined \def \showLCCN      #1{\unskip}     \fi
\ifx \shownote     \undefined \def \shownote      #1{#1}          \fi
\ifx \showarticletitle \undefined \def \showarticletitle #1{#1}   \fi
\ifx \showURL      \undefined \def \showURL       {\relax}        \fi
\providecommand\bibfield[2]{#2}
\providecommand\bibinfo[2]{#2}
\providecommand\natexlab[1]{#1}
\providecommand\showeprint[2][]{arXiv:#2}

\bibitem[\protect\citeauthoryear{Barrett, Sebastiani, Seshia, and
  Tinelli}{Barrett et~al\mbox{.}}{2009}]%
        {barrett2009satisfiability}
\bibfield{author}{\bibinfo{person}{Clark~W Barrett}, \bibinfo{person}{Roberto
  Sebastiani}, \bibinfo{person}{Sanjit~A Seshia}, {and} \bibinfo{person}{Cesare
  Tinelli}.} \bibinfo{year}{2009}\natexlab{}.
\newblock \showarticletitle{{Satisfiability Modulo Theories}}.
\newblock \bibinfo{journal}{\emph{Handbook of satisfiability}}
  \bibinfo{volume}{185} (\bibinfo{year}{2009}), \bibinfo{pages}{825--885}.
\newblock


\bibitem[\protect\citeauthoryear{Berger et~al\mbox{.}}{Berger
  et~al\mbox{.}}{2006}]%
        {RIGHT}
\bibfield{author}{\bibinfo{person}{Ronald~D Berger} {et~al\mbox{.}}}
  \bibinfo{year}{2006}\natexlab{}.
\newblock \showarticletitle{The {Rhythm ID} Going Head to Head Trial
  {(RIGHT)}}.
\newblock \bibinfo{journal}{\emph{Journal of cardiovascular electrophysiology}}
  \bibinfo{volume}{17}, \bibinfo{number}{7} (\bibinfo{year}{2006}),
  \bibinfo{pages}{749--753}.
\newblock


\bibitem[\protect\citeauthoryear{Biere et~al\mbox{.}}{Biere
  et~al\mbox{.}}{1999}]%
        {biere1999symbolic}
\bibfield{author}{\bibinfo{person}{Armin Biere} {et~al\mbox{.}}}
  \bibinfo{year}{1999}\natexlab{}.
\newblock \showarticletitle{Symbolic model checking without {BDDs}}. In
  \bibinfo{booktitle}{\emph{Tools and Algorithms for the Construction and
  Analysis of Systems}}. \bibinfo{pages}{193--207}.
\newblock


\bibitem[\protect\citeauthoryear{Bj{\o}rner, Phan, and Fleckenstein}{Bj{\o}rner
  et~al\mbox{.}}{2015}]%
        {bjorner2015nuz}
\bibfield{author}{\bibinfo{person}{Nikolaj Bj{\o}rner},
  \bibinfo{person}{Anh-Dung Phan}, {and} \bibinfo{person}{Lars Fleckenstein}.}
  \bibinfo{year}{2015}\natexlab{}.
\newblock \showarticletitle{$\nu$Z-An Optimizing {SMT} Solver}. In
  \bibinfo{booktitle}{\emph{TACAS}}, Vol.~\bibinfo{volume}{15}.
  \bibinfo{pages}{194--199}.
\newblock


\bibitem[\protect\citeauthoryear{{Boston Scientific Corporation}}{{Boston
  Scientific Corporation}}{2017}]%
        {bs_reference_guide1}
\bibfield{author}{\bibinfo{person}{{Boston Scientific Corporation}}.}
  \bibinfo{year}{2017}\natexlab{}.
\newblock \bibinfo{title}{{Implantable Cardioverter Defibrillator}, reference
  guide (part number: 359407-003)}.
\newblock   (\bibinfo{year}{2017}).
\newblock


\bibitem[\protect\citeauthoryear{Eberz et~al\mbox{.}}{Eberz
  et~al\mbox{.}}{2017}]%
        {eberz2017broken}
\bibfield{author}{\bibinfo{person}{Simon Eberz} {et~al\mbox{.}}}
  \bibinfo{year}{2017}\natexlab{}.
\newblock \showarticletitle{Broken hearted: How to attack {ECG} biometrics}. In
  \bibinfo{booktitle}{\emph{Network and Distributed System Security Symposium
  (NDSS) 2017}}. Internet Society.
\newblock


\bibitem[\protect\citeauthoryear{{Electrogram}}{{Electrogram}}{2018}]%
        {AAEL}
\bibfield{author}{\bibinfo{person}{{Electrogram}}.}
  \bibinfo{year}{2018}\natexlab{}.
\newblock \bibinfo{title}{{Ann Arbor Electrogram Libraries}}.
\newblock   (\bibinfo{year}{2018}).
\newblock
\urldef\tempurl%
\url{http://electrogram.com/}
\showURL{%
\tempurl}


\bibitem[\protect\citeauthoryear{Fawzi, Tabuada, and Diggavi}{Fawzi
  et~al\mbox{.}}{2014}]%
        {fawzi2014secure}
\bibfield{author}{\bibinfo{person}{Hamza Fawzi}, \bibinfo{person}{Paulo
  Tabuada}, {and} \bibinfo{person}{Suhas Diggavi}.}
  \bibinfo{year}{2014}\natexlab{}.
\newblock \showarticletitle{Secure estimation and control for cyber-physical
  systems under adversarial attacks}.
\newblock \bibinfo{journal}{\emph{IEEE Trans. Automat. Control}}
  \bibinfo{volume}{59}, \bibinfo{number}{6} (\bibinfo{year}{2014}),
  \bibinfo{pages}{1454--1467}.
\newblock


\bibitem[\protect\citeauthoryear{{Food and Drug Administration}}{{Food and Drug
  Administration}}{2017}]%
        {fda_recall}
\bibfield{author}{\bibinfo{person}{{Food and Drug Administration}}.}
  \bibinfo{year}{2017}\natexlab{}.
\newblock \bibinfo{title}{Implantable Cardiac Pacemakers by {Abbott}: Safety
  Communication}.
\newblock   (\bibinfo{year}{2017}).
\newblock
\urldef\tempurl%
\url{https://www.fda.gov/safety/medwatch/safetyinformation/safetyalertsforhumanmedicalproducts/ucm573854.htm}
\showURL{%
\tempurl}


\bibitem[\protect\citeauthoryear{Halperin et~al\mbox{.}}{Halperin
  et~al\mbox{.}}{2008}]%
        {halperin2008pacemakers}
\bibfield{author}{\bibinfo{person}{Daniel Halperin} {et~al\mbox{.}}}
  \bibinfo{year}{2008}\natexlab{}.
\newblock \showarticletitle{Pacemakers and implantable cardiac defibrillators:
  Software radio attacks and zero-power defenses}. In
  \bibinfo{booktitle}{\emph{IEEE Security and Privacy Symposium}}.
  \bibinfo{pages}{129--142}.
\newblock


\bibitem[\protect\citeauthoryear{Hei, Du, Lin, Lee, and Sokolsky}{Hei
  et~al\mbox{.}}{2015}]%
        {hei2015patient}
\bibfield{author}{\bibinfo{person}{Xiali Hei}, \bibinfo{person}{Xiaojiang Du},
  \bibinfo{person}{Shan Lin}, \bibinfo{person}{Insup Lee}, {and}
  \bibinfo{person}{Oleg Sokolsky}.} \bibinfo{year}{2015}\natexlab{}.
\newblock \showarticletitle{Patient infusion pattern based access control
  schemes for wireless insulin pump system}.
\newblock \bibinfo{journal}{\emph{IEEE Transactions on Parallel and Distributed
  Systems}} \bibinfo{volume}{26}, \bibinfo{number}{11} (\bibinfo{year}{2015}),
  \bibinfo{pages}{3108--3121}.
\newblock


\bibitem[\protect\citeauthoryear{Inverso, Bemporad, and Tribastone}{Inverso
  et~al\mbox{.}}{2018}]%
        {inverso2018sat}
\bibfield{author}{\bibinfo{person}{Omar Inverso}, \bibinfo{person}{Alberto
  Bemporad}, {and} \bibinfo{person}{Mirco Tribastone}.}
  \bibinfo{year}{2018}\natexlab{}.
\newblock \showarticletitle{SAT-based synthesis of spoofing attacks in
  cyber-physical control systems}. In \bibinfo{booktitle}{\emph{Proceedings of
  the 9th ACM/IEEE International Conference on Cyber-Physical Systems}}.
  \bibinfo{pages}{1--9}.
\newblock


\bibitem[\protect\citeauthoryear{Jiang et~al\mbox{.}}{Jiang
  et~al\mbox{.}}{2014}]%
        {jiang2014heart}
\bibfield{author}{\bibinfo{person}{Zhihao Jiang} {et~al\mbox{.}}}
  \bibinfo{year}{2014}\natexlab{}.
\newblock \showarticletitle{{Heart-on-a-Chip}: a closed-loop testing platform
  for implantable pacemakers}.
\newblock  (\bibinfo{year}{2014}).
\newblock


\bibitem[\protect\citeauthoryear{Jiang et~al\mbox{.}}{Jiang
  et~al\mbox{.}}{2016}]%
        {jiang2016silico}
\bibfield{author}{\bibinfo{person}{Zhihao Jiang} {et~al\mbox{.}}}
  \bibinfo{year}{2016}\natexlab{}.
\newblock \showarticletitle{In-silico pre-clinical trials for implantable
  cardioverter defibrillators}. In \bibinfo{booktitle}{\emph{EMBC}}. IEEE,
  \bibinfo{pages}{169--172}.
\newblock


\bibitem[\protect\citeauthoryear{Jiang, Pajic, and Mangharam}{Jiang
  et~al\mbox{.}}{2012}]%
        {jiang2012cyber}
\bibfield{author}{\bibinfo{person}{Zhihao Jiang}, \bibinfo{person}{Miroslav
  Pajic}, {and} \bibinfo{person}{Rahul Mangharam}.}
  \bibinfo{year}{2012}\natexlab{}.
\newblock \showarticletitle{Cyber--physical modeling of implantable cardiac
  medical devices}.
\newblock \bibinfo{journal}{\emph{Proc. IEEE}} \bibinfo{volume}{100},
  \bibinfo{number}{1} (\bibinfo{year}{2012}), \bibinfo{pages}{122--137}.
\newblock


\bibitem[\protect\citeauthoryear{Kune et~al\mbox{.}}{Kune
  et~al\mbox{.}}{2013}]%
        {kune2013ghost}
\bibfield{author}{\bibinfo{person}{Denis~Foo Kune} {et~al\mbox{.}}}
  \bibinfo{year}{2013}\natexlab{}.
\newblock \showarticletitle{Ghost talk: Mitigating {EMI} signal injection
  attacks against analog sensors}. In \bibinfo{booktitle}{\emph{IEEE Security
  and Privacy Symposium}}. \bibinfo{pages}{145--159}.
\newblock


\bibitem[\protect\citeauthoryear{Kwiatkowska et~al\mbox{.}}{Kwiatkowska
  et~al\mbox{.}}{2015}]%
        {KMPP15}
\bibfield{author}{\bibinfo{person}{Marta Kwiatkowska} {et~al\mbox{.}}}
  \bibinfo{year}{2015}\natexlab{}.
\newblock \showarticletitle{Synthesising robust and optimal parameters for
  cardiac pacemakers using symbolic and evolutionary computation techniques}.
  In \bibinfo{booktitle}{\emph{Hybrid Systems and Biology}}
  \emph{(\bibinfo{series}{LNCS/LNBI})}, Vol.~\bibinfo{volume}{9271}.
  \bibinfo{publisher}{Springer}, \bibinfo{pages}{119--140}.
\newblock


\bibitem[\protect\citeauthoryear{Li, Raghunathan, and Jha}{Li
  et~al\mbox{.}}{2011}]%
        {li2011hijacking}
\bibfield{author}{\bibinfo{person}{Chunxiao Li}, \bibinfo{person}{Anand
  Raghunathan}, {and} \bibinfo{person}{Niraj~K Jha}.}
  \bibinfo{year}{2011}\natexlab{}.
\newblock \showarticletitle{Hijacking an insulin pump: Security attacks and
  defenses for a diabetes therapy system}. In \bibinfo{booktitle}{\emph{IEEE
  Healthcom}}. \bibinfo{pages}{150--156}.
\newblock


\bibitem[\protect\citeauthoryear{Moss et~al\mbox{.}}{Moss
  et~al\mbox{.}}{2012}]%
        {moss2012reduction}
\bibfield{author}{\bibinfo{person}{Arthur~J Moss} {et~al\mbox{.}}}
  \bibinfo{year}{2012}\natexlab{}.
\newblock \showarticletitle{Reduction in inappropriate therapy and mortality
  through ICD programming}.
\newblock \bibinfo{journal}{\emph{New England Journal of Medicine}}
  \bibinfo{volume}{367}, \bibinfo{number}{24} (\bibinfo{year}{2012}),
  \bibinfo{pages}{2275--2283}.
\newblock


\bibitem[\protect\citeauthoryear{Pajic et~al\mbox{.}}{Pajic
  et~al\mbox{.}}{2014}]%
        {pajic2014robustness}
\bibfield{author}{\bibinfo{person}{Miroslav Pajic} {et~al\mbox{.}}}
  \bibinfo{year}{2014}\natexlab{}.
\newblock \showarticletitle{Robustness of attack-resilient state estimators}.
  In \bibinfo{booktitle}{\emph{ACM/IEEE 5th International Conference on
  Cyber-Physical Systems}}. \bibinfo{pages}{163--174}.
\newblock


\bibitem[\protect\citeauthoryear{Pasqualetti, D{\"o}rfler, and
  Bullo}{Pasqualetti et~al\mbox{.}}{2013}]%
        {pasqualetti2013attack}
\bibfield{author}{\bibinfo{person}{Fabio Pasqualetti}, \bibinfo{person}{Florian
  D{\"o}rfler}, {and} \bibinfo{person}{Francesco Bullo}.}
  \bibinfo{year}{2013}\natexlab{}.
\newblock \showarticletitle{Attack detection and identification in
  cyber-physical systems}.
\newblock \bibinfo{journal}{\emph{IEEE Trans. Automat. Control}}
  \bibinfo{volume}{58}, \bibinfo{number}{11} (\bibinfo{year}{2013}),
  \bibinfo{pages}{2715--2729}.
\newblock


\bibitem[\protect\citeauthoryear{Peterson}{Peterson}{2013}]%
        {peterson2013yes}
\bibfield{author}{\bibinfo{person}{Andrea Peterson}.}
  \bibinfo{year}{2013}\natexlab{}.
\newblock \showarticletitle{Yes, terrorists could have hacked {Dick Cheney's}
  heart}.
\newblock \bibinfo{journal}{\emph{Washington Post}} (\bibinfo{year}{2013}).
\newblock


\bibitem[\protect\citeauthoryear{Rasmussen et~al\mbox{.}}{Rasmussen
  et~al\mbox{.}}{2009}]%
        {rasmussen2009proximity}
\bibfield{author}{\bibinfo{person}{Kasper~Bonne Rasmussen} {et~al\mbox{.}}}
  \bibinfo{year}{2009}\natexlab{}.
\newblock \showarticletitle{Proximity-based access control for implantable
  medical devices}. In \bibinfo{booktitle}{\emph{CCS}}. ACM,
  \bibinfo{pages}{410--419}.
\newblock


\bibitem[\protect\citeauthoryear{Rios and Butts}{Rios and Butts}{2018}]%
        {rios_black_hat}
\bibfield{author}{\bibinfo{person}{Billy Rios} {and} \bibinfo{person}{Jonathan
  Butts}.} \bibinfo{year}{2018}\natexlab{}.
\newblock \bibinfo{title}{Understanding and Exploiting Implanted Medical
  Devices}.
\newblock \bibinfo{howpublished}{{Black Hat USA conference}}.
  (\bibinfo{year}{2018}).
\newblock


\bibitem[\protect\citeauthoryear{Shoukry et~al\mbox{.}}{Shoukry
  et~al\mbox{.}}{2017}]%
        {shoukry2017secure}
\bibfield{author}{\bibinfo{person}{Yasser Shoukry} {et~al\mbox{.}}}
  \bibinfo{year}{2017}\natexlab{}.
\newblock \showarticletitle{Secure State Estimation for Cyber-Physical Systems
  Under Sensor Attacks: A {Satisfiability Modulo Theory} Approach}.
\newblock \bibinfo{journal}{\emph{IEEE Trans. Automat. Control}}
  \bibinfo{volume}{62}, \bibinfo{number}{10} (\bibinfo{year}{2017}),
  \bibinfo{pages}{4917--4932}.
\newblock


\bibitem[\protect\citeauthoryear{Singer}{Singer}{2001}]%
        {interv_EP}
\bibfield{author}{\bibinfo{person}{Igor Singer}.}
  \bibinfo{year}{2001}\natexlab{}.
\newblock \bibinfo{booktitle}{\emph{Interventional electrophysiology}}.
\newblock


\bibitem[\protect\citeauthoryear{Tiwari et~al\mbox{.}}{Tiwari
  et~al\mbox{.}}{2014}]%
        {tiwari2014safety}
\bibfield{author}{\bibinfo{person}{Ashish Tiwari} {et~al\mbox{.}}}
  \bibinfo{year}{2014}\natexlab{}.
\newblock \showarticletitle{Safety envelope for security}. In
  \bibinfo{booktitle}{\emph{Proceedings of the 3rd international conference on
  High confidence networked systems}}. ACM, \bibinfo{pages}{85--94}.
\newblock


\bibitem[\protect\citeauthoryear{Xu et~al\mbox{.}}{Xu et~al\mbox{.}}{2011}]%
        {xu2011imdguard}
\bibfield{author}{\bibinfo{person}{Fengyuan Xu} {et~al\mbox{.}}}
  \bibinfo{year}{2011}\natexlab{}.
\newblock \showarticletitle{{IMDGuard}: Securing implantable medical devices
  with the external wearable guardian}. In \bibinfo{booktitle}{\emph{IEEE
  Infocom}}. \bibinfo{pages}{1862--1870}.
\newblock


\bibitem[\protect\citeauthoryear{Zanker et~al\mbox{.}}{Zanker
  et~al\mbox{.}}{2016}]%
        {zanker2016tachycardia}
\bibfield{author}{\bibinfo{person}{Norbert Zanker} {et~al\mbox{.}}}
  \bibinfo{year}{2016}\natexlab{}.
\newblock \showarticletitle{{Tachycardia detection in ICDs by Boston
  Scientific}}.
\newblock \bibinfo{journal}{\emph{Herzschrittmachertherapie+
  Elektrophysiologie}} \bibinfo{volume}{27}, \bibinfo{number}{3}
  (\bibinfo{year}{2016}), \bibinfo{pages}{186--192}.
\newblock


\end{thebibliography}

\clearpage
\appendix

\section{Supplementary material}
\begin{table}
\centering
\begin{footnotesize}
\begin{tabular}{rl | llllllllll}
& & \multicolumn{10}{c}{\textbf{Effectiveness $\geq$}}\\
\multicolumn{2}{r|}{\textbf{Condition}} & 0.1 & 0.2 & 0.3 & 0.4 & 0.5 & 0.6 & 0.7 & 0.8 & 0.9 & 1\\ \hline
1 & SVT & 15 & 16 & 16 & 17 & 17 & 18 & 18 & 18 & NA &  NA  \\
2 & SVT & 15 & 15 & 16 & 17 & 17 & 17 & 17 & 18 & 18 &  NA  \\
3 & VT & 4 & 6 & 6 & 7 & 7 & 8 & 8 & 9 & 9 & 13  \\
4 & VT & 6 & 7 & 8 & 9 & 9 & 9 & 10 & 11 & 13 & 16  \\
5 & SVT & 3 & 3 & 5 & 11 & 13 & 13 & 14 & 14 & 15 & 17  \\
6 & SVT & 6 & 6 & 8 & 17 & 18 &  NA &  NA &  NA &  NA & NA   \\
7 & VT & 6 & 7 & 8 & 9 & 9 & 10 & 11 & 13 & 16 & 16  \\
8 & SVT & 10 & 11 & 11 & 12 &  NA &  NA & NA  & NA  &  NA & NA   \\
10 & VT & 4 & 5 & 6 & 7 & 7 & 7 & 8 & 8 & 9 & 13  \\
11 & SVT &  NA &  NA &  NA &  NA &  NA &  NA &  NA &  NA &  NA &   NA \\
12 & SVT & 8 & 8 & 10 & 15 & 17 & 18 & 18 &  NA & NA  &  NA  \\
13 & SVT & 17 & 18 &  NA &  NA &  NA & NA  &  NA & NA  &  NA &  NA  \\
14 & SVT & 17 &  NA & NA  & NA  &  NA & NA  &  NA & NA  &  NA &   NA \\
15 & SVT & 13 & 13 & 13 & 14 & 14 & 14 & 14 & 15 & 16 & NA   \\
16 & VT & 3 & 5 & 6 & 7 & 7 & 8 & 8 & 9 & 9 &  NA  \\
17 & VT & 7 & 8 & 9 & 10 & 11 & 12 & 13 & 16 & 16 & 16  \\
18 & VT & 7 & 8 & 9 & 10 & 11 & 12 & 13 & 13 & 16 & 16  \\
19 & VT & 7 & 8 & 9 & 10 & 11 & 13 & 13 & 13 & 16 & 16 \\ \hline
\end{tabular}
\end{footnotesize}
\caption{Distance of reprogramming attack required to achieve given effectiveness levels. NA entries indicate that the no parameter exists yielding that effectiveness level.}
\label{tbl:app_results_thresholds}
\end{table}

\begin{table}
\centering
\begin{footnotesize}
\begin{tabular}{r | lllllll | ll}
\multicolumn{10}{c}{\textbf{Condition 1}}\\
\# & $\mathrm{VF_{th}}$ & $\mathrm{VT_{th}}$ & $\mathrm{VFdur}$ & $\mathrm{VTdur}$ & $\mathrm{NSRcor_{th}}$ & $\mathrm{AFib_{th}}$ & $\mathrm{stb}$ & Eff. & Dist.\\ \hline
1 & 445 & 546 & 1.0 & 2.0 & 0.96 & 546 & 18 & 0.02 & 13\\ 
2 & 462 & 546 & 1.0 & 2.0 & 0.96 & 546 & 18 & 0.04 & 14\\ 
3 & 480 & 500 & 1.0 & 8.0 & 0.94 & 375 & 18 & 0.17 & 15\\ 
4 & 500 & 522 & 1.0 & 9.0 & 0.88 & 375 & 18 & 0.34 & 16\\ 
5 & 522 & 546 & 1.0 & 3.5 & 0.94 & 375 & 18 & 0.59 & 17\\ 
6 & 546 & 600 & 1.0 & 7.0 & 0.96 & 546 & 18 & 0.87 & 18\\ \hline
\end{tabular}
\end{footnotesize}
\caption{Pareto-optimal reprogramming attacks for condition 1. Detection thresholds are expressed as duration in milliseconds instead of BPM frequency.}
\label{tbl_app:cond1}
\end{table}

\begin{table}
\centering
\begin{footnotesize}
\begin{tabular}{r | lllllll | ll}
\multicolumn{10}{c}{\textbf{Condition 2}}\\
\# & $\mathrm{VF_{th}}$ & $\mathrm{VT_{th}}$ & $\mathrm{VFdur}$ & $\mathrm{VTdur}$ & $\mathrm{NSRcor_{th}}$ & $\mathrm{AFib_{th}}$ & $\mathrm{stb}$ & Eff. & Dist.\\ \hline
1 & 445 & 462 & 1.0 & 1.0 & 0.96 & 600 & 8 & 0.04 & 13\\ 
2 & 462 & 572 & 1.0 & 14.0 & 0.96 & 600 & 8 & 0.09 & 14\\ 
3 & 480 & 572 & 1.0 & 14.0 & 0.96 & 600 & 8 & 0.22 & 15\\ 
4 & 500 & 572 & 1.0 & 14.0 & 0.96 & 600 & 8 & 0.39 & 16\\ 
5 & 522 & 572 & 1.0 & 25.0 & 0.96 & 600 & 8 & 0.72 & 17\\ 
6 & 546 & 572 & 1.0 & 14.0 & 0.96 & 600 & 8 & 0.92 & 18\\ \hline
\end{tabular}
\end{footnotesize}
\caption{Pareto-optimal reprogramming attacks for condition 2. Units are as in Table~\ref{tbl_app:cond1}.}
\end{table}

\begin{table}
\centering
\begin{footnotesize}
\begin{tabular}{r | lllllll | ll}
\multicolumn{10}{c}{\textbf{Condition 3}}\\
\# & $\mathrm{VF_{th}}$ & $\mathrm{VT_{th}}$ & $\mathrm{VFdur}$ & $\mathrm{VTdur}$ & $\mathrm{NSRcor_{th}}$ & $\mathrm{AFib_{th}}$ & $\mathrm{stb}$ & Eff. & Dist.\\ \hline
1 & 293 & 388 & 1.5 & 3.0 & 0.94 & 353 & 18 & 0.01 & 1\\ 
2 & 286 & 388 & 2.0 & 3.5 & 0.94 & 375 & 20 & 0.04 & 2\\ 
3 & 286 & 364 & 2.5 & 3.5 & 0.91 & 429 & 24 & 0.09 & 3\\ 
4 & 261 & 353 & 3.0 & 4.5 & 0.90 & 462 & 12 & 0.13 & 4\\ 
5 & 293 & 343 & 3.5 & 5.0 & 0.94 & 462 & 28 & 0.18 & 5\\ 
6 & 261 & 334 & 4.0 & 6.0 & 0.91 & 546 & 18 & 0.34 & 6\\ 
7 & 240 & 325 & 2.0 & 7.0 & 0.91 & 500 & 10 & 0.54 & 7\\ 
8 & 240 & 316 & 2.0 & 8.0 & 0.91 & 600 & 35 & 0.76 & 8\\ 
9 & 261 & 308 & 6.0 & 9.0 & 0.90 & 231 & 45 & 0.91 & 9\\ 
10 & 261 & 300 & 3.5 & 10.0 & 0.88 & 223 & 50 & 0.97 & 10\\ 
11 & 240 & 293 & 8.0 & 11.0 & 0.88 & 215 & 50 & 0.99 & 11\\ 
12 & 240 & 273 & 10.0 & 13.0 & 0.81 & 200 & 70 & 1.00 & 13\\ \hline
\end{tabular}
\end{footnotesize}
\caption{Pareto-optimal reprogramming attacks for condition 3. Units are as in Table~\ref{tbl_app:cond1}.}
\end{table}

\begin{table}
\centering
\begin{footnotesize}
\begin{tabular}{r | lllllll | ll}
\multicolumn{10}{c}{\textbf{Condition 4}}\\
\# & $\mathrm{VF_{th}}$ & $\mathrm{VT_{th}}$ & $\mathrm{VFdur}$ & $\mathrm{VTdur}$ & $\mathrm{NSRcor_{th}}$ & $\mathrm{AFib_{th}}$ & $\mathrm{stb}$ & Eff. & Dist.\\ \hline
1 & 286 & 353 & 2.5 & 4.0 & 0.90 & 286 & 28 & 0.01 & 4\\ 
2 & 293 & 353 & 3.5 & 5.0 & 0.89 & 316 & 30 & 0.05 & 5\\ 
3 & 250 & 334 & 2.0 & 6.0 & 0.89 & 429 & 30 & 0.12 & 6\\ 
4 & 250 & 325 & 4.5 & 7.0 & 0.89 & 316 & 35 & 0.22 & 7\\ 
5 & 250 & 316 & 4.5 & 8.0 & 0.89 & 334 & 40 & 0.38 & 8\\ 
6 & 250 & 308 & 4.0 & 9.0 & 0.90 & 334 & 45 & 0.68 & 9\\ 
7 & 250 & 300 & 1.0 & 9.0 & 0.90 & 316 & 50 & 0.76 & 10\\ 
8 & 240 & 293 & 2.5 & 9.0 & 0.84 & 300 & 55 & 0.80 & 11\\ 
9 & 250 & 286 & 9.0 & 12.0 & 0.90 & 300 & 60 & 0.86 & 12\\ 
10 & 240 & 273 & 1.0 & 10.0 & 0.81 & 300 & 70 & 0.92 & 13\\ 
11 & 250 & 273 & 10.0 & 10.0 & 0.80 & 429 & 80 & 0.93 & 14\\ 
12 & 250 & 462 & 2.5 & 20.0 & 0.78 & 200 & 55 & 1.00 & 16\\ \hline
\end{tabular}
\end{footnotesize}
\caption{Pareto-optimal reprogramming attacks for condition 4. Units are as in Table~\ref{tbl_app:cond1}.}
\end{table}

\begin{table}
\centering
\begin{footnotesize}
\begin{tabular}{r | lllllll | ll}
\multicolumn{10}{c}{\textbf{Condition 5}}\\
\# & $\mathrm{VF_{th}}$ & $\mathrm{VT_{th}}$ & $\mathrm{VFdur}$ & $\mathrm{VTdur}$ & $\mathrm{NSRcor_{th}}$ & $\mathrm{AFib_{th}}$ & $\mathrm{stb}$ & Eff. & Dist.\\ \hline
1 & 308 & 414 & 1.0 & 2.0 & 0.95 & 353 & 22 & 0.01 & 1\\ 
2 & 316 & 429 & 1.0 & 1.5 & 0.96 & 400 & 24 & 0.09 & 2\\ 
3 & 273 & 445 & 1.0 & 1.0 & 0.91 & 316 & 26 & 0.24 & 3\\ 
4 & 300 & 462 & 1.0 & 1.0 & 0.90 & 316 & 28 & 0.27 & 4\\ 
5 & 334 & 480 & 1.0 & 1.0 & 0.89 & 316 & 30 & 0.34 & 5\\ 
6 & 240 & 500 & 1.0 & 1.0 & 0.88 & 316 & 28 & 0.39 & 6\\ 
7 & 414 & 546 & 1.0 & 1.0 & 0.85 & 316 & 28 & 0.42 & 11\\ 
8 & 429 & 546 & 1.0 & 1.0 & 0.86 & 316 & 28 & 0.45 & 12\\ 
9 & 445 & 522 & 1.0 & 1.0 & 0.85 & 316 & 60 & 0.60 & 13\\ 
10 & 462 & 667 & 1.0 & 2.0 & 0.85 & 316 & 80 & 0.82 & 14\\ 
11 & 480 & 522 & 1.0 & 1.0 & 0.85 & 316 & 26 & 0.95 & 15\\ 
12 & 500 & 522 & 1.0 & 1.0 & 0.85 & 316 & 45 & 0.98 & 16\\ 
13 & 522 & 546 & 1.0 & 1.0 & 0.85 & 316 & 35 & 1.00 & 17\\  \hline
\end{tabular}
\end{footnotesize}
\caption{Pareto-optimal reprogramming attacks for condition 5. Units are as in Table~\ref{tbl_app:cond1}.}
\end{table}

\begin{table}
\centering
\begin{footnotesize}
\begin{tabular}{r | lllllll | ll}
\multicolumn{10}{c}{\textbf{Condition 6}}\\
\# & $\mathrm{VF_{th}}$ & $\mathrm{VT_{th}}$ & $\mathrm{VFdur}$ & $\mathrm{VTdur}$ & $\mathrm{NSRcor_{th}}$ & $\mathrm{AFib_{th}}$ & $\mathrm{stb}$ & Eff. & Dist.\\ \hline
1 & 334 & 414 & 1.0 & 1.5 & 0.96 & 462 & 26 & 0.03 & 4\\ 
2 & 343 & 429 & 1.0 & 1.5 & 0.95 & 375 & 18 & 0.08 & 5\\ 
3 & 353 & 500 & 1.0 & 1.5 & 0.96 & 546 & 30 & 0.20 & 6\\ 
4 & 364 & 522 & 1.0 & 7.0 & 0.88 & 375 & 18 & 0.29 & 7\\ 
5 & 375 & 429 & 1.0 & 2.0 & 0.95 & 375 & 18 & 0.34 & 8\\ 
6 & 388 & 500 & 1.0 & 1.5 & 0.96 & 600 & 45 & 0.38 & 9\\ 
7 & 500 & 572 & 1.0 & 9.0 & 0.79 & 375 & 18 & 0.39 & 16\\ 
8 & 522 & 572 & 1.0 & 25.0 & 0.79 & 400 & 18 & 0.42 & 17\\ 
9 & 546 & 667 & 1.0 & 30.0 & 0.77 & 375 & 18 & 0.55 & 18\\  \hline
\end{tabular}
\end{footnotesize}
\caption{Pareto-optimal reprogramming attacks for condition 6. Units are as in Table~\ref{tbl_app:cond1}.}
\end{table}

\begin{table}
\centering
\begin{footnotesize}
\begin{tabular}{r | lllllll | ll}
\multicolumn{10}{c}{\textbf{Condition 7}}\\
\# & $\mathrm{VF_{th}}$ & $\mathrm{VT_{th}}$ & $\mathrm{VFdur}$ & $\mathrm{VTdur}$ & $\mathrm{NSRcor_{th}}$ & $\mathrm{AFib_{th}}$ & $\mathrm{stb}$ & Eff. & Dist.\\ \hline
1 & 286 & 375 & 1.5 & 1.5 & 0.92 & 334 & 24 & 0.01 & 2\\ 
2 & 273 & 364 & 2.5 & 4.0 & 0.91 & 334 & 26 & 0.04 & 3\\ 
3 & 273 & 364 & 2.0 & 5.0 & 0.90 & 286 & 28 & 0.07 & 5\\ 
4 & 261 & 334 & 1.0 & 6.0 & 0.91 & 546 & 32 & 0.12 & 6\\ 
5 & 273 & 325 & 4.5 & 6.0 & 0.90 & 375 & 35 & 0.22 & 7\\ 
6 & 261 & 316 & 4.5 & 6.0 & 0.86 & 353 & 40 & 0.36 & 8\\ 
7 & 250 & 308 & 5.0 & 9.0 & 0.90 & 300 & 45 & 0.54 & 9\\ 
8 & 240 & 300 & 1.0 & 10.0 & 0.84 & 316 & 50 & 0.63 & 10\\ 
9 & 261 & 293 & 8.0 & 11.0 & 0.83 & 316 & 55 & 0.72 & 11\\ 
10 & 250 & 286 & 2.5 & 10.0 & 0.83 & 286 & 55 & 0.76 & 12\\ 
11 & 250 & 273 & 10.0 & 10.0 & 0.81 & 600 & 70 & 0.85 & 13\\ 
12 & 250 & 273 & 9.0 & 9.0 & 0.83 & 600 & 80 & 0.86 & 14\\ 
13 & 261 & 273 & 2.5 & 15.0 & 0.83 & 286 & 40 & 0.87 & 15\\ 
14 & 250 & 546 & 13.0 & 20.0 & 0.82 & 334 & 100 & 1.00 & 16\\  \hline
\end{tabular}
\end{footnotesize}
\caption{Pareto-optimal reprogramming attacks for condition 7. Units are as in Table~\ref{tbl_app:cond1}.}
\end{table}

\begin{table}
\centering
\begin{footnotesize}
\begin{tabular}{r | lllllll | ll}
\multicolumn{10}{c}{\textbf{Condition 8}}\\
\# & $\mathrm{VF_{th}}$ & $\mathrm{VT_{th}}$ & $\mathrm{VFdur}$ & $\mathrm{VTdur}$ & $\mathrm{NSRcor_{th}}$ & $\mathrm{AFib_{th}}$ & $\mathrm{stb}$ & Eff. & Dist.\\ \hline
1 & 308 & 522 & 2.0 & 2.0 & 0.89 & 273 & 6 & 0.01 & 7\\ 
2 & 353 & 546 & 1.0 & 1.0 & 0.86 & 273 & 6 & 0.03 & 8\\ 
3 & 308 & 572 & 1.0 & 1.0 & 0.85 & 273 & 6 & 0.05 & 9\\ 
4 & 293 & 600 & 1.0 & 1.0 & 0.96 & 600 & 8 & 0.14 & 10\\ 
5 & 414 & 632 & 1.0 & 1.0 & 0.92 & 223 & 55 & 0.31 & 11\\ 
6 & 343 & 667 & 1.0 & 1.0 & 0.96 & 273 & 6 & 0.48 & 12\\  \hline
\end{tabular}
\end{footnotesize}
\caption{Pareto-optimal reprogramming attacks for condition 8. Units are as in Table~\ref{tbl_app:cond1}.}
\end{table}

\begin{table}
\centering
\begin{footnotesize}
\begin{tabular}{r | lllllll | ll}
\multicolumn{10}{c}{\textbf{Condition 10}}\\
\# & $\mathrm{VF_{th}}$ & $\mathrm{VT_{th}}$ & $\mathrm{VFdur}$ & $\mathrm{VTdur}$ & $\mathrm{NSRcor_{th}}$ & $\mathrm{AFib_{th}}$ & $\mathrm{stb}$ & Eff. & Dist.\\ \hline
1 & 300 & 375 & 2.0 & 3.5 & 0.92 & 316 & 22 & 0.01 & 2\\ 
2 & 273 & 364 & 2.0 & 4.0 & 0.91 & 300 & 14 & 0.08 & 3\\ 
3 & 273 & 353 & 3.0 & 4.5 & 0.91 & 375 & 28 & 0.16 & 4\\ 
4 & 261 & 343 & 3.0 & 5.0 & 0.89 & 286 & 22 & 0.24 & 5\\ 
5 & 261 & 334 & 3.5 & 6.0 & 0.88 & 286 & 32 & 0.37 & 6\\ 
6 & 250 & 325 & 4.0 & 7.0 & 0.87 & 600 & 35 & 0.62 & 7\\ 
7 & 273 & 316 & 4.5 & 8.0 & 0.86 & 375 & 6 & 0.84 & 8\\ 
8 & 273 & 308 & 6.0 & 9.0 & 0.85 & 231 & 45 & 0.94 & 9\\ 
9 & 250 & 300 & 4.5 & 10.0 & 0.90 & 546 & 6 & 0.96 & 10\\ 
10 & 286 & 293 & 8.0 & 11.0 & 0.90 & 215 & 8 & 0.99 & 11\\ 
11 & 240 & 293 & 10.0 & 13.0 & 0.86 & 600 & 35 & 1.00 & 13\\  \hline
\end{tabular}
\end{footnotesize}
\caption{Pareto-optimal reprogramming attacks for condition 10. Units are as in Table~\ref{tbl_app:cond1}.}
\label{tbl_app:cond10}
\end{table}

\begin{table}
\centering
\begin{footnotesize}
\begin{tabular}{r | lllllll | ll}
\multicolumn{10}{c}{\textbf{Condition 11}}\\
\# & $\mathrm{VF_{th}}$ & $\mathrm{VT_{th}}$ & $\mathrm{VFdur}$ & $\mathrm{VTdur}$ & $\mathrm{NSRcor_{th}}$ & $\mathrm{AFib_{th}}$ & $\mathrm{stb}$ & Eff. & Dist.\\ \hline
1 & 400 & 600 & 1.0 & 1.0 & 0.85 & 223 & 18 & 0.01 & 10\\ 
2 & 375 & 632 & 1.0 & 1.5 & 0.96 & 215 & 6 & 0.03 & 11\\ 
3 & 375 & 667 & 1.0 & 1.0 & 0.96 & 215 & 6 & 0.06 & 12\\ \hline
\end{tabular}
\end{footnotesize}
\caption{Pareto-optimal reprogramming attacks for condition 11. Units are as in Table~\ref{tbl_app:cond1}.}
\end{table}

\begin{table}
\centering
\begin{footnotesize}
\begin{tabular}{r | lllllll | ll}
\multicolumn{10}{c}{\textbf{Condition 12}}\\
\# & $\mathrm{VF_{th}}$ & $\mathrm{VT_{th}}$ & $\mathrm{VFdur}$ & $\mathrm{VTdur}$ & $\mathrm{NSRcor_{th}}$ & $\mathrm{AFib_{th}}$ & $\mathrm{stb}$ & Eff. & Dist.\\ \hline
1 & 325 & 375 & 1.0 & 1.0 & 0.96 & 353 & 18 & 0.01 & 3\\ 
2 & 353 & 375 & 1.0 & 1.0 & 0.96 & 286 & 10 & 0.02 & 6\\ 
3 & 364 & 388 & 1.0 & 1.0 & 0.96 & 429 & 8 & 0.09 & 7\\ 
4 & 375 & 388 & 1.0 & 6.0 & 0.95 & 429 & 8 & 0.20 & 8\\ 
5 & 388 & 400 & 1.0 & 1.0 & 0.96 & 231 & 45 & 0.29 & 9\\ 
6 & 400 & 414 & 1.0 & 1.5 & 0.96 & 231 & 50 & 0.33 & 10\\ 
7 & 429 & 667 & 1.0 & 7.0 & 0.82 & 231 & 10 & 0.35 & 12\\ 
8 & 445 & 546 & 1.0 & 13.0 & 0.81 & 231 & 10 & 0.36 & 13\\ 
9 & 462 & 667 & 1.0 & 2.5 & 0.96 & 231 & 10 & 0.38 & 14\\ 
10 & 480 & 667 & 1.0 & 2.5 & 0.96 & 231 & 10 & 0.41 & 15\\ 
11 & 500 & 667 & 1.0 & 2.5 & 0.96 & 231 & 10 & 0.47 & 16\\ 
12 & 522 & 667 & 1.0 & 1.0 & 0.92 & 231 & 10 & 0.58 & 17\\ 
13 & 546 & 600 & 1.0 & 2.5 & 0.96 & 231 & 10 & 0.75 & 18\\ \hline
\end{tabular}
\end{footnotesize}
\caption{Pareto-optimal reprogramming attacks for condition 12. Units are as in Table~\ref{tbl_app:cond1}.}
\end{table}

\begin{table}
\centering
\begin{footnotesize}
\begin{tabular}{r | lllllll | ll}
\multicolumn{10}{c}{\textbf{Condition 13}}\\
\# & $\mathrm{VF_{th}}$ & $\mathrm{VT_{th}}$ & $\mathrm{VFdur}$ & $\mathrm{VTdur}$ & $\mathrm{NSRcor_{th}}$ & $\mathrm{AFib_{th}}$ & $\mathrm{stb}$ & Eff. & Dist.\\ \hline
1 & 462 & 546 & 1.0 & 14.0 & 0.96 & 600 & 32 & 0.01 & 14\\ 
2 & 480 & 500 & 1.0 & 14.0 & 0.96 & 600 & 32 & 0.02 & 15\\ 
3 & 500 & 522 & 1.0 & 14.0 & 0.96 & 600 & 32 & 0.06 & 16\\ 
4 & 522 & 546 & 1.0 & 14.0 & 0.96 & 600 & 32 & 0.13 & 17\\ 
5 & 546 & 572 & 1.0 & 3.5 & 0.96 & 429 & 120 & 0.20 & 18\\ \hline
\end{tabular}
\end{footnotesize}
\caption{Pareto-optimal reprogramming attacks for condition 13. Units are as in Table~\ref{tbl_app:cond1}.}
\end{table}

\begin{table}
\centering
\begin{footnotesize}
\begin{tabular}{r | lllllll | ll}
\multicolumn{10}{c}{\textbf{Condition 14}}\\
\# & $\mathrm{VF_{th}}$ & $\mathrm{VT_{th}}$ & $\mathrm{VFdur}$ & $\mathrm{VTdur}$ & $\mathrm{NSRcor_{th}}$ & $\mathrm{AFib_{th}}$ & $\mathrm{stb}$ & Eff. & Dist.\\ \hline
1 & 325 & 667 & 1.0 & 1.5 & 0.96 & 261 & 28 & 0.01 & 12\\ 
2 & 462 & 667 & 1.0 & 1.5 & 0.80 & 600 & 24 & 0.02 & 14\\ 
3 & 480 & 667 & 1.0 & 2.0 & 0.96 & 261 & 28 & 0.03 & 15\\ 
4 & 500 & 667 & 1.0 & 1.5 & 0.96 & 261 & 28 & 0.06 & 16\\ 
5 & 522 & 667 & 1.0 & 1.5 & 0.96 & 261 & 28 & 0.12 & 17\\ 
6 & 546 & 667 & 1.0 & 2.0 & 0.96 & 261 & 28 & 0.16 & 18\\  \hline
\end{tabular}
\end{footnotesize}
\caption{Pareto-optimal reprogramming attacks for condition 14. Units are as in Table~\ref{tbl_app:cond1}.}
\end{table}

\begin{table}
\centering
\begin{footnotesize}
\begin{tabular}{r | lllllll | ll}
\multicolumn{10}{c}{\textbf{Condition 15}}\\
\# & $\mathrm{VF_{th}}$ & $\mathrm{VT_{th}}$ & $\mathrm{VFdur}$ & $\mathrm{VTdur}$ & $\mathrm{NSRcor_{th}}$ & $\mathrm{AFib_{th}}$ & $\mathrm{stb}$ & Eff. & Dist.\\ \hline
1 & 414 & 445 & 1.0 & 2.5 & 0.96 & 500 & 18 & 0.01 & 11\\ 
2 & 429 & 445 & 1.0 & 2.5 & 0.96 & 500 & 18 & 0.09 & 12\\ 
3 & 445 & 462 & 1.0 & 4.0 & 0.96 & 500 & 18 & 0.33 & 13\\ 
4 & 462 & 667 & 1.0 & 14.0 & 0.81 & 375 & 8 & 0.75 & 14\\ 
5 & 480 & 572 & 1.0 & 15.0 & 0.96 & 375 & 90 & 0.89 & 15\\ 
6 & 500 & 572 & 1.5 & 20.0 & 0.96 & 375 & 80 & 0.92 & 16\\ \hline
\end{tabular}
\end{footnotesize}
\caption{Pareto-optimal reprogramming attacks for condition 15. Units are as in Table~\ref{tbl_app:cond1}.}
\end{table}

\begin{table}
\centering
\begin{footnotesize}
\begin{tabular}{r | lllllll | ll}
\multicolumn{10}{c}{\textbf{Condition 16}}\\
\# & $\mathrm{VF_{th}}$ & $\mathrm{VT_{th}}$ & $\mathrm{VFdur}$ & $\mathrm{VTdur}$ & $\mathrm{NSRcor_{th}}$ & $\mathrm{AFib_{th}}$ & $\mathrm{stb}$ & Eff. & Dist.\\ \hline
1 & 308 & 388 & 1.0 & 3.0 & 0.95 & 375 & 18 & 0.02 & 1\\ 
2 & 293 & 375 & 2.0 & 3.5 & 0.92 & 316 & 16 & 0.07 & 2\\ 
3 & 273 & 364 & 2.5 & 4.0 & 0.91 & 300 & 16 & 0.11 & 3\\ 
4 & 261 & 353 & 3.0 & 4.5 & 0.90 & 286 & 28 & 0.19 & 4\\ 
5 & 250 & 353 & 1.5 & 5.0 & 0.91 & 300 & 30 & 0.20 & 5\\ 
6 & 240 & 334 & 4.0 & 6.0 & 0.90 & 546 & 14 & 0.34 & 6\\ 
7 & 273 & 325 & 4.0 & 7.0 & 0.90 & 286 & 18 & 0.58 & 7\\ 
8 & 240 & 316 & 1.5 & 8.0 & 0.90 & 546 & 14 & 0.77 & 8\\ 
9 & 261 & 308 & 4.0 & 9.0 & 0.85 & 400 & 32 & 0.91 & 9\\ 
10 & 286 & 300 & 7.0 & 10.0 & 0.91 & 375 & 18 & 0.97 & 10\\ 
11 & 286 & 293 & 7.0 & 11.0 & 0.91 & 375 & 18 & 0.99 & 11\\  \hline
\end{tabular}
\end{footnotesize}
\caption{Pareto-optimal reprogramming attacks for condition 16. Units are as in Table~\ref{tbl_app:cond1}.}
\end{table}

\begin{table}
\centering
\begin{footnotesize}
\begin{tabular}{r | lllllll | ll}
\multicolumn{10}{c}{\textbf{Condition 17}}\\
\# & $\mathrm{VF_{th}}$ & $\mathrm{VT_{th}}$ & $\mathrm{VFdur}$ & $\mathrm{VTdur}$ & $\mathrm{NSRcor_{th}}$ & $\mathrm{AFib_{th}}$ & $\mathrm{stb}$ & Eff. & Dist.\\ \hline
1 & 240 & 334 & 2.5 & 6.0 & 0.88 & 316 & 32 & 0.05 & 6\\ 
2 & 300 & 325 & 4.0 & 7.0 & 0.87 & 353 & 35 & 0.12 & 7\\ 
3 & 308 & 316 & 2.0 & 8.0 & 0.86 & 546 & 40 & 0.25 & 8\\ 
4 & 300 & 308 & 2.5 & 9.0 & 0.85 & 546 & 45 & 0.36 & 9\\ 
5 & 250 & 300 & 1.0 & 7.0 & 0.84 & 546 & 50 & 0.44 & 10\\ 
6 & 273 & 293 & 4.0 & 11.0 & 0.83 & 600 & 45 & 0.52 & 11\\ 
7 & 261 & 286 & 1.5 & 12.0 & 0.82 & 600 & 50 & 0.61 & 12\\ 
8 & 250 & 273 & 1.0 & 13.0 & 0.86 & 546 & 40 & 0.77 & 13\\ 
9 & 250 & 273 & 9.0 & 14.0 & 0.89 & 600 & 60 & 0.78 & 14\\ 
10 & 250 & 400 & 5.0 & 20.0 & 0.78 & 207 & 60 & 1.00 & 16\\ \hline
\end{tabular}
\end{footnotesize}
\caption{Pareto-optimal reprogramming attacks for condition 17. Units are as in Table~\ref{tbl_app:cond1}.}
\end{table}

\begin{table}
\centering
\begin{footnotesize}
\begin{tabular}{r | lllllll | ll}
\multicolumn{10}{c}{\textbf{Condition 18}}\\
\# & $\mathrm{VF_{th}}$ & $\mathrm{VT_{th}}$ & $\mathrm{VFdur}$ & $\mathrm{VTdur}$ & $\mathrm{NSRcor_{th}}$ & $\mathrm{AFib_{th}}$ & $\mathrm{stb}$ & Eff. & Dist.\\ \hline
1 & 300 & 334 & 4.0 & 6.0 & 0.88 & 546 & 32 & 0.04 & 6\\ 
2 & 316 & 325 & 4.0 & 7.0 & 0.87 & 600 & 35 & 0.16 & 7\\ 
3 & 300 & 316 & 1.0 & 8.0 & 0.86 & 462 & 40 & 0.28 & 8\\ 
4 & 300 & 308 & 3.0 & 9.0 & 0.86 & 462 & 45 & 0.34 & 9\\ 
5 & 293 & 300 & 4.0 & 10.0 & 0.86 & 462 & 50 & 0.48 & 10\\ 
6 & 261 & 293 & 5.0 & 9.0 & 0.86 & 429 & 50 & 0.57 & 11\\ 
7 & 273 & 286 & 1.0 & 12.0 & 0.84 & 429 & 12 & 0.67 & 12\\ 
8 & 240 & 273 & 9.0 & 13.0 & 0.85 & 500 & 50 & 0.81 & 13\\ 
9 & 240 & 273 & 9.0 & 15.0 & 0.85 & 500 & 50 & 0.82 & 15\\ 
10 & 250 & 546 & 1.0 & 20.0 & 0.78 & 429 & 100 & 1.00 & 16\\ \hline
\end{tabular}
\end{footnotesize}
\caption{Pareto-optimal reprogramming attacks for condition 18. Units are as in Table~\ref{tbl_app:cond1}.}
\end{table}

\begin{table}
\centering
\begin{footnotesize}
\begin{tabular}{r | lllllll | ll}
\multicolumn{10}{c}{\textbf{Condition 19}}\\
\# & $\mathrm{VF_{th}}$ & $\mathrm{VT_{th}}$ & $\mathrm{VFdur}$ & $\mathrm{VTdur}$ & $\mathrm{NSRcor_{th}}$ & $\mathrm{AFib_{th}}$ & $\mathrm{stb}$ & Eff. & Dist.\\ \hline
1 & 250 & 334 & 1.5 & 6.0 & 0.88 & 546 & 32 & 0.04 & 6\\ 
2 & 316 & 325 & 1.0 & 7.0 & 0.87 & 500 & 35 & 0.16 & 7\\ 
3 & 261 & 316 & 1.5 & 8.0 & 0.86 & 316 & 40 & 0.29 & 8\\ 
4 & 273 & 308 & 4.0 & 8.0 & 0.85 & 300 & 45 & 0.35 & 9\\ 
5 & 273 & 300 & 7.0 & 10.0 & 0.84 & 429 & 50 & 0.47 & 10\\ 
6 & 261 & 293 & 7.0 & 10.0 & 0.83 & 250 & 45 & 0.52 & 11\\ 
7 & 273 & 286 & 2.0 & 11.0 & 0.82 & 286 & 60 & 0.58 & 12\\ 
8 & 250 & 273 & 6.0 & 13.0 & 0.81 & 316 & 50 & 0.82 & 13\\ 
9 & 261 & 273 & 2.5 & 14.0 & 0.80 & 286 & 80 & 0.83 & 14\\ 
10 & 250 & 286 & 8.0 & 20.0 & 0.78 & 546 & 100 & 1.00 & 16\\ \hline
\end{tabular}
\end{footnotesize}
\caption{Pareto-optimal reprogramming attacks for condition 19. Units are as in Table~\ref{tbl_app:cond1}.}
\label{tbl_app:cond19}
\end{table}

\begin{table}
\centering
\begin{footnotesize}
\begin{tabular}{r | lllllll | ll}
\multicolumn{10}{c}{\textbf{SVT conditions}}\\
\# & $\mathrm{VF_{th}}$ & $\mathrm{VT_{th}}$ & $\mathrm{VFdur}$ & $\mathrm{VTdur}$ & $\mathrm{NSRcor_{th}}$ & $\mathrm{AFib_{th}}$ & $\mathrm{stb}$ & Eff. & Dist.\\ \hline
1 & 316 & 429 & 1.5 & 1.5 & 0.96 & 400 & 24 & 0.005 & 2\\ 
2 & 273 & 445 & 1.5 & 1.5 & 0.96 & 353 & 22 & 0.010 & 3\\ 
3 & 334 & 462 & 1.0 & 1.0 & 0.96 & 353 & 22 & 0.015 & 4\\ 
4 & 343 & 480 & 1.0 & 1.0 & 0.96 & 400 & 30 & 0.035 & 5\\ 
5 & 353 & 500 & 1.0 & 1.0 & 0.96 & 353 & 32 & 0.045 & 6\\ 
6 & 364 & 500 & 1.0 & 1.0 & 0.96 & 353 & 24 & 0.060 & 7\\ 
7 & 375 & 546 & 1.0 & 1.5 & 0.96 & 400 & 30 & 0.085 & 8\\ 
8 & 388 & 572 & 1.0 & 1.0 & 0.85 & 353 & 24 & 0.090 & 9\\ 
9 & 400 & 600 & 1.0 & 1.0 & 0.85 & 353 & 24 & 0.095 & 10\\ 
10 & 388 & 632 & 1.0 & 1.0 & 0.96 & 353 & 24 & 0.115 & 11\\ 
11 & 429 & 667 & 1.0 & 1.0 & 0.82 & 353 & 24 & 0.135 & 12\\ 
12 & 445 & 667 & 1.0 & 1.0 & 0.81 & 353 & 24 & 0.200 & 13\\ 
13 & 462 & 667 & 1.0 & 1.0 & 0.80 & 353 & 24 & 0.230 & 14\\ 
14 & 480 & 667 & 1.0 & 1.0 & 0.81 & 353 & 24 & 0.285 & 15\\ 
15 & 500 & 667 & 1.0 & 1.0 & 0.80 & 353 & 24 & 0.320 & 16\\ 
16 & 522 & 667 & 1.0 & 1.0 & 0.96 & 353 & 24 & 0.380 & 17\\ 
17 & 546 & 667 & 1.0 & 1.0 & 0.80 & 353 & 24 & 0.490 & 18\\   \hline
\end{tabular}
\end{footnotesize}
\caption{Pareto-optimal reprogramming attacks for the group of SVT conditions. Units are as in Table~\ref{tbl_app:cond1}.}
\label{tbl_app:params_SVT}
\end{table}

\begin{table}
\centering
\begin{footnotesize}
\begin{tabular}{r | lllllll | ll}
\multicolumn{10}{c}{\textbf{VT conditions}}\\
\# & $\mathrm{VF_{th}}$ & $\mathrm{VT_{th}}$ & $\mathrm{VFdur}$ & $\mathrm{VTdur}$ & $\mathrm{NSRcor_{th}}$ & $\mathrm{AFib_{th}}$ & $\mathrm{stb}$ & Eff. & Dist.\\ \hline
1 & 300 & 388 & 1.5 & 3.0 & 0.95 & 334 & 18 & 0.005 & 1\\ 
2 & 293 & 375 & 2.0 & 3.0 & 0.92 & 334 & 24 & 0.010 & 2\\ 
3 & 273 & 364 & 2.5 & 4.0 & 0.93 & 300 & 26 & 0.040 & 3\\ 
4 & 261 & 353 & 3.0 & 4.5 & 0.96 & 375 & 28 & 0.060 & 4\\ 
5 & 250 & 343 & 2.0 & 5.0 & 0.92 & 316 & 30 & 0.095 & 5\\ 
6 & 273 & 334 & 4.0 & 6.0 & 0.88 & 316 & 32 & 0.195 & 6\\ 
7 & 250 & 325 & 3.5 & 7.0 & 0.87 & 600 & 35 & 0.340 & 7\\ 
8 & 250 & 316 & 3.0 & 8.0 & 0.96 & 300 & 40 & 0.515 & 8\\ 
9 & 261 & 308 & 3.0 & 9.0 & 0.96 & 286 & 45 & 0.640 & 9\\ 
10 & 261 & 300 & 7.0 & 10.0 & 0.87 & 375 & 50 & 0.745 & 10\\ 
11 & 261 & 293 & 7.0 & 11.0 & 0.87 & 286 & 55 & 0.795 & 11\\ 
12 & 250 & 286 & 3.0 & 12.0 & 0.86 & 546 & 55 & 0.865 & 12\\ 
13 & 250 & 273 & 7.0 & 12.0 & 0.86 & 546 & 45 & 0.940 & 13\\ 
14 & 250 & 293 & 3.0 & 20.0 & 0.78 & 429 & 100 & 1.000 & 16\\   \hline
\end{tabular}
\end{footnotesize}
\caption{Pareto-optimal reprogramming attacks for the group of VT conditions. Units are as in Table~\ref{tbl_app:cond1}.}
\label{tbl_app:params_VT}
\end{table}

\begin{figure*}
\centering
\subfloat[Cond. 1, validation score -0.0217]{\includegraphics[width = .25\textwidth]{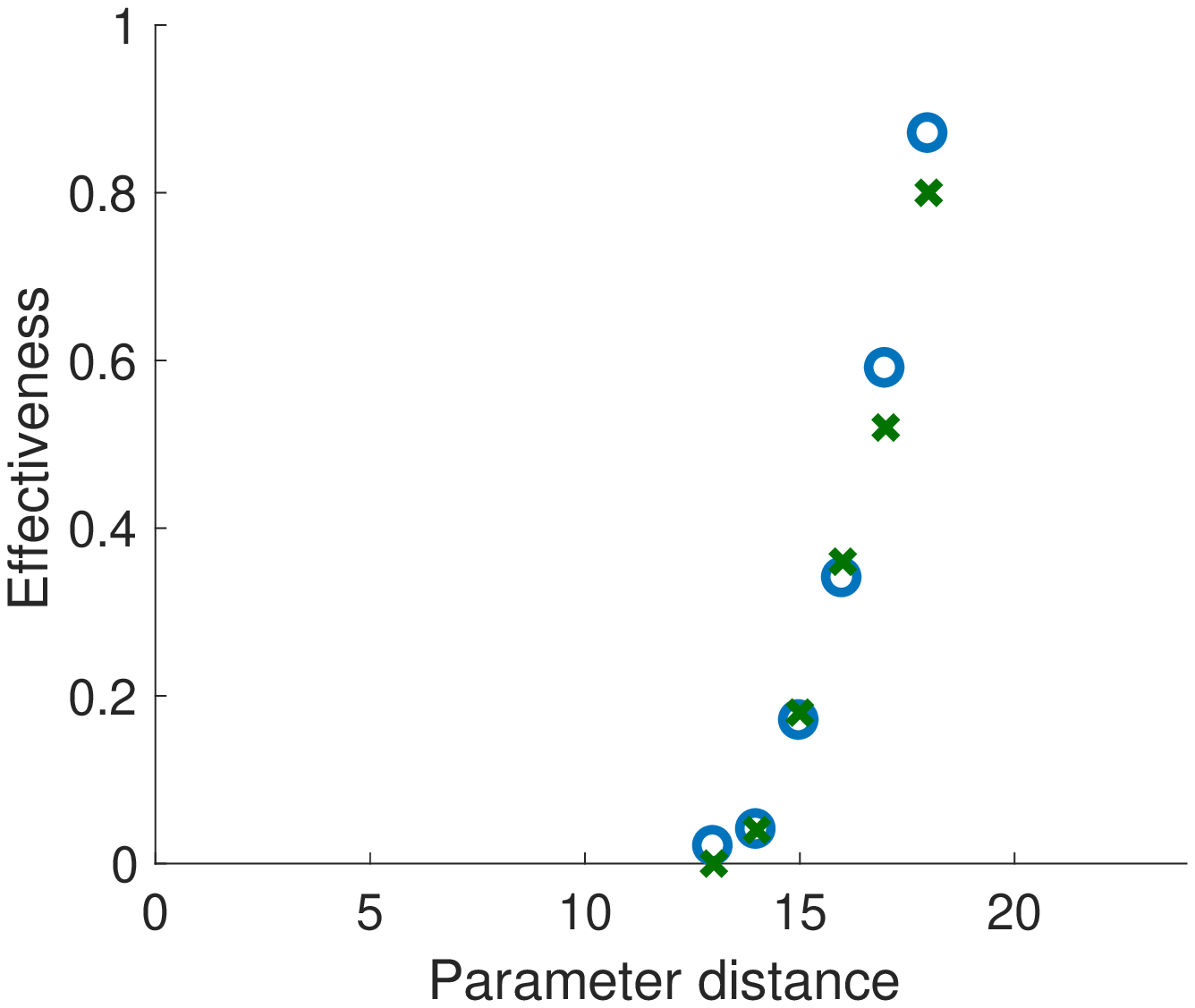}} \hfill 
\subfloat[Cond. 2, validation score -0.0433]{\includegraphics[width = .25\textwidth]{img/fronts/sigs_100_cond_2_Pareto_preciseBounds.eps}} \hfill 
\subfloat[Cond. 3, validation score -0.0033]{\includegraphics[width = .25\textwidth]{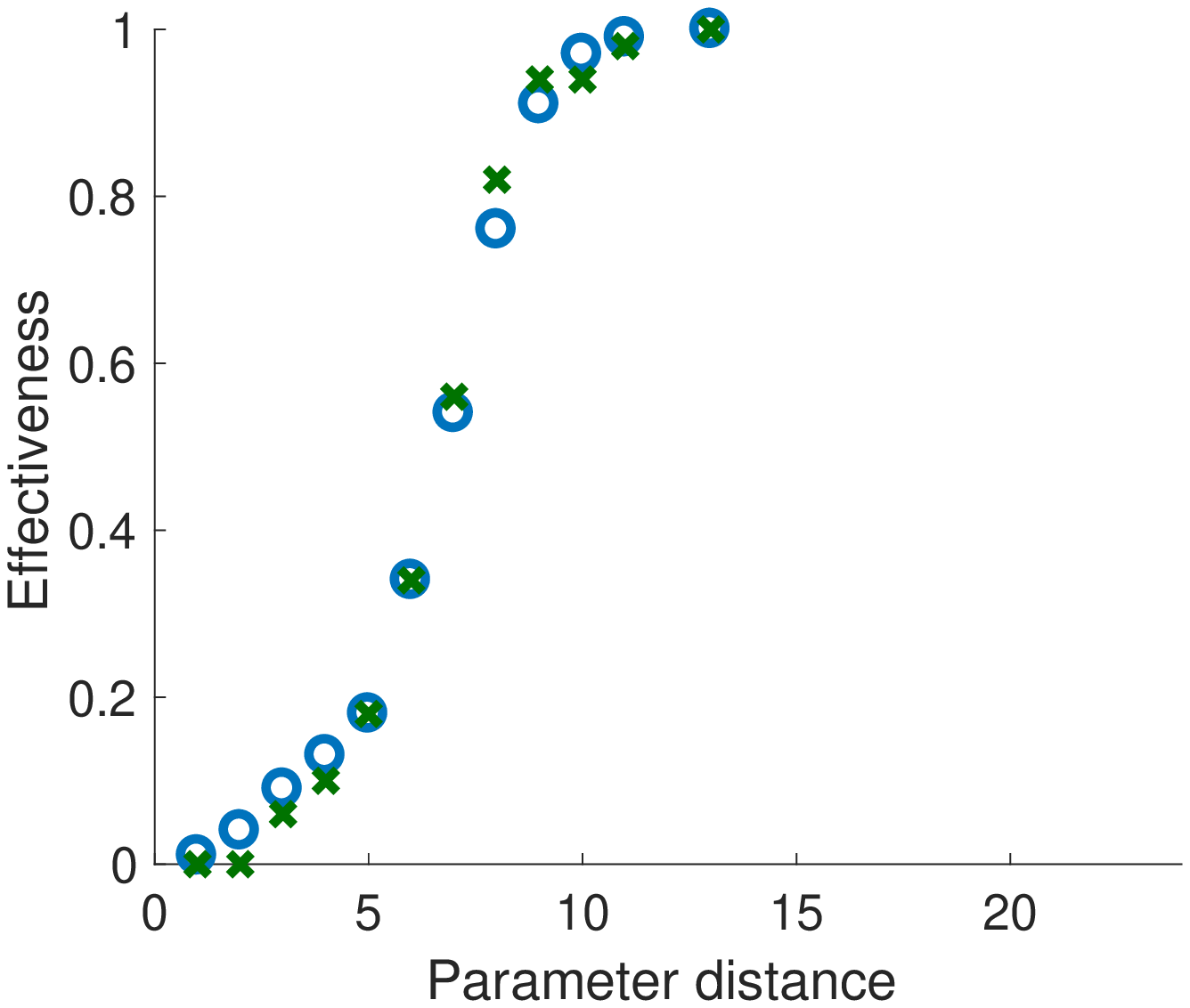}} \hfill 
\subfloat[Cond. 4, validation score 0.0025]{\includegraphics[width = .25\textwidth]{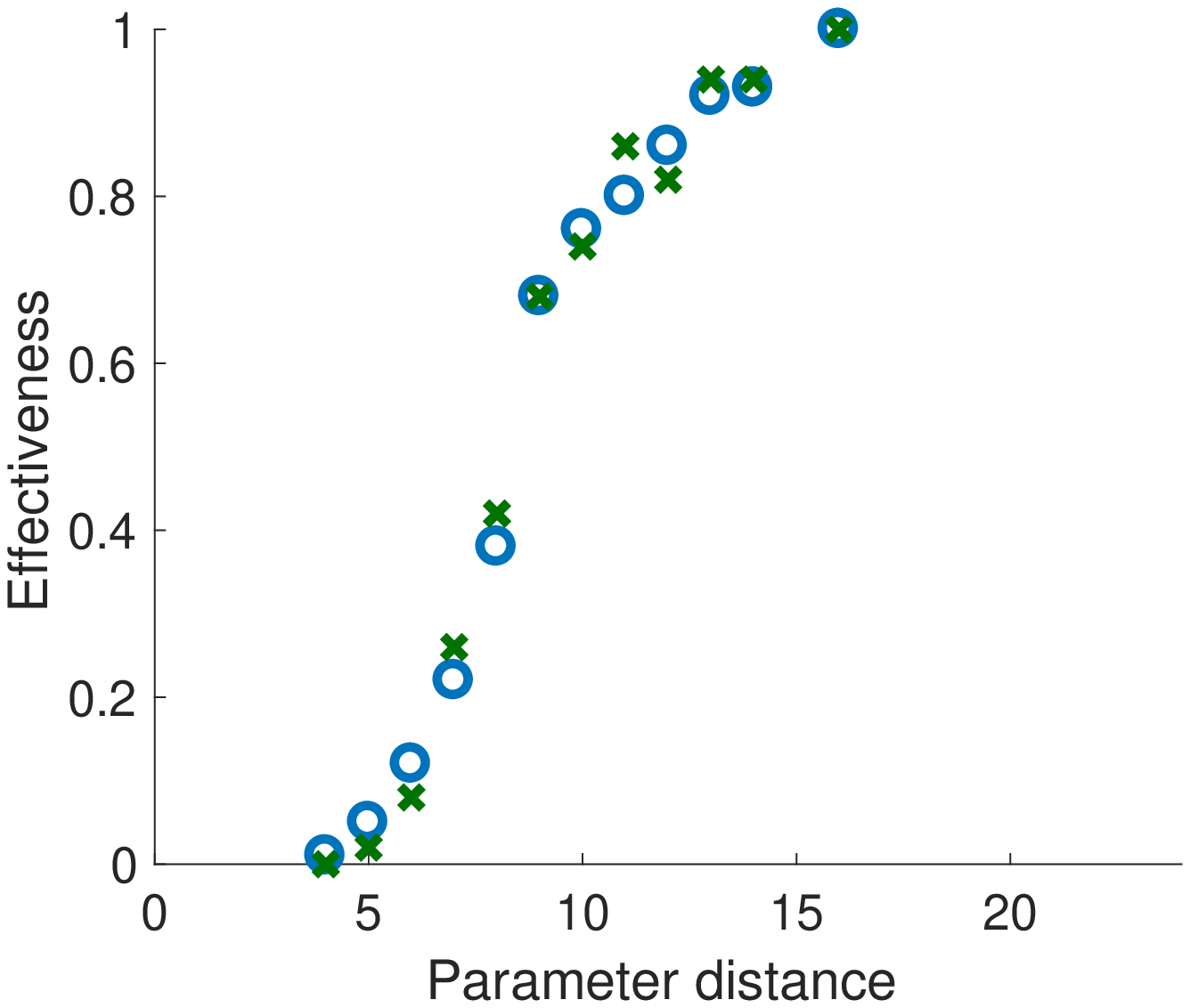}} \\
\subfloat[Cond. 5, validation score -0.0523]{\includegraphics[width = .25\textwidth]{img/fronts/sigs_100_cond_5_Pareto_preciseBounds.eps}} \hfill 
\subfloat[Cond. 6, validation score 0.02]{\includegraphics[width = .25\textwidth]{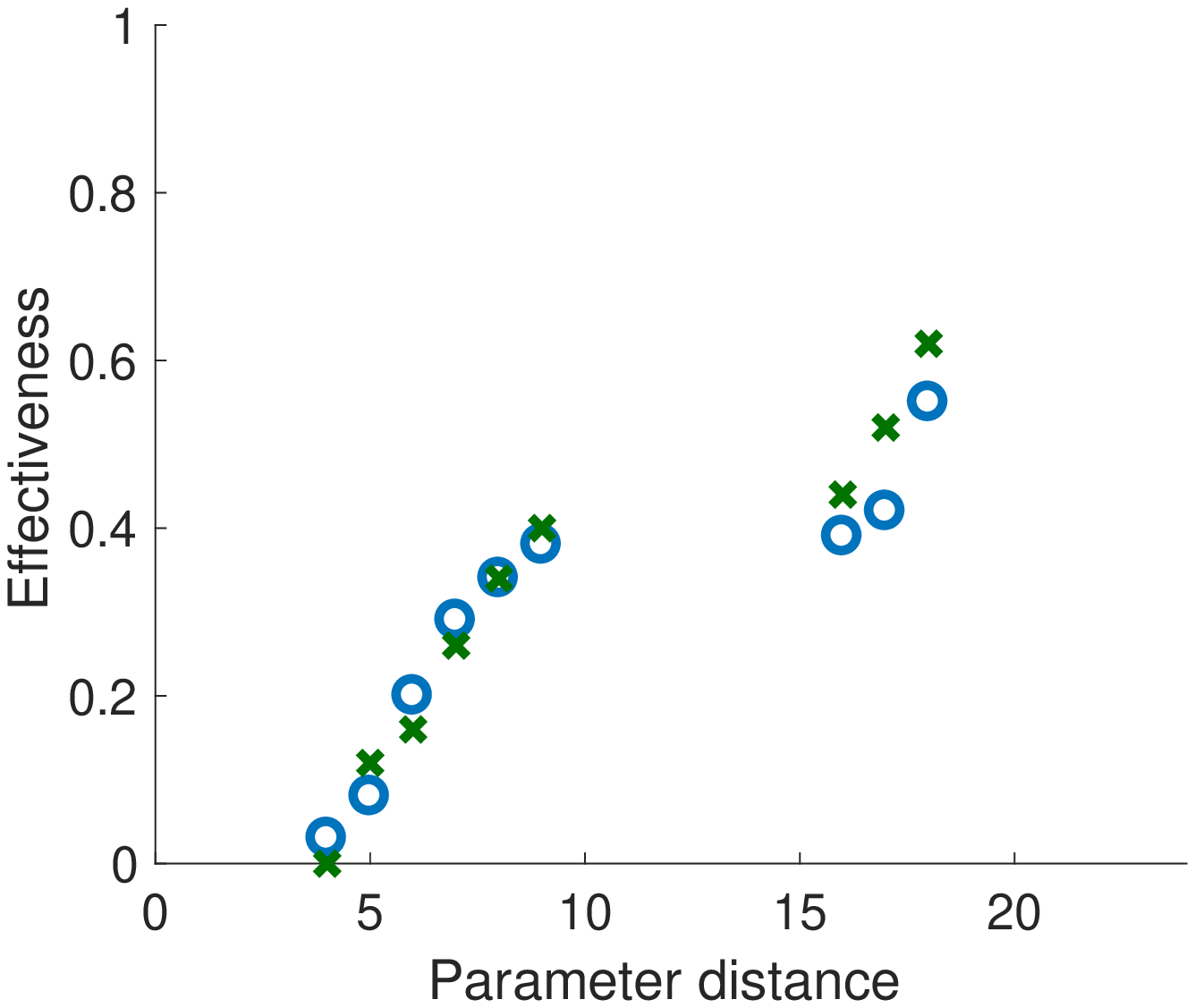}} \hfill 
\subfloat[Cond. 7, validation score -0.0593]{\includegraphics[width = .25\textwidth]{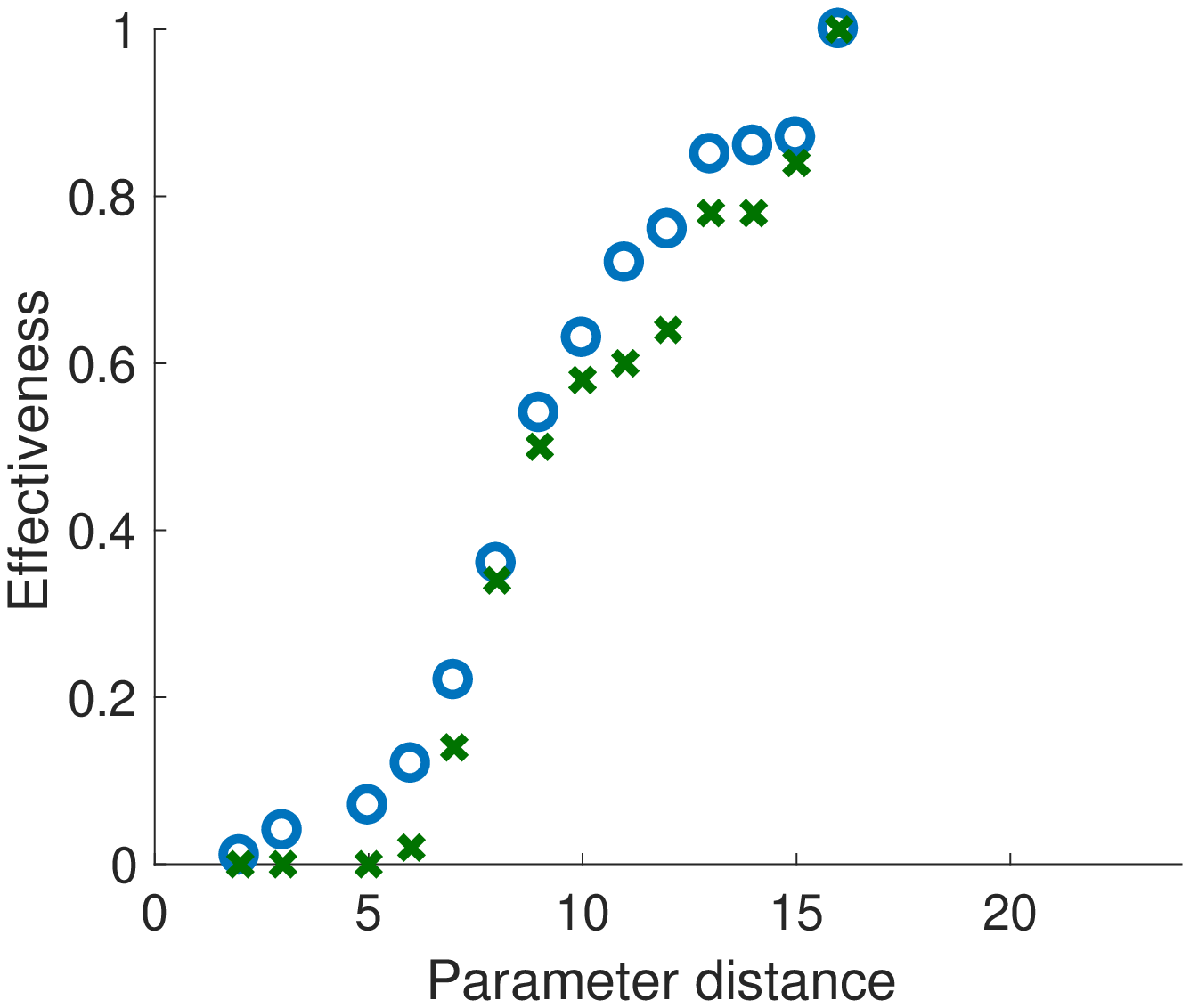}} \hfill 
\subfloat[Cond. 8, validation score -0.005]{\includegraphics[width = .25\textwidth]{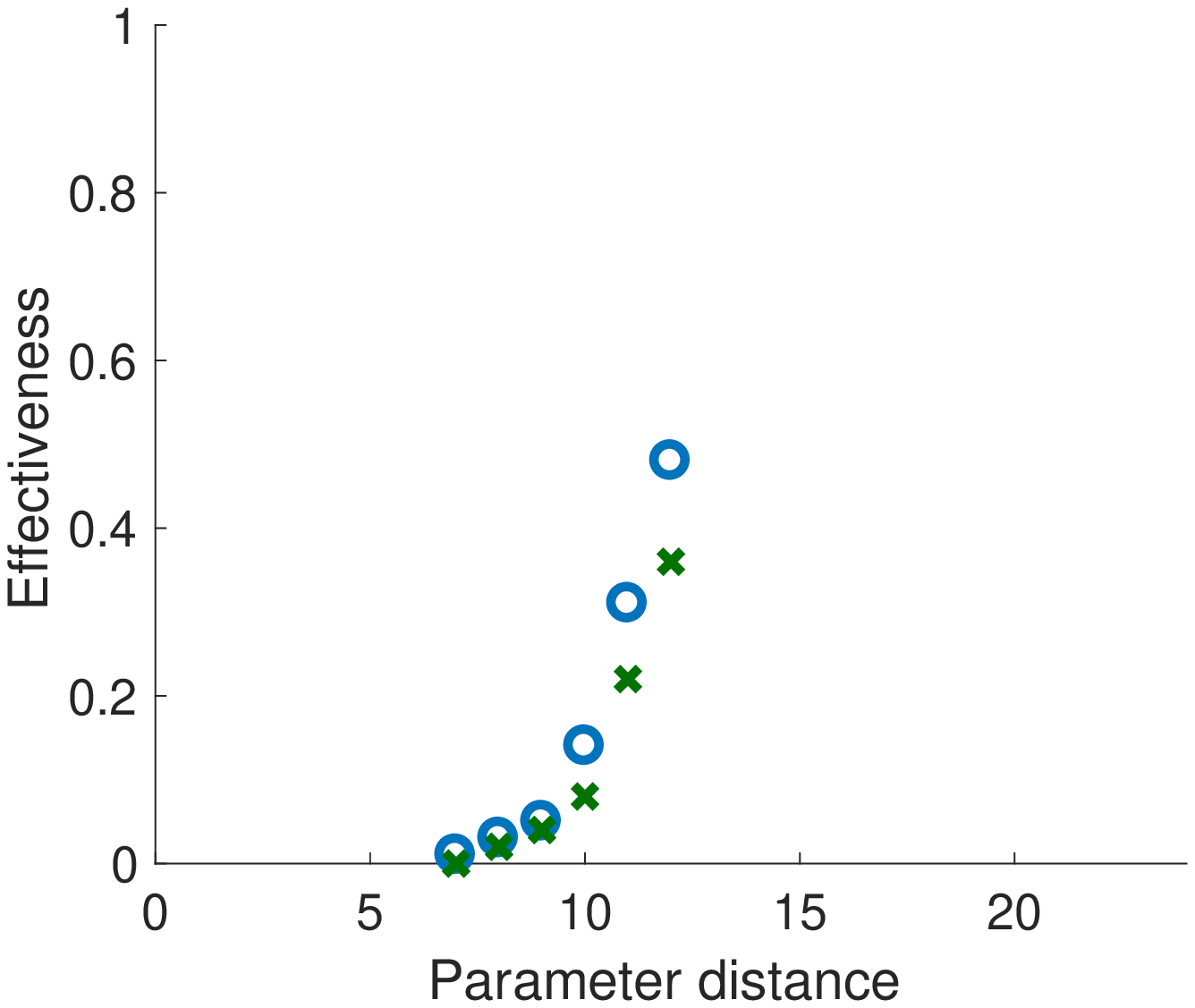}} \\
\subfloat[Cond. 10, validation score -0.0518]{\includegraphics[width = .25\textwidth]{img/fronts/sigs_100_cond_10_Pareto_preciseBounds.eps}} \hfill 
\subfloat[Cond. 11, validation score -0.0267]{\includegraphics[width = .25\textwidth]{img/fronts/sigs_100_cond_11_Pareto_preciseBounds.eps}} \hfill 
\subfloat[Cond. 12, validation score -0.0077]{\includegraphics[width = .25\textwidth]{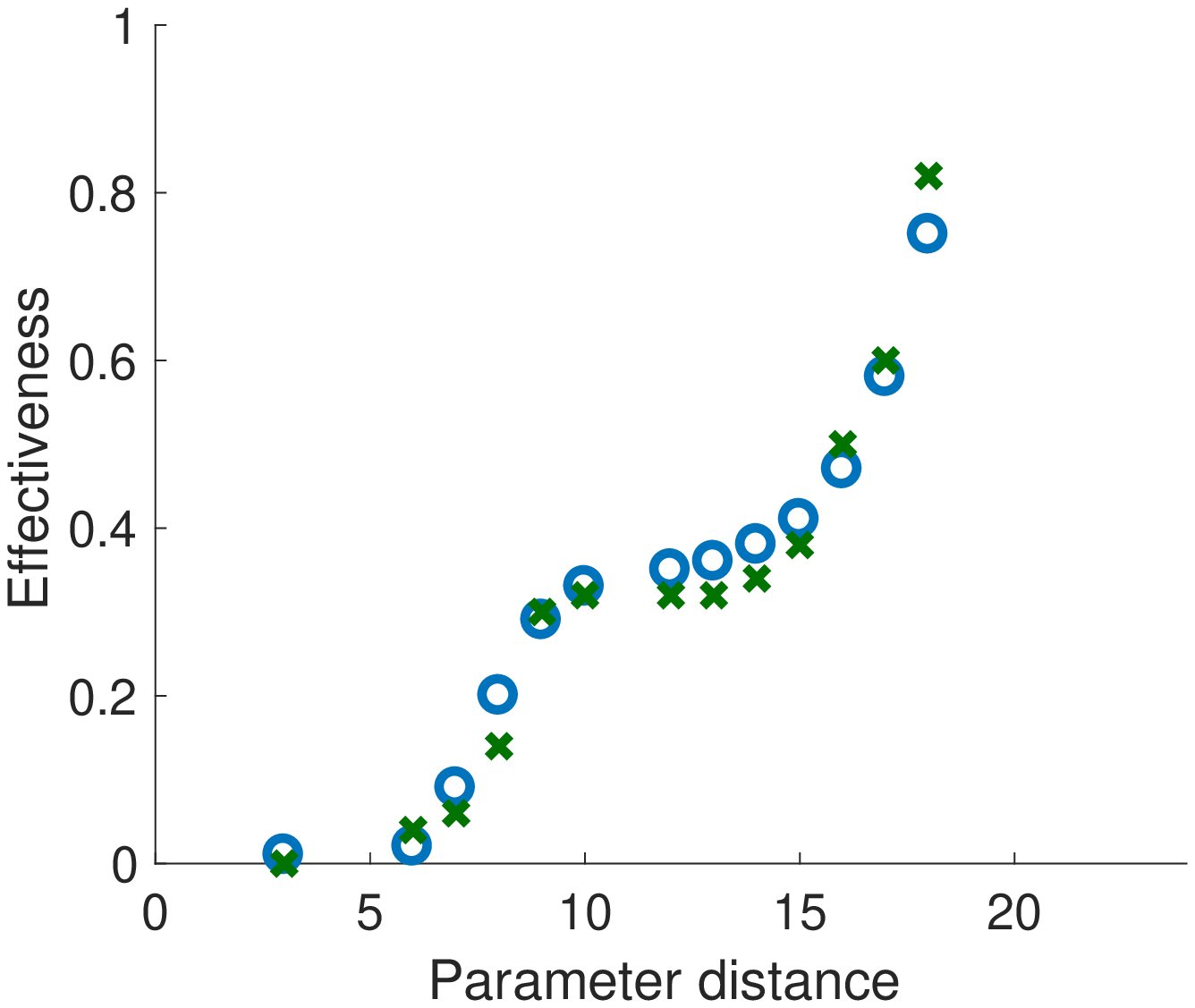}} \hfill 
\subfloat[Cond. 13, validation score -0.036]{\includegraphics[width = .25\textwidth]{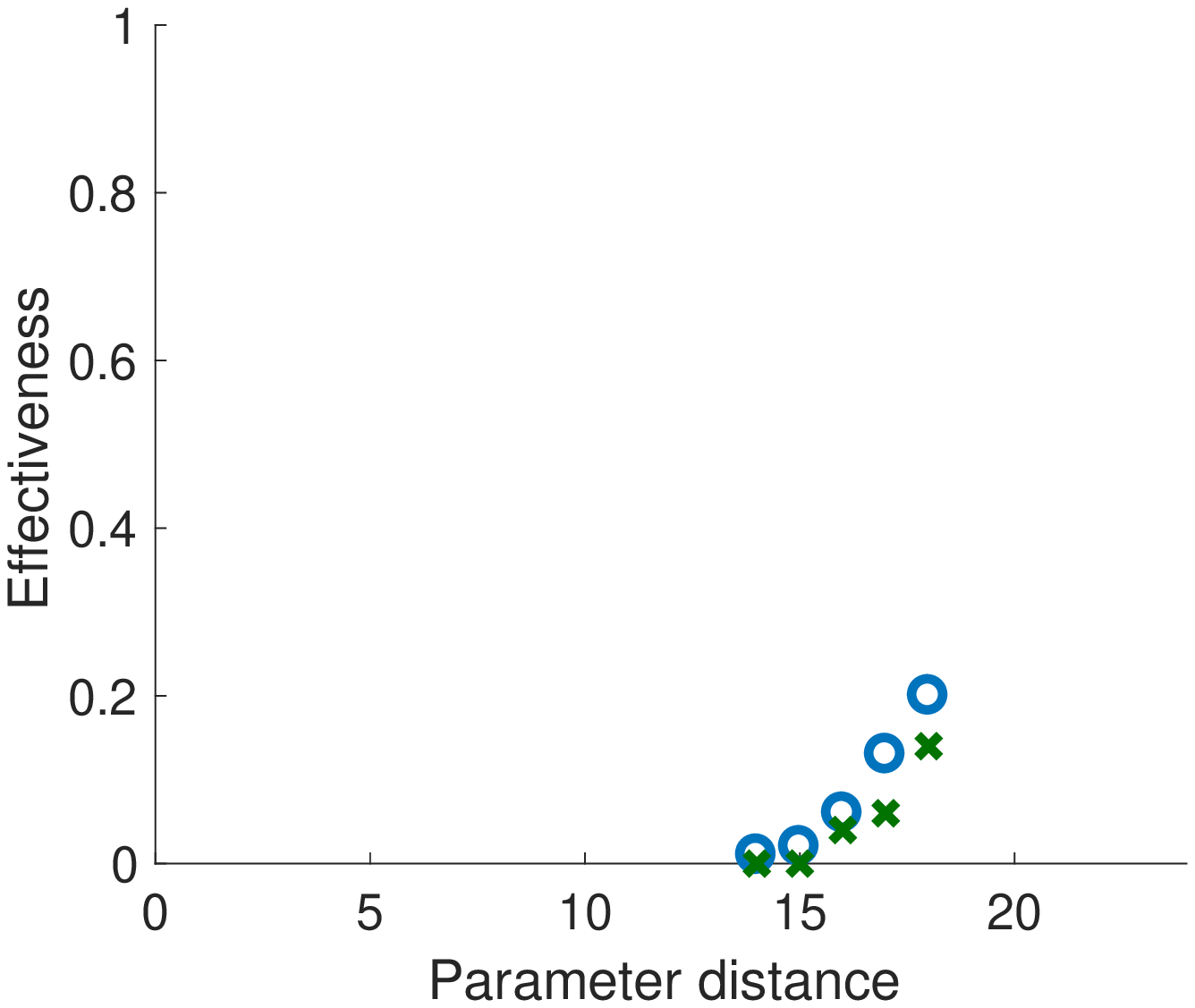}} \\ 
\subfloat[Cond. 14, validation score -0.01]{\includegraphics[width = .25\textwidth]{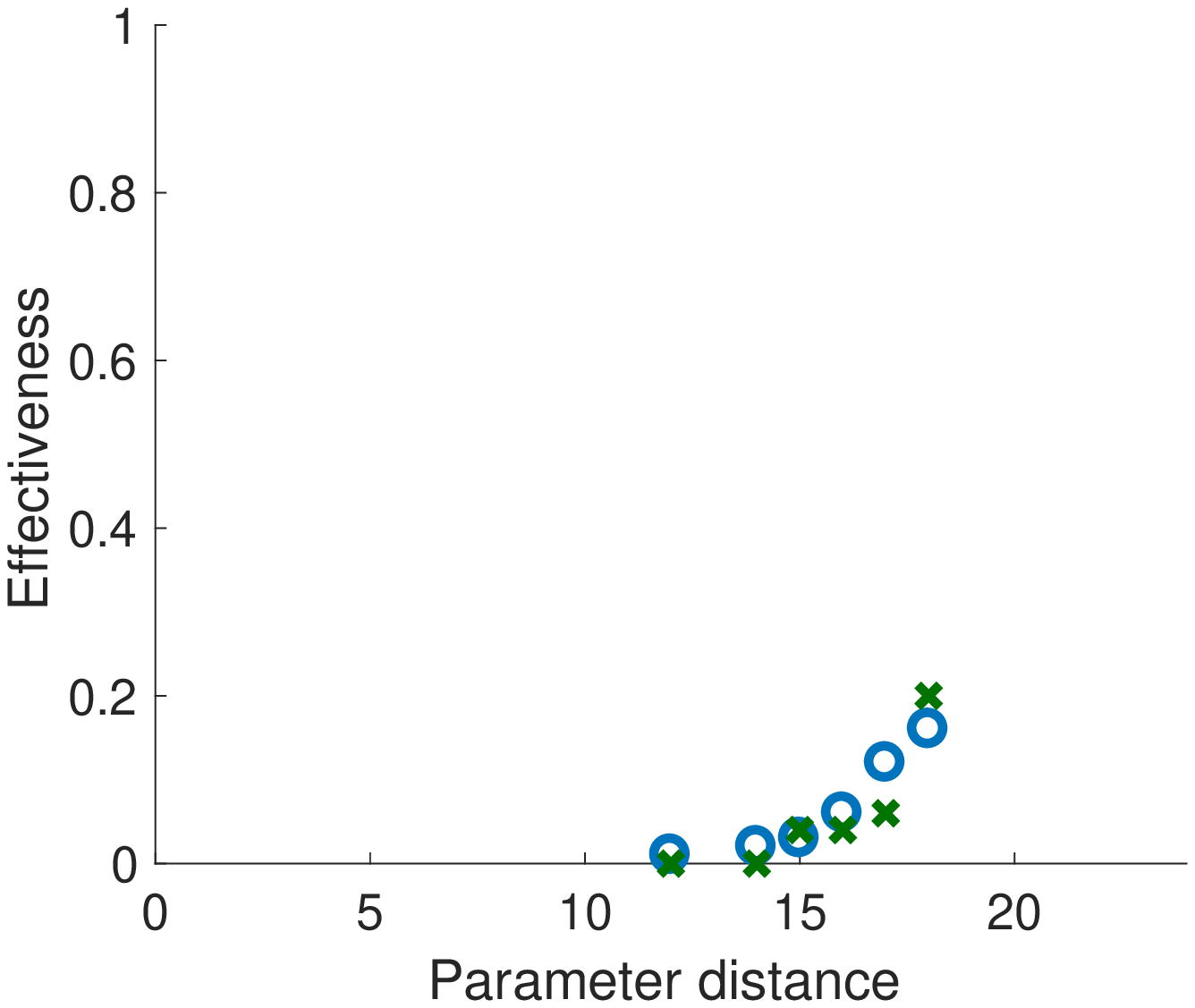}} \hfill
\subfloat[Cond. 15, validation score 0.0083]{\includegraphics[width = .25\textwidth]{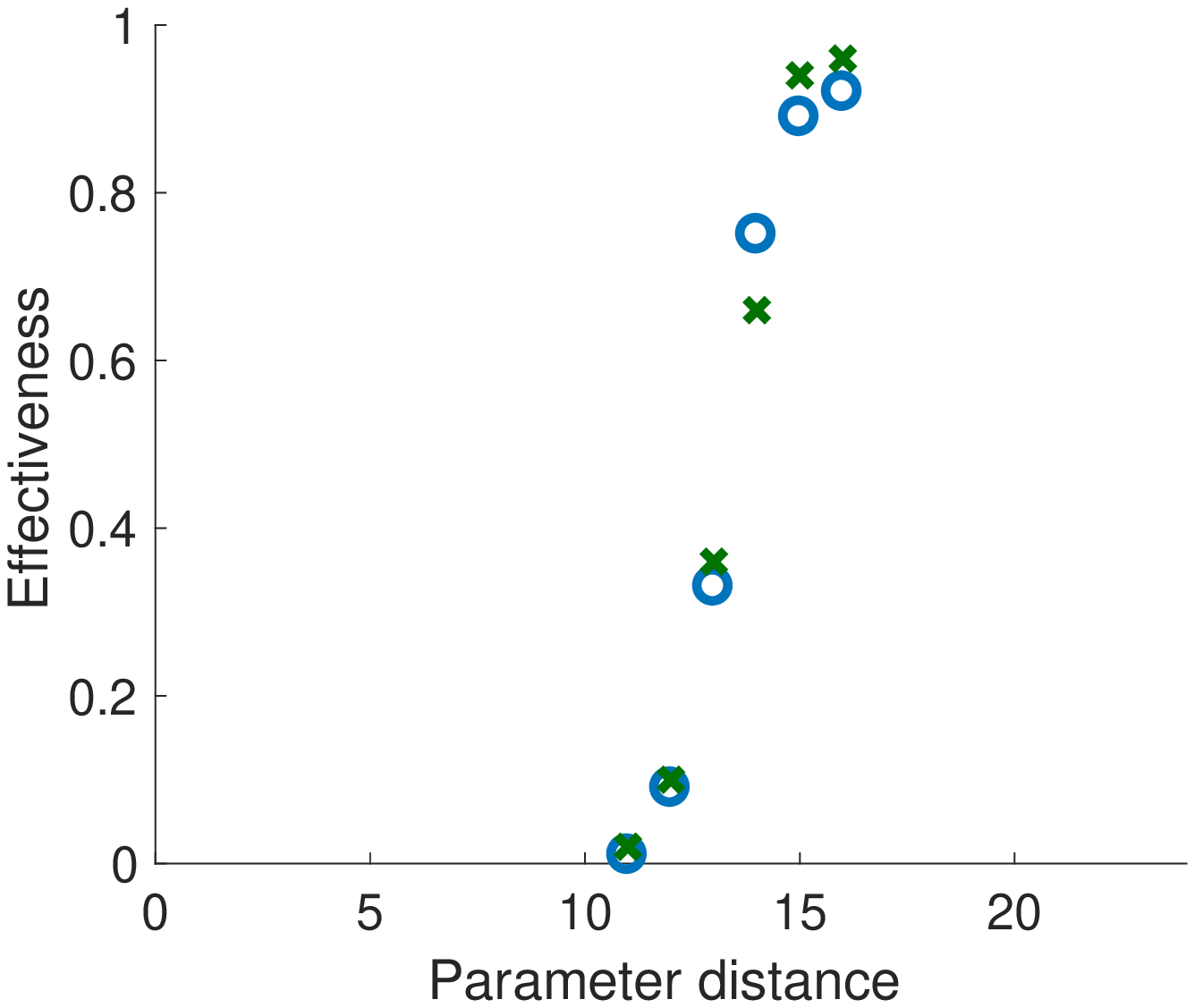}} \hfill 
\subfloat[Cond. 16, validation score -0.0064]{\includegraphics[width = .25\textwidth]{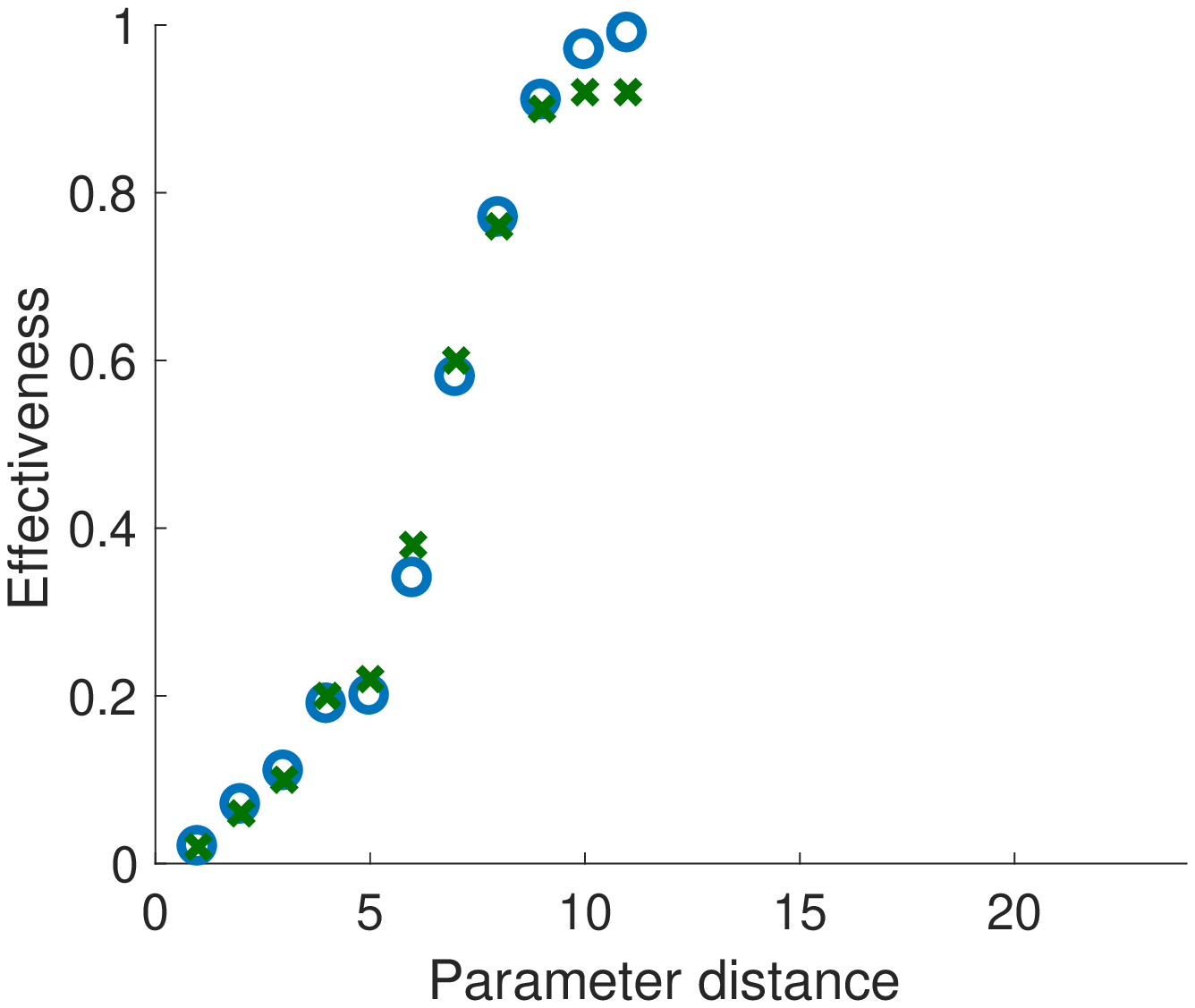}} \hfill 
\subfloat[Cond. 17, validation score -0.004]{\includegraphics[width = .25\textwidth]{img/fronts/sigs_100_cond_17_Pareto_preciseBounds.eps}} \\ 
\subfloat[Cond. 18, validation score -0.009]{\includegraphics[width = .25\textwidth]{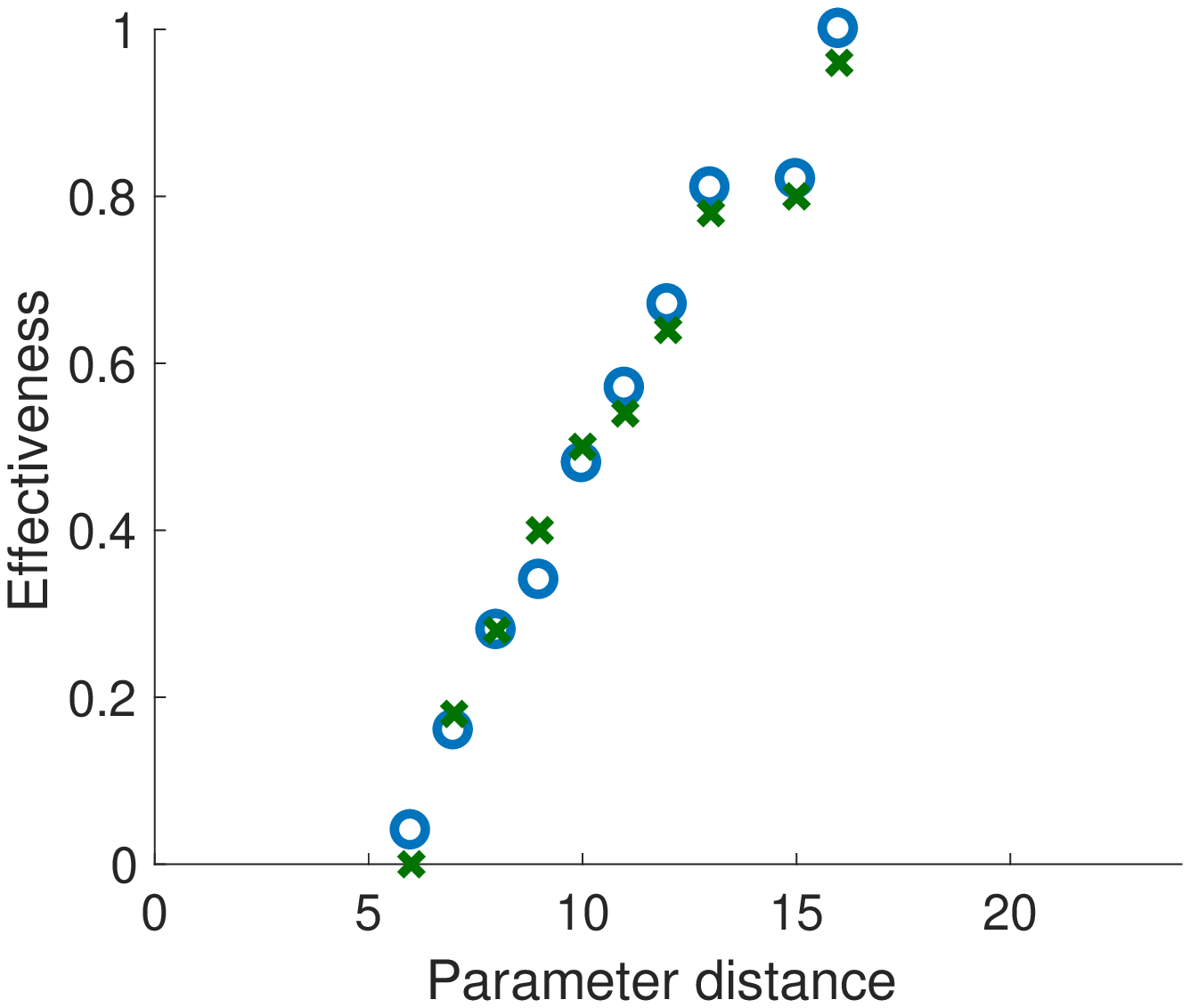}} \hspace{1cm}
\subfloat[Cond. 19, validation score -0.02]{\includegraphics[width = .25\textwidth]{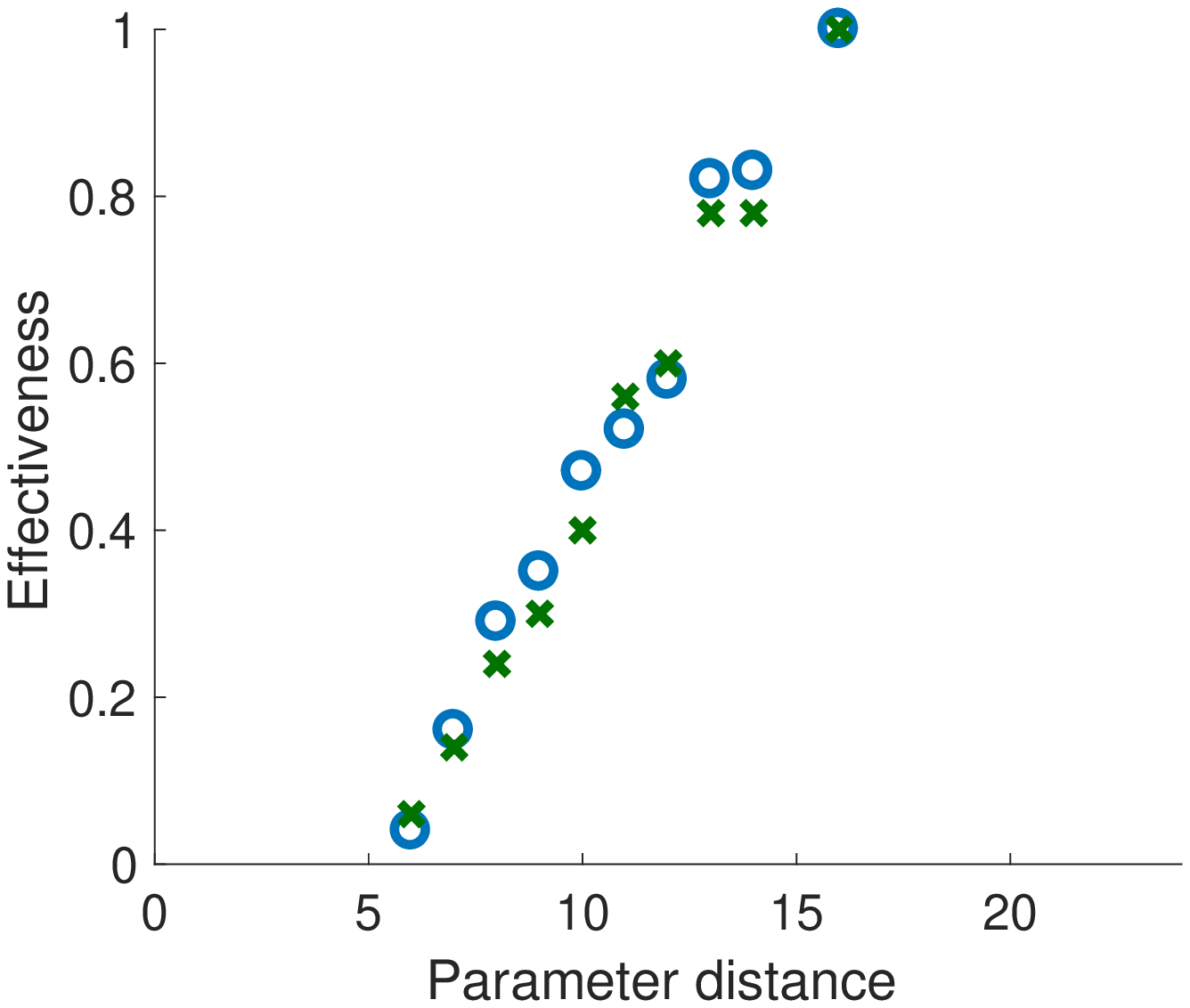}}
\caption{Pareto fronts for condition-specific reprogramming attacks. Blue dots represent the Pareto front obtained with training signals. Green crosses indicate the performance of the synthesized parameters on the test signals. Front for condition 9 is omitted as it only contains the trivial solution (0,0).}
\label{fig:app_pareto_fronts}
\end{figure*}

\begin{table}
\centering
\begin{tabular}{c|cc|cc}
& \multicolumn{2}{c|}{\textbf{Training EGMs}} & \multicolumn{2}{c}{\textbf{Test EGMs}}\\
\textbf{Cond.} & \textbf{OMT} & \textbf{RS} & \textbf{OMT} & \textbf{RS}\\ \hline
1 & \textbf{6.935} & 6.775 & \textbf{6.3} & 6.260\\ 
2 & \textbf{7.680} & 7.260 & 6.520 & 6.520\\ 
3 & \textbf{17.455} & 17.325 & \textbf{17.41} & 17.16\\ 
4 & \textbf{15.21} & 13.595 & \textbf{15.23 }& 14.28\\ 
5 & \textbf{14.68} & 14.135 & \textbf{13.58} & 13.26\\ 
6 & \textbf{8.060} & 7.675 & \textbf{8.790} & 8.770\\ 
7 & \textbf{14.61} & 14.30 & \textbf{13.72} & 13.54\\ 
8 & \textbf{6.570} & 6.045 & 4.860 & \textbf{4.910}\\ 
10 & \textbf{17.71} & 17.605 & \textbf{17.14} & 16.67\\ 
11 & \textbf{0.8350} & 0.7450 & 0.25 & 0.25\\ 
12 & \textbf{8.745} & 7.580 & \textbf{9.010} & 8.930\\ 
13 & \textbf{1.585} & 1.575 & 1.010 & 1.010\\ 
14 &\textbf{ 1.350} & 1.240 & 1.440 & \textbf{1.460}\\ 
15 & \textbf{9.940} & 9.9 & 10.34 & 10.34\\ 
16 & \textbf{17.525} & 17.305 & \textbf{16.58} & 16.24\\ 
17 & \textbf{13.415} & 13.385 & \textbf{13.32} & 13.24\\ 
18 & 13.585 & 13.585 & 13.07 & \textbf{13.24}\\ 
19 & 13.575 & 13.575 & \textbf{13.40} & 13.37\\ \hline
\end{tabular}
\caption{Area under the curve (AUC) of Pareto fronts synthesized by OMT and random search (RS) evaluated on training and test data. The best value between OMT and RS is highlighted in bold.}
\end{table}


\begin{figure}
\centering
\includegraphics[width=.8\columnwidth]{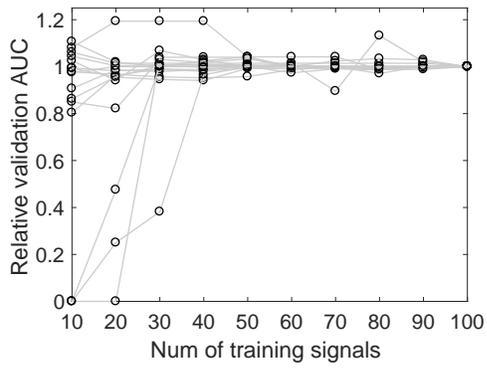}
\caption{Area under the curve (AUC) of Pareto fronts synthesized by OMT with different sizes of the training set, and evaluated on test signals. Each line represents the AUCs of a specific condition. Reported AUC values are relative to the AUC obtained with 100 training signals.}
\end{figure}

\end{document}